\documentclass[12pt]{article}
\usepackage{fullpage}
\usepackage{amsmath,amssymb,amsthm,amsfonts}
\usepackage{hyperref}
\usepackage{stmaryrd}
\SetSymbolFont{stmry}{bold}{U}{stmry}{m}{n}
\usepackage{subcaption}
\usepackage[style=alphabetic,backref,giveninits,maxbibnames=99,doi=true,isbn=false,url=true]{biblatex}
\usepackage{parskip}
\usepackage{geometry}
\usepackage{tikz}
\usepackage{mathtools}
\mathtoolsset{showonlyrefs}

\numberwithin{equation}{section}

\renewcommand\hat\widehat
\renewcommand\tilde\widetilde

\newcommand\be{\begin{equation}}
\newcommand\ee{\end{equation}}
\newcommand\ii{\mathrm{i}}
\newcommand\dif{\mathop{}\!\mathrm{d}}
\newcommand\eu{\mathrm{e}}
\newcommand\QJK{\mathrm{QJK}}
\newcommand\disk{\mathrm{disk}}
\newcommand\e{\epsilon}
\newcommand\BR{\mathbb{R}}
\newcommand\BP{\mathbb{P}}
\newcommand\BC{\mathbb{C}}
\newcommand\BZ{\mathbb{Z}}
\newcommand\BQ{\mathbb{Q}}
\newcommand\cF{\mathcal{F}}
\newcommand\cG{\mathcal{G}}
\newcommand\cH{\mathcal{H}}
\newcommand\cK{\mathcal{K}}
\newcommand\eGLSM{{$\e$-GLSM}}
\newcommand\vol{\mathrm{vol}}
\newcommand\XId{X_1}
\newcommand\XIId{X_2}
\newcommand\XIIId[1][]{X_3^{#1}}
\newcommand\XIVd{X_4}
\newcommand\XVd{X_5}
\newcommand\ie{\textit{i.e.}}
\newcommand\eg{\textit{e.g.}}
\renewcommand\Re{\mathrm{Re}}

\newcommand\Beta{\mathrm{B}}

\DeclareMathOperator{\cD}{\mathcal{D}}
\DeclareMathOperator{\Ga}{\Gamma}
\DeclareMathOperator{\Li}{Li}

\DeclareMathOperator\PF{\mathcal{L}}

\newtheorem{theorem}{Theorem}[section]

\newtheorem{remark}[theorem]{Remark}
\theoremstyle{definition}

\newcommand\preprint[1]{
 \hfill{\small\scshape #1}\par
 \vspace{25pt}
}

\begin{document}
\preprint{UUITP-22/24}

\begin{center} \Large
{\bf Symplectic cuts and open/closed strings II} \\[12mm] \normalsize
{\bf Luca Cassia${}^{a}$, Pietro Longhi${}^{b,c,d}$ and Maxim Zabzine${}^{b,c}$} \\[8mm]
{\small\textit{${}^a$School of Mathematics and Statistics, The University of Melbourne,}\\ \textit{Parkville, VIC, 3010, Australia}\\
e-mail: \ttfamily{luca.cassia@unimelb.edu.au}\\}
{\small\it ${}^b$Department of Physics and Astronomy, Uppsala University,\\ Box 516, SE-75120 Uppsala, Sweden\\}
{\small\it ${}^c$Centre for Geometry and Physics, Uppsala University,\\
Box 480, SE-75106 Uppsala, Sweden\\}
{\small\it ${}^d$Department of Mathematics, Uppsala University,\\
Box 480, SE-75106 Uppsala, Sweden\\}
\end{center}
\vspace{25pt}

\begin{abstract}
In \cite{Cassia:2023uvd} we established a connection between symplectic cuts of Calabi--Yau threefolds and open topological strings, and used this to introduce an equivariant deformation of the disk potential of toric branes.
In this paper we establish a connection to higher-dimensional Calabi--Yau geometries by showing that the equivariant disk potential arises as an equivariant period of certain Calabi--Yau fourfolds and fivefolds, which encode moduli spaces of one and two symplectic cuts (the maximal case) by a construction of Braverman \cite{Braverman+1999+85+98}.
Extended Picard--Fuchs equations for toric branes, capturing dependence on both open and closed string moduli, are derived from a suitable limit of the equivariant quantum cohomology rings of the higher Calabi--Yau geometries.
\end{abstract}

\eject

\tableofcontents

\section{Introduction}

The computation of disk potentials of toric branes by means of $B$-model chain integrals is a fundamental example of open-string mirror symmetry \cite{Aganagic:2000gs}, which motivated and contributed to many developments on counts of open curves with Lagrangian boundaries, including higher-genus open-Gromov--Witten theory, refined topological strings, homological invariants, skeins on branes, and much more.

Recently, a deformation of the disk potential by a collection of equivariant parameters $\e_i$ was proposed \cite{Cassia:2023uvd}.
The definition of \emph{equivariant} disk potentials hinges on a connection between toric branes and the operation of symplectic cutting \cite{lerman1995symplectic}.
The cut of a Calabi--Yau threefold $\XIIId$ is a singular geometry composed of two half spaces glued along a common divisor $\XIId$, which in our setting is a Calabi--Yau twofold
\be\label{eq:symplectic-cut-intro}
 \XIIId\quad\rightsquigarrow\quad\XIIId[<]\cup_{\XIId}\XIIId[>]\,.
\ee
The quantization of the cut is defined by an equivariant Gauged Linear Sigma Model (\eGLSM{}) with target $\XIId$,
whose quantum volume $\cH^D$ defines the equivariant disk potential via
\be\label{eq:W-disk-eq-intro}
 W(t,c,\e) = \frac{1}{2\pi\ii} \int^{c+\pi\ii}_{c-\pi\ii} \cH^D (t,c',\e) \dif c' \,,
\ee
where $t$ collectively refers to the K\"ahler moduli of $\XIIId$ while $c$ corresponds to the additional K\"ahler modulus associated to the cut which leads to $\XIId$. In the l.h.s.\ of this equation, $c$ enters the equivariant disk potential as the open-string modulus.
The standard non-equivariant disk potential is recovered from the small $\e_i$ expansion of $W(t,c,\e)$.

In this paper, we build on the relation between toric branes and symplectic cuts to uncover new connections between equivariant disk potentials and higher dimensional Calabi--Yau geometries.
A description of symplectic cuts developed by Braverman \cite{Braverman+1999+85+98}
involves a family of CY threefolds parameterized by the open-string moduli,
which features the singular manifold \eqref{eq:symplectic-cut-intro} as a distinguished fiber.
For the case of a single toric brane, corresponding to a single cut, Braverman's construction gives rise to a Calabi--Yau fourfold $\XIVd$, while for two branes it gives rise to a Calabi--Yau fivefold $\XVd$.

For a single toric brane, the half-spaces defined by the cut \eqref{eq:symplectic-cut-intro} are found to descend from distinguished divisors $D_\pm$ in $\XIVd$
\be
	\begin{array}{c}
	\text{CY3 with toric brane}\\
	(\XIIId,L)
	\end{array}
	\quad\leftrightarrow \quad
	\begin{array}{c}
	\text{symplectic cut}\\
	\XIIId[<] \cup_{\XIId} \XIIId[>]
	\end{array}
	\quad\leftrightarrow \quad
	\begin{array}{c}
	\text{CY4 divisors}\\
	\XIVd, D_\pm
	\end{array}
\ee
The equivariant disk potential coincides with the equivariant period of the mutual intersection of these divisors
\be
\begin{array}{c}
	\text{disk potential} \\
	W_{\disk}(\XIIId,L)
\end{array}
\quad\leftarrow\quad
\begin{array}{c}
	\text{equivariant disk potential} \\
	W(t,c,\e)
\end{array}
\quad\leftarrow\quad
\begin{array}{c}
	\text{4d equivariant period} \\
	\Pi(D_+\cap D_-)
\end{array}
\ee
where arrows correspond to turning off some of the equivariant parameters.

A pair of toric branes in $\XIIId$ is encoded by a pair of symplectic cuts.
Braverman's construction provides a description in terms of a Calabi--Yau fivefold $\XVd$, presented as a fibration of CY3's over the parameter space of the two branes.
The fivefold geometry encodes the equivariant disk potentials of both branes, in addition to the equivariant quantum cohomology ring of $\XIIId$ itself. We therefore find a hierarchy of CY geometries
\be
  \XIIId\quad\to\quad\XIVd\quad\to\quad\XVd
\ee
where the arrows represent inclusions of the generic fibers.
The corresponding GLSM's also form a hierarchy where the arrows, loosely speaking, correspond to successive symplectic quotients (or \emph{gaugings}).
In the case of two branes, the divisor $\XIId$ of $\XIIId$ is replaced by a CY1 $\XId$ (arising as the intersection of two such divisors, see Figure \ref{fig:double-cut-intro}), whose quantum volume $\cK^D$ generalizes the role of $\cH^D$ in \eqref{eq:W-disk-eq-intro}, by encoding $\cH^D$ for each of the underlying cuts
\be
	\cH_1^D(t,c_1,\e) = \int \cK^D(t,c_1,c_2,\e) \, \dif c_2\,,
	\qquad
	\cH_2^D(t,c_2,\e) = \int \cK^D(t,c_1,c_2,\e) \, \dif c_1\,,
\ee
and therefore encoding the equivariant disk potentials of both branes at once.

We therefore have the following hierarchy of quantum cuts and higher-dimensional CY geometries
\be
\begin{array}{rcl}
	(\XIIId, L^{(1)}) \ \  \to\ \  \text{CY4} \ \XIVd^{(1)} \ \  & & \\
	& \searrow & \\
	& & \text{CY5} \ \XVd\\
	& \nearrow & \\
	(\XIIId, L^{(2)}) \ \  \to\ \  \text{CY4} \ \XIVd^{(2)}  \ \  & & \\
\end{array}\ee

Both in the case of one and of two branes, the higher-dimensional Calabi--Yau geometry
elegantly encodes a set of extended (equivariant) Picard--Fuchs equations
that capture the dependence of equivariant disk potentials on open and closed moduli.%
\footnote{A connection between equivariant enumerative geometry of CY fivefolds and CY threefolds was recently established from a seemingly different perspective in \cite{Brini:2024gtn}.}

We now summarize the main results of this paper in more detail.

\subsubsection*{Quantum cuts and mirror curves}
We provide a derivation of the connection between symplectic cuts and disk potentials of toric branes observed in our previous work \cite{Cassia:2023uvd}. More precisely, we show that the quantum Lebesgue measure $\cH^D$ admits a closed form expression in terms of Lauricella's $D$-type hypergeometric function
\begin{multline}
\label{eq:cHD-FD-intro}
 \cH^D (t,c,\e)
 = \Ga\Big(\sum_i\e_i\Big)\,\eu^{-t\cdot M\cdot\e}\, x^{A_1\cdot\e}
 \Beta\Big(A_2\cdot\e,k\sum_i\e_i-A_2\cdot\e\Big)
 \left(\prod_{j=1}^k (-y_j)^{-\sum_i\e_i} \right) \\
 \times F^{(k)}_D\Big(A_2\cdot\e,\sum_i\e_i,\dots,\sum_i\e_i,k\sum_i\e_i;
 1+y_1^{-1},\dots,1+y_k^{-1}\Big)
\end{multline}
whose arguments are given by the sheets $y_j\equiv y_j(x)$ of the mirror curve of the toric brane, in the sense of \cite{Aganagic:2013jpa}, where $x=\eu^{-c}$.
The $y_j$ are directly related to the usual non-equivariant disk potential of the brane by the Abel--Jacobi map \cite{Aganagic:2000gs}, namely $W_{\disk} =\int \log y(x) \dif\log x$.
Using this direct correspondence between $\cH^D$ and $W_\disk$ we obtain an exact relation between the monodromy of $\cH^D$ and the disk potential, which holds for all toric Calabi--Yau threefolds.

\subsubsection*{Connections to Calabi--Yau fourfolds}
It was observed in \cite{Braverman+1999+85+98} that there exists a complex manifold $\XIVd$
which contains the symplectic cut \eqref{eq:symplectic-cut-intro} of $\XIIId$ as a complex codimension-one subspace.
The manifold $\XIVd$ is defined by the symplectic quotient
\be
	\XIVd = (\XIIId\times \BC_+ \times \BC_-)\sslash U(1)\,,
\ee
with respect to a canonical extension of the moment map from $\XIIId$ to $\XIIId\times \BC_+ \times \BC_-$. The resulting space $\XIVd$ can be regarded as a fibration over $\BC$, such that the generic fiber is complex isomorphic to the original space $\XIIId$, while the fiber over the origin is complex isomorphic to the singular (reducible) space \eqref{eq:symplectic-cut-intro}.
If we require that the moment map used to define the symplectic cut satisfies the Calabi--Yau condition, then it follows that $\XIVd$ is itself a CY manifold.

We study a quantization of the equivariant volume of the fourfold, defined by the hemisphere partition function of the \eGLSM{} with target $\XIVd$, and find that the equivariant disk potential \eqref{eq:W-disk-eq-intro} admits a natural uplift as an equivariant period associated to $\XIVd$, which recovers the former in the limit where two additional equivariant parameters vanish (\ie{} those associated to the homogeneous coordinates on $\BC_\pm$).
An important property of the four-dimensional uplift is that it arises as the equivariant period associated with an intersection of toric divisors of $\XIVd$, implying that it obeys the fourfold's equivariant Picard--Fuchs (PF) equations.
In the limit $\e_\pm \to 0$, these reduce to \emph{extended} Picard--Fuchs equations for $W(t,c,\e)$ and for $\cF_{\lessgtr}(t,c,\e)$ in the original CY3, with dependence on both open and closed string moduli.
In the fully non-equivariant limit $\e_i\to0$, in which $W(t,c,\e)$ reduces to the standard disk potential $W_\disk$, we show that the extended Picard--Fuchs equations become inhomogeneous.

The relation between symplectic cuts and CY4 makes contact with earlier results in the literature on several fronts. In particular, the fact that disk potentials obey inhomogeneous Picard--Fuchs equations was already observed in the case of the real quintic \cite{Walcher:2006rs, Walcher:2008gho, Morrison:2007bm}. Moreover, a relation between open strings and periods of Calabi--Yau fourfolds was observed in \cite{Mayr:2001xk, Lerche:2001cw}. Our results provide a derivation of this correspondence, and generalize it to the equivariant setting.

\begin{figure}[!ht]
\centering
\includegraphics[width=.4\textwidth]{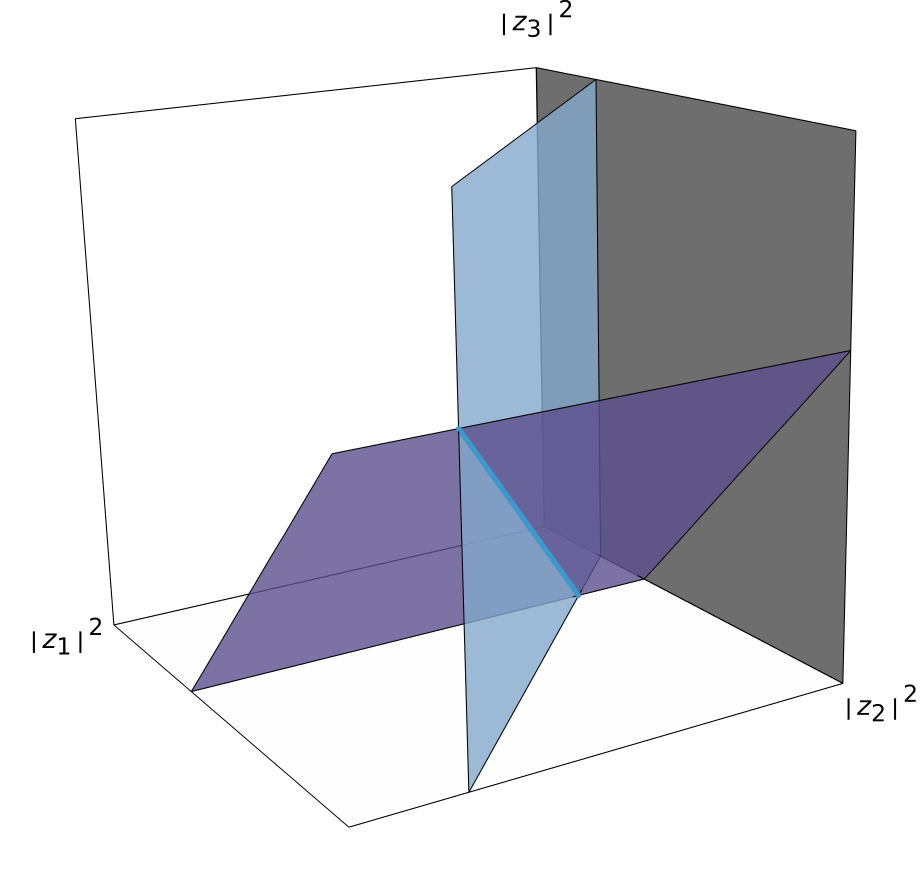}
\caption{Hyperplanes for a double symplectic cut of $\BC^3$.}
\label{fig:double-cut-intro}
\end{figure}

\subsubsection*{Double cuts and Calabi--Yau fivefolds}
Finally, we consider the possibility of performing multiple symplectic cuts on $\XIIId$. We assume that for each cut, the intersection of the two half-spaces, $\XIIId[<]$ and $\XIIId[>]$, forms a Calabi--Yau manifold. This corresponds to requiring that the hyperplane associated with the cut contains an affine line parallel to the vector $(1,1,\dots,1)$ in the base of the toric fibration. Under this condition, Braverman's construction ensures that each resulting fourfold remains Calabi--Yau.
We refer to such symplectic cuts as ``\emph{Calabi--Yau type cuts}'', or simply ``\emph{Calabi--Yau cuts}''.

If, furthermore, the hyperplanes are in generic position\footnote{This excludes parallel, overlapping hyperplanes, which would lead to quotients by trivially acting $U(1)$ groups and thus to Artin stacks that are not of Deligne--Mumford type.} with nontrivial intersection, we deduce that at most two CY symplectic cuts can be performed simultaneously. Moreover, the intersection of all resulting half-spaces defines a complex one-dimensional submanifold in~$\XIIId$.\footnote{For a general CY $d$-fold $X_d$, the maximum number of CY cuts would be $d-1$. An additional cut along one more hyperplane would reduce the intersection to a zero-dimensional locus, but this final cut would necessarily break the Calabi--Yau condition.}

A double cut of $\XIIId$ then consists of the singular space made up of four top-dimensional strata $\XIIId[\gtrless,\gtrless]$ glued along common divisors, see Figure \ref{fig:double-cut-intro}.
The four divisors are further glued along a common divisor, which is a Calabi--Yau onefold $\XId$.
The quantum volume of $\XId$, which we denote as $\cK^D(t,c_1,c_2,\e)$, is defined by the \eGLSM{} partition function with target $\XId$.
Here $c_1, c_2$ parameterize the location of the two cuts, and correspond to open string moduli for two toric Lagrangians in $\XIIId$.
We find that the quantum volume $\cK^D$ admits a universal expression
\be\label{eq:KD-B-int-intro}
\begin{aligned}
 \cK^D(t,c_1,c_2,\e)
 & = \Ga\Big(\sum_i\e_i\Big)\, \eu^{-t\cdot M\cdot\e}
 \frac{x_1^{A_1\cdot\e} x_2^{A_2\cdot\e}}
 {H(x_1,x_2,z)^{\sum_i\e_i}}\,,
\end{aligned}
\ee
where $H(x_1,x_2,z)$ is a Laurent polynomial in the exponentiated variables $x_\alpha=\eu^{-c_\alpha}$, $z_a=\eu^{-t^a}$ which is closely related to the Hori--Vafa mirror curve of $\XIIId$, and $A_\alpha\cdot\e$ and $t\cdot M\cdot\e$ are certain linear combinations of the equivariant parameters $\e_i$ (and K\"ahler moduli $t^a$), which we define explicitly in Appendix~\ref{sec:B-model-stuff}.
It turns out that $\cK^D$ is the most fundamental among quantum volumes, as it encodes both quantum Lebesgue measures $\cH^D_\alpha$ with $\alpha=1,2$ associated to the two symplectic cuts, as well as the quantum volume of $\XIIId$ itself.

Similarly to the construction of Braverman \cite{Braverman+1999+85+98}, the double cut can be realized as a codimension-two submanifold inside a two-parameter family of spaces isomorphic to $\XIIId$. The total space of this fibration is defined as a fivefold $\XVd$ together with a projection to $\BC^2$ such that the generic fiber is $\XIIId$, the fiber over either coordinate axis is isomorphic to one of the two CY4 associated to a single cut, and the fiber over the origin is isomorphic to the singular double cut space.
The space $\XVd$ is CY only in the case when both cuts are CY as described in the beginning of this section.

Once again, we observe that equivariant volumes of the top strata of the double cut, $\cF^D_{\gtrless,\gtrless}$ descend from equivariant periods of divisors of the CY5. We use this fact to derive extended Picard--Fuchs equations for the quantum double symplectic cut of $\XIIId$. See the main text for details.

\subsubsection*{Organization of the paper}

In Section~\ref{sec:quantum-cuts}, we review the definition of quantum symplectic cuts and their relation to disk potentials for toric branes. We then give a general formula for the monodromy of the quantum Lebesgue measure, and show that under suitable assumptions its regular part reproduces the prediction of open string mirror symmetry for the disk potential.
The Calabi--Yau fourfold description of symplectic cuts is discussed in Section~\ref{sec:CY4}. Here we also discuss the \eGLSM{} quantization of the fourfold and how it leads to extended Picard--Fuchs equations for quantum cuts of the CY3.
In Section~\ref{sec:double-cut}, we discuss double cuts, providing a description for these and for their quantization in terms of a Calabi--Yau fivefold geometry.
Section~\ref{sec:conclusions} contains a summary of the main results and concluding remarks.

\subsubsection*{Acknowledgements}

It is a pleasure to thank Ga\"etan Borot, Andrea Brini and Davide Scazzuso for illuminating discussions.
We also extend our gratitude to the organizers of the \emph{BPS Dynamics and Quantum Mathematics} workshop, held at GGI, where part of this work was conducted.
L.C.\ thanks Johanna Knapp, Joseph McGovern and Emanuel Scheidegger for helpful conversations on the matters of this paper.
L.C.\ also thanks the organizers of \emph{The Geometry of Moduli Spaces in String Theory} scientific program, hosted by the MATRIX Institute, where this work was completed.
We would also like to thank both referees for their careful reading of the manuscript, and especially for the constructive and insightful comments.

The work of L.C.\ was supported by the ARC Discovery Grant DP210103081.
The work of P.\ L.\ is supported by the Knut and Alice Wallenberg Foundation grant KAW2020.0307.
The research of M.\ Z.\ is supported by the VR excellence center grant ``Geometry and Physics'' 2022-06593.

\subsubsection*{Data Availability}

Data sharing not applicable to this article as no datasets were generated or analyzed during the current study.

\subsubsection*{Conflict of Interest}

The authors have no conflict of interest to declare.

\section{Holomorphic disks and quantum symplectic cuts}\label{sec:quantum-cuts}

\subsection{Toric branes, symplectic cuts and equivariance}\label{sec:review}

In \cite{Cassia:2023uvd}, we formulated a proposal for modeling $A$-branes in the framework of equivariant gauged linear sigma models (henceforth \eGLSM) developed in the earlier work by two of the authors \cite{Cassia:2022lfj}.
As argued in \cite{Witten:1993yc}, the nonlinear sigma model with target a toric Calabi--Yau threefold $\XIIId$ can be studied by considering instead a gauged linear sigma model for the symplectic quotient $\BC^{r+3}\sslash U(1)^r$.
Let $Q^a_i\in\BZ$, with $a=1,\dots, r$ and $i=1,\dots, r+3$, denote the matrix of charges for the $U(1)^r$ action on $\BC^{r+3}$. The symplectic quotient is defined by
\be\label{eq:CY3-symp-quot}
 \XIIId=\BC^{r+3}\sslash U(1)^r
 = \left\{ \sum_{i=1}^{r+3} Q^{a}_i |z_i|^2 = t^a \right\} \Big/ U(1)^{r}
\ee
for a regular choice of K\"ahler moduli $t^{a}$.
The \eGLSM{} partition function on the disk with a space-filling brane \cite{Hori:2013ika} is an integral over Coulomb branch moduli
\be\label{eq:disk-general-contour}
    \cF^D(t,\e,\lambda) = \lambda^{-r-3} \oint_{\QJK}
    \prod_{a=1}^r \frac{\dif\phi_a}{2\pi\ii}
    \,\eu^{\sum_{a=1}^r \phi_a t^a} \prod_{i=1}^{r+3}
    \Ga\Big( \frac{\e_i+\sum_{a=1}^r\phi_a Q^a_i}{\lambda} \Big)\,,
\ee
where $\lambda^{-1}$ is an equivariant parameter for the disk, and the integrand is the product of exponentiated Fayet--Ilioupoulos couplings (corresponding to K\"ahler moduli) and one-loop determinants of massive charged chiral multiplets. Here the contour of integration is a quantum deformation of the Jeffrey--Kirwan (JK) residue prescription compatible with a given choice of chamber in the extended K\"ahler cone. See also \cite{Bonelli:2013mma} for the analogous partition function in the case of a \eGLSM{} on the two-sphere.

The classical limit of the partition function, defined by taking $\lambda\to \infty$, computes the equivariant volume of the symplectic quotient
\be
 \lim_{\lambda\to\infty} \cF^D(t,\e,\lambda) = \int_{\XIIId}\eu^{\omega-H_\e}
 =: \vol_\e(\XIIId)\,.
\ee
where $\omega$ is the induced symplectic form on the quotient $\XIIId$ and $H_\e$ is the Hamiltonian for the $U(1)^{r+3}$-action.
At finite values of $\lambda$, the partition function therefore computes a notion of \emph{quantum volume} of the Calabi--Yau threefold $\XIIId$.

The disk function $\cF^D$ is homogeneous with respect to an overall rescaling of its parameters
\be
\label{eq:homogeneity}
 \cF^D(\xi^{-1}t,\xi\e,\xi\lambda) = \xi^{-3}\cF^D(t,\e,\lambda)
\ee
for any $\xi\in\BC^\times$.
We can therefore use this property to rescale away the dependence on the parameter $\lambda$. Moreover, in the limit $\xi\to1$ this implies the differential equation
\be
 \left( \sum_{a=1}^r t^a\frac{\partial}{\partial t^a}
 - \sum_{i=1}^{r+3} \e_i\frac{\partial}{\partial\e_i}
 -\lambda\frac{\partial}{\partial\lambda}
 -3
 \right) \cF^D(t,\e,\lambda) = 0
\ee

From now on, we will assume that the value of $\lambda$ is set to one for convenience and we will drop it from the arguments of the disk function $\cF^D$.

An important property of the partition function $\cF^D$ is that it is a solution to the equivariant Picard--Fuchs equations
\be\label{eq:equiv-PF}
 \left(\prod_{\{i|\sum_a \gamma_a Q^a_i>0\}} \left(\cD_i\right)_{\sum_a \gamma_a Q^a_i}
 - \eu^{-\sum_a \gamma_a t^a}
 \prod_{\{i|\sum_a \gamma_a Q^a_i\leq 0\}} \left(\cD_i\right)_{-\sum_a \gamma_a Q^a_i}
 \right)\cF^D(t,\e) = 0
\ee
where $\cD_i$ are differential operators defined as
\be
\label{eq:divisor-operator}
 \cD_i = \e_i+\sum_{a=1}^rQ^a_i\frac{\partial}{\partial t^a}\,,
\ee
$(z)_n$ is the Pochhammer symbol defined as in \eqref{eq:pochhammer} and $\gamma_a\in\BZ$ such that $\sum_{a=1}^r\gamma_at^a\geq0$ for any value of K\"ahler moduli $t^a$ within the chosen chamber. These equations arise as Ward identities for the integral representation in \eqref{eq:disk-general-contour} and they give a representation of the equivariant quantum cohomology relations for the target $\XIIId$.

Solutions of the non-equivariant PF equations are referred to as ``\emph{periods}'' due to the fact that by mirror symmetry they can be regarded as periods of the top holomorphic form of the mirror CY. In the equivariant setting, we refer to the solutions of \eqref{eq:equiv-PF} as ``\emph{equivariant periods}'' even though the precise details of the statement of mirror symmetry are yet to be worked out. An explicit basis of such solutions, is labeled by the torus fixed points in the target $\XIIId$ as discussed in \cite{Cassia:2022lfj}, however certain special linear combinations of the basis elements appear naturally by applying difference operators in the K\"ahler moduli on the quantum volume $\cF^D$. For each toric divisor in $\XIIId$, we define the difference operator
\be
\label{eq:difference-operators}
 \frac{1-\eu^{2\pi\ii\cD_i}}{2\pi\ii}\,,
\ee
where the exponential of the differential operator $\cD_i$ acts as a finite shift in the variables $t^a$,
\be
 \eu^{2\pi\ii\cD_i}f(t^a) = \eu^{2\pi\ii\e_i}f(t^a + 2\pi\ii Q_i^a)\,.
\ee
We will then define the equivariant period associated to the $i$-th toric divisor as the function
\be
\label{eq:equiv-period}
 \Pi(D_i) := \frac{1-\eu^{2\pi\ii\cD_i}}{2\pi\ii} \cdot \cF^D(t,\e)
\ee
and similarly, we can define equivariant periods for intersections of multiple divisors by applying multiple difference operators at once. The motivation for such a definition stems from the fact that periods (both equivariant and not) are fully determined by their semiclassical behavior \cite{Cassia:2022lfj}, and in the case of the function $\Pi(D_i)$, it is straightforward to show that the semiclassical behavior is that of the equivariant volume of the divisor $D_i$. In particular, one can show that the difference operator \eqref{eq:difference-operators} satisfies two crucial properties. Firstly, it commutes with the PF operators in \eqref{eq:equiv-PF}, which means that acting with it on a solution automatically gives back another solution. Secondly, it can be expanded as a power series in $\cD_i$ with vanishing constant term, which implies that it annihilates all the basis elements associated to fixed points which are not in the divisor $D_i$.

The disk function $\cF^D$ should not be confused with the Gromov--Witten free energy. Rather, it computes the equivariant period (or central charge) of a space-filling brane, which is equivalent to imposing Neumann boundary conditions for all the chiral fields of the 2d gauge theory.
Equivariant periods associated to other types of $B$-branes can be obtained by a similar \eGLSM{} computation, by imposing a suitable combination of Neumann and Dirichlet boundary conditions on the fields \cite{Hori:2013ika, Cassia:2022lfj}.
As in \cite{Cassia:2023uvd}, this paper will be devoted to the problem of modeling $A$-branes instead.\footnote{See also \cite{Govindarajan:2000ef, Govindarajan:2001zk, Katz:2001vm, Graber:2001dw, Brini:2010sw, Brini:2011ij, Brini:2013zsa, Brini:2014fea} for other approaches to this question.
}

At weak string coupling, $A$-branes admit a geometric description involving a special Lagrangian submanifold of the Calabi--Yau threefold together with an Abelian local system.
We will focus on a class of special Lagrangian manifolds that can be defined in any toric Calabi--Yau threefold, corresponding to certain resolutions of Harvey--Lawson cones over $T^2$ \cite{harvey1982calibrated} known as toric branes in physics \cite{Aganagic:2000gs}.

Let $(|z_i|,\theta_i = \arg z_i)$ denote local coordinates for the ambient space $\BC^{r+3}$, regarded as a $T^{r+3}$ torus fibration over $\BR^{r+3}_{\geq 0}$.
In the base, we consider the intersection of two hyperplanes $h^1$ and $h^2$ defined by the solutions to two moment map equations
\be
\label{eq:toric-brane-hyperplanes}
     h^1:\ \sum_{i=1}^{r+3} q^1_i |z_i|^2 = c\,,
     \qquad
     h^2:\ \sum_{i=1}^{r+3} q^2_i |z_i|^2 = 0\,.
\ee
where $q^\alpha_i\in\BZ$ with $i=1,\dots,r+3$ and $\alpha=1,2$, are the charges for the two $U(1)$-actions.
Due to the Calabi--Yau condition, \ie{} $\sum_i Q^a_i = \sum_i q^\alpha_i = 0$, the equations \eqref{eq:toric-brane-hyperplanes} and the moment map condition in \eqref{eq:CY3-symp-quot} define an affine line with slope $(1,1,\dots,1)$ in $\BR^{r+3}_{\geq 0}$. A special Lagrangian is then obtained by fibering the dual torus defined by $\theta_1+\dots+\theta_{r+3}=0$.
This descends to a special Lagrangian $L$ in the symplectic quotient \eqref{eq:CY3-symp-quot}, and is the total space of a $T^2$-fibration over an affine line in the three-dimensional moment polytope (\ie{} the intersection of $\BR^{r+3}_{\geq 0}$ with the solutions of the moment map constraints associated to $Q^a_i$).

By construction, the special Lagrangian is invariant under a redefinition of the charges $q^1_i \mapsto q^1_i + f q^2_i$ for arbitrary $f\in \BZ$.
However, the definition of the $A$-brane also involves an Abelian local system, which is sensitive to these shifts.
At the quantum level this leads to a discrete degree of freedom for the $A$-brane, known as framing \cite{Aganagic:2001nx}.
In a given choice of framing, a toric brane therefore defines two codimension-one hyperplanes $h^1$, $h^2$ inside of the moment polytope of $\XIIId$.

\bigskip

The two hyperplanes associated to a framed toric brane are not on the same footing: while $h^1$ carries information about the open string modulus $c$, the hyperplane $h^2$ parametrizes the framing shift ambiguity.
Following \cite{Cassia:2023uvd}, let us introduce a Calabi--Yau twofold $\XIId$ defined by the symplectic quotient $\BC^{r+3}\sslash U(1)^{r+1}$ with charge matrix $Q^a_i$ augmented by $q^1_i$
\be\label{eq:CY2-symp-quot}
    \XIId = \left\{ \sum_{i=1}^{r+3} Q^{a}_i |z_i|^2 = t^a,\
    \sum_{i=1}^{r+3} q^{1}_i |z_i|^2 = c\right\} \Big/ U(1)^{r+1}
\ee
This manifold determines an operation on the original Calabi--Yau threefold $\XIIId$ known as symplectic cut~\cite{lerman1995symplectic}. See \cite{Cassia:2023uvd} and Section~\ref{sec:Braverman} for details.
The symplectic cut of $\XIIId$ along $\XIId$ is the connected sum
\be\label{eq:symplectic-cut}
 \XIIId[<] \cup_{\XIId} \XIIId[>]
\ee
where the two `half-spaces' $\XIIId[\lessgtr]$ (not necessarily Calabi--Yau) are glued along a common divisor isomorphic to $\XIId$.
There is a simple relation between the equivariant volume of $\XIId$ and that of $\XIIId$
\be\label{eq:classical-slicing}
    \vol_{\XIIId}(t,\e) = \int_{-\infty}^{+\infty} \vol_{\XIId} (t,c,\e) \, \dif c\,.
\ee
where $c$ is the modulus parametrizing the location of the hyperplane $h^1$.
In fact, this can further be refined to
\be
    \vol_{\XIIId}(t,\e) = \vol_{\XIIId[<]}(t,c,\e) + \vol_{\XIIId[>]}(t,c,\e)\,,
\ee
where we define
\be\label{eq:half-vol-int}
 \vol_{\XIIId[<]}(t,c,\e):=\int_{-\infty}^{c} \vol_{\XIId}(t,c',\e) \, \dif c'\,,
 \hspace{30pt}
 \vol_{\XIIId[>]}(t,c,\e):=\int_{c}^{+\infty} \vol_{\XIId}(t,c',\e) \, \dif c'\,.
\ee
In \cite{Cassia:2023uvd}, the quantum symplectic cut of $\XIIId$ was defined by considering the \eGLSM{} for the quotient \eqref{eq:CY2-symp-quot}. The partition function of the \eGLSM{} computes the quantum volume of $\XIId$
\be\label{eq:cHD-def}
    \cH^D(t,c,\e)
    =
    \oint
    \prod_{a=1}^r \frac{\dif\phi_a}{2\pi\ii}
    \oint \frac{\dif\varphi}{2\pi\ii}
    \,\eu^{\sum_a \phi_a t^a + \varphi c} \prod_{i=1}^{r+3}
    \Ga( \e_i+\sum_a\phi_a Q^a_i + \varphi q^1_i )\,.
\ee
for an appropriate choice of contour depending on the chamber.

Remarkably, a generalization of the relation~\eqref{eq:classical-slicing} holds in the full quantum theory. In that setting, we obtain an expression for the quantum volume of the threefold $\XIIId$ as an integral of the quantum volume of $\XIId$ over the open string modulus
\be\label{eq:q-lebesgue-integral}
    \cF^D(t,\e) = \int_{-\infty}^{+\infty} \cH^D(t,c,\e)\, \dif c\,.
\ee
For this reason, $\cH^D$ is known as a `quantum Lebesgue measure' defined by the hyperplane $h^1$.
In fact, also the splitting \eqref{eq:half-vol-int} admits a natural uplift to
\be\label{eq:quantum-half-volumes}
    \cF_<^D(t,c,\e) = \int_{-\infty}^{c} \cH^D(t,c',\e)\, \dif c'\,,
    \qquad
    \cF_>^D(t,c,\e) = \int_{c}^{+\infty} \cH^D(t,c',\e)\, \dif c'\,,
\ee
which add up to \eqref{eq:q-lebesgue-integral}.
Notice however, that while $\cH^D$ is a quantum volume, namely that of the hyperplane $h^1$, the functions $\cF^D_{<}$ and $\cF^D_{>}$ are not quantum volumes of the corresponding half-spaces, in the sense that they do not correspond to the partition functions of the \eGLSM{} with target $\XIIId[<]$ and $\XIIId[>]$, respectively. Nevertheless, it is possible to show that in the classical limit (\ie{} $\lambda\to\infty$) they do reduce to the classical equivariant volumes of the spaces $\XIIId[<]$ and $\XIIId[>]$, just like $\cH^D$ reduces to the equivariant volume of $\XIId$.

The main claim of \cite{Cassia:2023uvd} is that quantum symplectic cuts defined by hyperplanes associated to toric $A$-branes compute the disk potential of the corresponding brane.
To illustrate this claim, we introduce the ``\emph{equivariant disk potential}''
\be\label{eq:equiv-W}
 W(t,c,\e) = \frac{1}{2\pi\ii} \int^{c+\pi\ii}_{c-\pi\ii} \cH^D (t,c',\e) \dif c'  \,.
\ee
By introducing the ``\emph{monodromy operator}'' $\Delta_c$ defined on arbitrary functions $f(c)$ as
\be
 \label{eq:monodromy-operator}
 \Delta_c\, f(c) := \frac1{2\pi\ii}\Big(f(c+\pi\ii) - f(c-\pi\ii)\Big)\,,
\ee
it will also be useful in the following to rewrite the defining relation of $W(t,c,\e)$ as the differential-difference equation
\be\label{eq:equiv-W-deriv}
 \frac{\partial}{\partial c} W(t,c,\e) = \Delta_c \cH^D (t,c,\e) \,.
\ee

\begin{remark}[Conventions]
\label{rmk:c-shift}
The definition of disk potential in \eqref{eq:equiv-W} contains a shift of the open string modulus, compared to the one given in \cite{Cassia:2023uvd}
\be
	W^{\mathrm{(here)}}(t, c, \e) = W^{\text{(\cite{Cassia:2023uvd})}}(t, c-\pi \ii, \e)
\ee
This shift is purely a matter of convention, and we make this choice in this paper because $W(t,c,\e)$ written in this way will obey certain differential equations with a suggestive form (analogous equations can be obtained for the equivariant disk potential of \cite{Cassia:2023uvd} upon changing variables). This is the same exact shift that appeared between conventions of \cite{Aganagic:2000gs} and those of \cite{Aganagic:2001nx}.
\end{remark}

Next we consider an expansion of $W$ at small values of the equivariant parameters. In general the coefficients of the series depend on how one takes such an expansion since this function is not analytic at $\e_i=0$.\footnote{For example $\frac{1}{\e_1+\e_2}$ can be expanded in powers of either of $(\e_1/\e_2)^{\pm 1}$.}
Following \cite{Cassia:2023uvd}, we choose to take a coarse regularization prescription where all $\e_i\to 0$ at the same rate, by substituting $\e_i \mapsto \xi \e_i$ and expanding around $\xi=0$. This prescription leads to a decomposition of the equivariant disk potential into terms of degree $d$ in $\xi$
\be
\label{eq:series-expansion-of-W}
 W(t,c,\xi\e)
 = \sum_{d=-1}^{\infty} \left[W(t,c,\e)\right]_d \, \xi^d\,.
\ee
where, by definition, $\left[W(t,c,\e)\right]_d$ indicates the coefficient of $\xi^d$ in the Laurent expansion of $W(t,c,\xi\e)$ at $\xi=0$.
Focusing on $[W(t,c,\e)]_0$, we find two types of terms:
those with polynomial dependence on $c$, and those with exponential dependence
\be
 [W(t,c,\e)]_0 = \text{polynomial in $c$}
 + \sum_{n>0} w_n(\eu^{-t},\e) \, \eu^{-n\,c} \,.
\ee
The latter represent worldsheet instanton contributions, and in \cite{Cassia:2023uvd} it was shown in several cases that they agree with the disk potential for the toric brane predicted by open-string mirror symmetry \cite{Aganagic:2000gs,Aganagic:2001nx}
\be\label{eq:Wdisk2}
 \sum_{n>0} w_n(\eu^{-t},\e) \, \eu^{-n\,c}
 \propto \sum_{\beta} N_\beta \, \eu^{-\beta\cdot(t,c)}
 \equiv W_\disk(\eu^{-t},\eu^{-c})\,,
\ee
up to an overall constant coefficient.
Each term in $W_\disk$ represents the contribution from holomorphic worldsheet instantons in a certain relative homology class $\beta\in H_2^{\rm rel}(\XIIId,L)$ where $L$ is the toric Lagrangian.
The set of relevant instanton charges depends on the `phase' of the geometry, \ie{} the choice of hyperplanes $(h^1,h^2)$, and on the sign of relevant linear combinations of closed and open K\"ahler moduli $\beta\cdot(t,c) = \sum_{a=1}^r \beta_a t^a + \beta_{r+1} c$.

\subsection{Equivariant disk potential of general toric threefolds}\label{sec:Lauricella}

The computation of $A$-brane disk potentials by means of quantum symplectic cuts reviewed above is supported by nontrivial checks in several examples, and by a physical interpretation advanced in \cite{Cassia:2023uvd}.
In this section, we will argue that \eqref{eq:Wdisk2} is in fact true in general, and therefore quantum symplectic cuts compute disk potentials of toric $A$-branes in arbitrary Calabi--Yau threefolds.
On the one hand, this will provide an explanation for all the observations made so far. On the other hand, it will clarify how the equivariant disk potential \eqref{eq:equiv-W} is related to the disk potential computed via open string mirror symmetry \cite{Aganagic:2000gs,Aganagic:2001nx}.

In order to prove a relation similar to \eqref{eq:Wdisk2} valid for general toric CY3,
we will use \eqref{eq:equiv-W-deriv} to argue that, up to overall constants, the instanton part of the non-equivariant expansion of $\partial_c W(t,c,\e)$ matches with the derivative of the disk potential $W_\disk$. More precisely, we claim that
\be
\label{eq:DeltacH=dcW}
 \Delta_c\left[\cH^D(t,c,\e)\right]_0 =
 \text{polynomial in $c$}
 + \sum_j k_j \, \partial_c W^{(j)}_\disk(\eu^{-t},\eu^{-c})\,.
\ee
where $[\cH^D]_0$ is defined as the order zero term in $\xi$ in the series expansion of $\cH^D(t,c,\xi\e)$, analogously to \eqref{eq:series-expansion-of-W}, while the sum in the r.h.s.\ ranges over different branches of the mirror curve as explained below. Moreover, $k_j$ are certain proportionality constants fixed by the details of the cut.
In specific examples, this more general relation reduces to the simpler relation \eqref{eq:Wdisk2} observed in \cite{Cassia:2023uvd}.

We begin by observing that one can use the integral representation of $\Ga$ functions in the integrand of \eqref{eq:cHD-def}, to write the quantum Lebesgue measure in the following form
\be\label{eq:HD-B-int-main}
\begin{aligned}
 \cH^D(t,c,\e)
 = \Ga\Big(\sum_i\e_i\Big) \,\eu^{-t\cdot M\cdot\e}\,x^{A_1\cdot\e}
 \int_0^\infty \dif y\,
 \frac{y^{A_2\cdot\e-1}}{H(x,y,z)^{\sum_i\e_i}}
\end{aligned}
\ee
where $x=\eu^{-c}$, $z_a=\eu^{-t^a}$ for $a=1,\dots,r$, and the precise definitions of $A_\alpha$ and $M$ depend on the specific details of the geometry, see \eqref{eq:HD-B-int-app} and its derivation in Appendix~\ref{app:technical-identities} for further details.
Here it is important to observe that if we regard $H(x,y,z)=0$ as an equation for the variables $x$ and $y$, the locus of its solutions can be identified with the mirror curve of $\XIIId$ (below we sometimes omit the dependence on the complex moduli $z_a$).
\footnote{
More precisely \eqref{eq:HD-B-int-main} can be written in terms of the mirror curve for the toric brane, starting from its relation to $\cK^D$ given in \eqref{eq:HD-B-int-app} and by implementing a suitable change of the integration variable, see Appendix~\ref{sec:B-model-integrals}. In the present discussion it is implicit that these redefinitions have been implemented, and we treat $H(x,y)$ as the actual mirror curve. This means that $H(x,y)$ might not coincide with the one that appears in an expression for $\cK^D$ describing a double cut, in general.
}

In the integral expression in \eqref{eq:HD-B-int-main} we choose a parametrization of the integrand such that the function $H(x,y,z)$ that appears in the denominator is a monic polynomial in $y$ (with no negative powers of $y$, \ie{}
\be\label{eq:H-factorized-main}
 H(x,y,z) = \prod_{j=1}^k (y-y_j(x,z))\,.
\ee
Here $y_j(x,z)$ are the roots of $H$, regarded as a polynomial of degree $k$ in $y$.
Thanks to this and identity \eqref{eq:Lauricella-y-integral}, the quantum Lebesgue measure can be expressed in terms of Lauricella's hypergeometric function of type $D$
\begin{multline}
\label{eq:cHD-FD}
 \cH^D(t,c,\e)
 = \Ga\Big(\sum_i\e_i\Big)
 \Beta\Big(A_2\cdot\e,k\sum_i\e_i-A_2\cdot\e\Big)
 \frac{\eu^{-t\cdot M\cdot\e}\,x^{A_1\cdot\e}}{H(x,0,z)^{\sum_i\e_i}} \\
 \times F^{(k)}_D\Big(A_2\cdot\e,\sum_i\e_i,\dots,\sum_i\e_i,k\sum_i\e_i;
 1+y_1^{-1},\dots,1+y_k^{-1}\Big)
\end{multline}
where $\Beta(a,b)=\frac{\Ga(a)\Ga(b)}{\Ga(a+b)}$ is Euler's beta function and $H(x,0,z)=\prod_{j=1}^k(-y_j(x))$ is the product of all the roots, up to sign.
The definition of $F^{(k)}_D$ and its most relevant properties are collected in
Appendix~\ref{app:lauricella}. In particular, \eqref{eq:cHD-FD} follows directly from \eqref{eq:HD-B-int-main} after applying the integral identity \eqref{eq:Lauricella-y-integral}.

In general, $H(x,y,z)$ as given in \eqref{eq:technical-defs1} is not of the form \eqref{eq:H-factorized-main}.
To bring it into this form one needs to redefine $H$ by factoring out overall constants in $y$ and the lowest power of $y$ or by passing to a universal cover such that one can get rid of all fractional powers of $x,y,z$.
This introduces a certain redefinition of $\cH^D$ which is rather mild.
We give a few examples:
\begin{itemize}
\item
If $H(x,y,z)$ is polynomial in $y$ but not monic, then one can collect an overall function $f(x,z)y^r$ where $r$ is the lowest power of $y$ appearing in the polynomial, \ie{} $H(x,y,z)=f(x,z)y^r\prod_j(y-y_j(x,z))$.
In this case the normalization can be absorbed in a redefinition of $A_2\cdot\e\to A_2\cdot\e-r\sum_i\e_i$ and of the factor outside of the integral in \eqref{eq:HD-B-int-main}.
\footnote{This term appears as an overall multiplicative factor in the computation of the monodromy. This can be seen by repeating the steps of Appendix~\ref{app:Lauricella-D}.}
\item
If $H(x,y,z)$ has fractional powers of $y$, then we redefine the integration variable as $y=(y')^m$ where $m$ is the smallest positive integer such that $H(x,(y')^m,z)$ is polynomial in $y'$.
\end{itemize}

Physically, to give a meaning to worldsheet instantons it is necessary to specify a large volume chamber in the (open and closed) K\"ahler moduli space.
Mathematically, this translates into the observation that the monodromy $\Delta_c\cH^D$ features certain poles, and the integration contour in \eqref{eq:equiv-W} crosses some of these when switching from one large volume chamber to another, causing $W(t,c,\e)$ to jump.
For simplicity, in the following we will fix a choice of phase of the open string modulus in such a way that $x=\eu^{-c}$ is small.

Thanks to \eqref{eq:cHD-FD}, we are able to express the equivariant disk potential defined in \eqref{eq:equiv-W} directly in terms of the roots $y_j(x)$ of the mirror curve.
Let $\alpha_j\in\BQ$ denote the exponents controlling the asymptotic behavior of
the roots $y_j(x)$ of the mirror curve equation $H(x,y)=0$ in this phase, namely
\be\label{eq:yi-asympt}
 \alpha_j := \lim_{x\to 0} \frac{\log y_j(x)}{\log x}
\ee
In other words, $\alpha_j$ are the slopes of external legs in the toric diagram of $\XIIId$.
The monodromy of $\cH^D$ can be computed using the general expression in terms of Lauricella's function \eqref{eq:cHD-FD}, as explained in Appendix~\ref{app:monodromy-HD}.
Assuming, for the sake of concreteness, that the roots of the curve satisfy $|1+y_j^{-1}|<1$, we can use an explicit series expansion%
\footnote{Power series representations of Lauricella functions are generically not unique, and in fact depend on a choice of region for the arguments of the function within which the series is convergent. See the end of Appendix~\ref{app:monodromy-HD} for a discussion of this issue.} of the Lauricella to compute the monodromy and
we find that the regular part, in the phase $x\to 0$, is
\be\label{eq:delta-cHD-0-final-main}
 \Delta_c[\cH^D(t,c,\e)]_0
 = \frac1{A_2\cdot\e}
 \left( A_1\cdot\e - \sum_i\e_i \sum_{j=1}^k \alpha_j \right)
 \sum_{j=1}^k \log(-y_j(-x))
 + \sum_{j=1}^k \alpha_j \log(-y_{j}(-x)) + \dots
\ee
where the ellipses denote terms that are polynomial in $c$, which therefore do not contribute to the disk potential.

Because of identity \eqref{eq:equiv-W-deriv}, the expression we just derived also gives the regular instantonic part of $\partial_c W(t,c,\e)$, so it remains now to show that this matches with the disk potential as in \eqref{eq:DeltacH=dcW}.
This can be achieved by observing that $W_\disk$ is in fact defined by its relation to the roots of the mirror curve as follows
\be\label{eq:W-y-rel}
 \partial_c W^{(j)}_\disk = \log (-y_j(-x)) +\dots\,,
\ee
for a brane whose vacuum configuration is labeled by the $j$-th sheet of the mirror curve.
This is precisely the type of logarithmic contributions we have found in the r.h.s.\ of \eqref{eq:delta-cHD-0-final-main}. Interestingly, we can conclude that equation \eqref{eq:delta-cHD-0-final-main} actually leads to a generalization of both claims in \eqref{eq:Wdisk2} and \eqref{eq:DeltacH=dcW}, because instead of a single disk potential it features a linear combination of them, weighted by the asymptotic slopes $\alpha_j$.
Note however, that the disk instantons $W^{(j)}_\disk$ on different branches are actually related, since roots $y_j(x)$ are related to each other by analytic continuation around branch points of the mirror curve.\footnote{For example, if the mirror curve is quadratic in $y$ then $H(x,y) = y^2 + a_1(x) y+ a_2(x)$ where $a_2(x) = y_1 y_2$ is the product of the two roots. Equation \eqref{eq:W-y-rel} then implies that $W_\disk^{(2)} =  - W_\disk^{(1)}  + \int^c \, \log a_2(x) \, \dif\log x$.}
This explains why in all examples considered in \cite{Cassia:2023uvd}, the simpler statement \eqref{eq:Wdisk2} holds, with suitable proportionality coefficients.

\subsection{Examples}
Here we illustrate applications of the general monodromy formula \eqref{eq:delta-cHD-0-final-main}, showing that it computes the disk potential predicted by open string mirror symmetry.

\subsubsection{\texorpdfstring{$\BC^3$}{C3}}\label{sec:C3-lauricella}
We consider a family of toric branes in $\BC^3$ defined by the hyperplane normal to the charge vector
\be
 q^1 = (0,1,-1)\,.
\ee
This covers two distinct phases for the toric Lagrangian, characterized by the sign of $c$ \cite{Cassia:2023uvd}.

First, we check that the general expression of $\cH^D$ in terms of Lauricella's function reproduces the quantum Lebesgue measure, which was shown to be (see \cite[(3.11)]{Cassia:2023uvd} for a derivation)
\be\label{eq:C3-HD-1-main}
 \cH^D(c,\e) = \Ga(\e_1)\Ga(\e_2+\e_3)
 \frac{\eu^{-\e_2 c}}{(1+\eu^{-c})^{\e_2+\e_3}}\,.
\ee
To exemplify the general approach outlined in the last section,
we are now going to show how to match \eqref{eq:C3-HD-1-main} with the general formula in \eqref{eq:cHD-FD}. In order to do so, we start from the definition of $\cH^D$
as a \eGLSM{} hemisphere partition function as in \eqref{eq:cHD-def} and then rewrite each $\Ga$ in the integrand using
\be\label{eq:gamma-rescaling-identity-main}
 \int_0^\infty \frac{\dif u}{u^{1-\e}} \eu^{-u \alpha}
 = \alpha^{-\e}\Ga(\e) \,,
\ee
We then obtain
\be
\begin{aligned}
 \cH^D(c,\e) &= \int_{\ii\BR}\frac{\dif\varphi}{2\pi\ii}\,\eu^{c\varphi}
 \Ga(\e_1)\Ga(\e_2+\varphi)\Ga(\e_3-\varphi) \\
 &= \int_0^\infty\frac{\dif u_1}{u_1}
 \int_0^\infty\frac{\dif u_2}{u_2}
 \int_0^\infty\frac{\dif u_3}{u_3}
 \delta\Big(\log(\eu^{c}u_2u_3^{-1})\Big)
 \prod_{i=1}^3 u_i^{\e_i} \eu^{-u_i} \\
 &= \int_0^\infty\frac{\dif s_1}{s_1}
 \int_0^\infty\frac{\dif s_2}{s_2}
 \delta\Big(\log(\eu^{c}s_2)\Big)s_1^{\e_1}s_2^{\e_2}
 \int_0^\infty\frac{\dif u_3}{u_3}
 u_3^{\e_1+\e_2+\e_3} \eu^{-u_3(s_1+s_2+1)} \\
 &= \Ga(\e_1+\e_2+\e_3)\int_0^\infty\frac{\dif s_1}{s_1}
 \int_0^\infty\frac{\dif s_2}{s_2}
 \delta\Big(\log(\eu^{c}s_2)\Big)s_1^{\e_1}s_2^{\e_2}
 (1+s_1+s_2)^{-\e_1-\e_2-\e_3} \\
 &= \Ga(\e_1+\e_2+\e_3)\,x^{\e_2}\int_0^\infty\dif y
 \frac{y^{\e_1-1}}{(1+x+y)^{\e_1+\e_2+\e_3}} \\
\end{aligned}
\ee
where in the second line we have obtained a Dirac $\delta$-function after integrating over $\varphi$, and next we used a reparametrization of the integration variables%
\footnote{The choice of which variable $u_i$ to rescale by and integrate out first is not canonical and different choices lead to different but equivalent expressions for $\cH^D$. Similar manipulations can be performed in the general context of an arbitrary single or double cut of a CY threefold as discussed in Appendix~\ref{app:technical-identities}} $u_{1,2}=u_3 s_{1,2}$ and later the redefinition $s_1=y$, $s_2=x$. 
From this expression of $\cH^D$ we can read
\be
 A_1\cdot\e = \e_2\,,
 \hspace{30pt}
 A_2\cdot\e = \e_1\,,
 \hspace{30pt}
 H(x,y) = 1+x+y
\ee
and $M=0$.
Here, we have chosen to identify the variable $y$ dual to $x$ as the ratio $y=s_1=u_1/u_3$, corresponding to the choice of homogeneous hyperplane $q^2=(1,0,-1)$.
The unique root of the polynomial $H(x,y)$ is
\be
\label{eq:mirror-curve-sol-C3}
 y_1(x) = -(1+x)\,.
\ee
Making use of the integral identity \eqref{eq:Lauricella-y-integral} we find
\be
 \cH^D(c,\e) = \Ga(\e_1)\Ga(\e_2+\e_3)\frac{x^{\e_2}}{(1+x)^{\e_1+\e_2+\e_3}}
 F_D^{(1)}\Big(\e_1,\sum_{i=1}^3\e_i,\sum_{i=1}^3\e_i;1+y_1(x)^{-1}\Big)
\ee
which matches with the expression in \eqref{eq:cHD-FD}. The univariate Lauricella function in this case reduces to the Gauss hypergeometric ${}_2F_1$ as in \eqref{eq:Lauricella2Gauss}, and the identity ${}_2F_1(a,b,b;x)=(1-x)^{-a}$ then yields the desired expression of $\cH^D$ as in \eqref{eq:C3-HD-1-main}.

Next, we check that the regular part of the monodromy of $\cH^D$ agrees with the general formula~\eqref{eq:delta-cHD-0-final-main} in terms of roots of the mirror curve.
The computation depends on a choice of large volume phase. There are two distinct possibilities, characterized by the sign of $c$. We discuss each in turn.

If $c>0$, then $x=\eu^{-c}$ is within the unit circle, and the monodromy contains the following regular term (leaving aside polynomial terms in $c$)
\be\label{eq:C3-monodromy-true-main}
 \Delta_c [\cH^D]_0 = \frac{\e_2}{\e_1} \log(1-\eu^{-c})+ \dots
\ee
where the monodromy operator $\Delta_c$ is defined as in \eqref{eq:monodromy-operator}.
Noting that $\lim_{x\to 0}y_1(x)=-1$, we deduce that $\alpha_1=0$, which shows that
\eqref{eq:C3-monodromy-true-main} matches with the general formula \eqref{eq:delta-cHD-0-final-main}.

If $c<0$, the monodromy changes, and contains the following regular terms
\be\label{eq:C3-monodromy-true-phase2-main}
 \Delta_c[\cH^D]_0 = -\frac{\e_3}{\e_1} \log (1-\eu^{c})+ \dots
\ee
To compare this with the general formula \eqref{eq:delta-cHD-0-final-main}, we need to find the value of $\alpha_1$ compatible with this choice of chamber.
Observe that $y_1(x)\sim-x$ for $x\to\infty$, from which we deduce that $\alpha_1=1$.
Plugging this value in \eqref{eq:delta-cHD-0-final-main}, together with the appropriate identification of equivariant parameters, we
obtain the same regular terms as in \eqref{eq:C3-monodromy-true-phase2-main} (up to negligible polynomial terms in $c$).

\subsubsection{Local \texorpdfstring{$\BP^2$}{P2}}\label{sec:localP2-lauricella}
We next consider a symplectic cut of local $\BP^2$ defined by the charge matrix
\be\label{eq:P2-charges-main}
\left[\begin{array}{c}
    Q^a_i \\
    q^1_i \\
\end{array}\right] =
\left[\begin{array}{cccc}
    1 & 1 & 1 & -3 \\
    0 & 0 & -1 & 1 \\
\end{array}\right]\,.
\ee
For this choice of cut and assuming $t>0$, there exist three distinct phases for the associated toric Lagrangian: $c>0$, $0>c>-t$ and $-t>c$, respectively.
The quantum volume of local $\BP^2$ before the cut is given by the integral
\be
 \cF^D(t,\e) = \int_{\ii\BR}\frac{\dif\phi}{2\pi\ii}
 \eu^{t\phi}
 \Ga(\e_1+\phi)\Ga(\e_2+\phi)\Ga(\e_3+\phi)\Ga(\e_4-3\phi)
\ee
which was evaluated explicitly in \cite[eq. (C.3)]{Cassia:2023uvd}.
Similarly, the quantum Lebesgue measure was computed in \cite[eq. (5.23)]{Cassia:2023uvd} in the phase $0>-t>c$
\be\label{eq;P2Hd3-resum-main}
\begin{aligned}
 \cH^D
 = &
 \frac12 \eu^{\frac12\e_3(t+3c)+\frac12\e_4(t+c)}
 \Big\{
 \Ga\left(\e_1+\frac{\e_3}{2}+\frac{\e_4}{2}\right)
 \Ga\left(\e_2+\frac{\e_3}{2}+\frac{\e_4}{2}\right) \times \\
 &\quad\quad\times\, {}_2F_1\left(\e_1+\frac{\e_3}{2}+\frac{\e_4}{2},
 \e_2+\frac{\e_3}{2}+\frac{\e_4}{2};
 \frac12;\frac14 \eu^{t+c} (1+\eu^c)^2\right) \\
 & - \eu^{\frac12(t+c)} (1+\eu^c)
 \Ga\left(\e_1+\frac{\e_3}{2}+\frac{\e_4}{2}+1\right)
 \Ga\left(\e_2+\frac{\e_3}{2}+\frac{\e_4}{2}+1\right) \times \\
 &\quad\quad\times\, {}_2F_1\left(\e_1+\frac{\e_3}{2}+\frac{\e_4}{2}+1,
 \e_2+\frac{\e_3}{2}+\frac{\e_4}{2}+1;\frac32;\frac14 \eu^{t+c} (1+\eu^c)^2\right)
 \Big\}
\end{aligned}
\ee
The same expression holds in other phases, by analytic continuation.

\begin{remark}
\label{rmk:analyticity}
Mathematically, the fact that quantum volumes can be analytically continued across phases is a consequence of the ``Crepant Transformation Correspondence'' \cite{Horja:2000hyp, Coates:2008wal, Coates:2009com, Coates:2009wal, Iritani:2009ani, COATES20181002, Brini:2013zsa, Brini:2014fea}.
\end{remark}

This should be compared to the general formula \eqref{eq:cHD-FD}.
We repeat the derivation for clarity and to match variables,
starting from the \eGLSM{} integral encoded by \eqref{eq:P2-charges-main}
\be\label{eq:localP2-CHD-main}
\begin{aligned}
 & \cH^D
 =
 \int_{\ii\BR}\frac{\dif\phi}{2\pi\ii}\int_{\ii\BR}\frac{\dif\varphi}{2\pi\ii}
 \eu^{t\phi+c\varphi}
 \Ga(\e_1+\phi)\Ga(\e_2+\phi)\Ga(\e_3+\phi-\varphi)\Ga(\e_4-3\phi+\varphi) \\
 & =
 \int_0^\infty \frac{\dif u_1}{u_1^{1-\e_1}} \dots \frac{\dif u_4}{u_1^{1-\e_4}}
 \eu^{-(u_1+u_2+u_3+u_4)}
 \delta\left(\log\left(\eu^{t} u_1 u_2 u_3 u_4^{-3}\right)\right)
 \delta\left(\log\left(\eu^{c} u_3^{-1} u_4\right)\right) \\
 & = \int_0^\infty
 \frac{\dif s_1}{s_1^{1-\e_1}} \frac{\dif s_2}{s_2^{1-\e_2}}
 \frac{\dif s_4}{s_4^{1-\e_4}} \frac{\dif u_3}{u_3^{1-\e_1-\e_2-\e_3-\e_4}}
 \eu^{-u_3(1+s_1+s_2+s_4)}
 \delta\left(\log\left(\eu^{t} s_1 s_2 s_4^{-3}\right)\right)
 \delta\left(\log\left(\eu^{c} s_4 \right)\right) \\
 & =
 \Ga(\e_1+\e_2+\e_3+\e_4)\, \eu^{-c\e_4}
 \int_0^\infty
 \frac{\dif s_1}{s_1^{1-\e_1}} \frac{\dif s_2}{s_2^{1-\e_2}}
 \frac1{(1+s_1+s_2+\eu^{-c})^{\e_1+\e_2+\e_3+\e_4}}
 \delta\left(\log\left(\eu^{t} s_1 s_2 \eu^{3c}\right)\right) \\
 & =
 \Ga(\e_1+\e_2+\e_3+\e_4)\, x^{3\e_2+\e_4} z^{\e_2}
 \int_0^\infty \dif y
 \frac{y^{2\e_1+\e_3+\e_4-1}}{(y^2+y(1+x)+z x^3)^{\e_1+\e_2+\e_3+\e_4}}
\end{aligned}
\ee
where we made use of the identity \eqref{eq:gamma-rescaling-identity-main}
for $\Re(\alpha)>0$ and $\Re(\e)>0$, to rewrite every $\Ga$ function in the integrand and then we exchanged the order of the integrations.
In the third line, we introduced rescaled integration variables $s_i:=u_i/u_3$ for $i=1,2,4$ so that we could integrate out $u_3$\footnote{Such a rescaling is always possible due to the CY condition on the charges of the symplectic quotient, however, the choice of which integration variable to rescale away is not unique and each choice produces a different but equivalent expression for the integral.}.
Finally, in the last line we renamed the variables as $z=\eu^{-t}$, $x=\eu^{-c}$ and $y=s_1=u_1/u_3$.
This expression for $\cH^D$ matches with \eqref{eq:HD-B-int-main} upon identifying
\be
 A_1\cdot\e = 3\e_2+\e_4\,,\qquad
 A_2\cdot\e = 2\e_1+\e_3+\e_4\,,\qquad
 t\cdot M\cdot\e = t \e_2\,,
\ee
with $H(x,y,z)=y^2+y(1+x)+z x^3$.
In order to fully specify the identification between $\cH^D$ and the Lauricella's function expression in \eqref{eq:cHD-FD}, we need to determine the roots $y_j$ of the polynomial $H(x,y,z)$, which is straightforward.
The curve $H(x,y,z)=0$ (which coincides with \cite[eq. (5.27)]{Cassia:2023uvd}) has two roots given by
\be\label{eq:localP2-mirror-sheets}
\begin{aligned}
  y_1(x,z) &= -\left(\frac12(1+x)+\sqrt{-zx^{3}+\tfrac14(1+x)^2}\right) \,, \\
  y_2(x,z) &= -\left(\frac12(1+x)-\sqrt{-zx^{3}+\tfrac14(1+x)^2}\right) \,.
\end{aligned}
\ee
Next we consider the monodromy in different phases, studying the applicability of \eqref{eq:delta-cHD-0-final-main} based on the data we just determined.

Starting with the phase $c\to -\infty$, corresponding to $x\to \infty$, the arguments of Lauricella's function behave as follows
\be
 1+(y_{1,2})^{-1} = 1\mp \sqrt{- z^{-1}x^{-3}} + \dots
\ee
In particular, this expression makes it clear that if one of the arguments lies within the unit disk, the other must lie outside.
Therefore this is an example in which condition \eqref{eq:validity} is not satisfied, and formula \eqref{eq:delta-cHD-0-final-main} is not expected to hold.
\footnote{
Indeed, it is easy to see that applying this formula would lead to a wrong result. Since $y_1 y_2 = z x^{3}$ the first term in \eqref{eq:delta-cHD-0-final-main} would not give any interesting contribution, while the second term would vanish since $\alpha_1 = \alpha_2$ and because of this simple relation between the roots. We emphasize that the expression \eqref{eq:cHD-FD} of $\cH^D$ in terms of Lauricella's function still holds. Its monodromy however needs to be computed by other means, such as those adopted in \cite{Cassia:2023uvd}, or by relying on known analytic continuation formulae (see \eg{} \cite{Bezrodnykh_2018} and references therein).
}

In the phase $c\to+\infty$, corresponding to $x\to 0$, the two roots behave instead as follows
\be
 y_1\sim -1\,,
 \qquad
 y_2\sim -z x^3\,.
\ee
for which we read the exponents $\alpha_1=0$ and $\alpha_2=3$.
In this case, both $1+(y_{1,2})^{-1}$ lie within the unit disk.
Plugging the exponents $\alpha_j$ into \eqref{eq:delta-cHD-0-final-main} yields the following expression for the monodromy
\be\label{eq:P2-case1}
\begin{aligned}
 \Delta_c[\cH^D]_0
 & = -\frac{3\e_1+3\e_3+2\e_4}{2\e_1+\e_3+\e_4} \log(y_1(-x) y_2(-x))
 + 3 \log (-y_2(-x)) + \dots \\
 & = \frac{3\e_1+\e_4}{2\e_1+\e_3+\e_4} \log(z x^3) - 3 \log (-y_1(-x)) + \dots \\
 & = - 3 \log (-y_1(-x)) + \dots
\end{aligned}
\ee
where we dropped all terms polynomial in $c$.
We therefore obtain that the regular part of the monodromy of $\cH^D$
includes the contribution $\log(-y_1(-x))\equiv\partial_c W_\disk$, which is the logarithmic derivative of the disk potential defined in \cite{Aganagic:2000gs,Aganagic:2001nx}, up to an overall constant that is controlled by the coefficients in~\eqref{eq:delta-cHD-0-final-main}.

\section{Branes and cuts from CY4}\label{sec:CY4}

\subsection{Braverman's construction}\label{sec:Braverman}

The symplectic cut of $\XIIId$ is the singular space \eqref{eq:symplectic-cut} consisting of the union of two `half-spaces' $\XIIId[\lessgtr]$ glued along a common divisor $\XIId$.
In this section, we review a useful viewpoint introduced by
Braverman \cite{Braverman+1999+85+98},
who constructed a one-parameter family of manifolds realizing the singular space $\XIIId[<]\cup_{\XIId}\XIIId[>]$ as a degeneration of $\XIIId$.

The symplectic cut is defined by the charge vector $q:=q^1$ in \eqref{eq:toric-brane-hyperplanes} encoding the moment map of a Hamiltonian $U(1)$ action \eqref{eq:CY2-symp-quot}.\footnote{We restrict to the case where $\XIIId$ is a toric Calabi--Yau threefold. The construction is more general, as it only requires that $\XIIId$ is a K\"ahler manifold with a Hamiltonian $U(1)$ action.}
Instead of the symplectic quotient of $\BC^{r+3}$ as defined earlier, we consider a larger ambient space $\BC^{r+3}\times \BC_+\times\BC_-$, with K\"ahler form
$\tilde\omega = \frac{\ii}{2} \left(\sum_{i=1}^{r+3} \dif z_i\wedge \dif \bar z_i
+ \dif z_+\wedge \dif\bar z_+ + \dif z_-\wedge \dif\bar z_-\right)$.
We extend the $U(1)$ action to $\BC_+\times\BC_-$ with weights $(-1,+1)$, so that the corresponding moment map is
\be\label{eq:Braverman-moment-map}
 \tilde\mu_q = \sum_{i=1}^{r+3} q_i |z_i|^2 - |z_+|^2 + |z_-|^2\,,
\ee
while the original $U(1)^r$ action is extended trivially over $\BC_+\times \BC_-$.
The $U(1)^{r+1}$ symplectic quotient is defined by the extended charge matrix,
\be\label{eq:CY4-charges}
 \tilde{Q} =
 \left[
  \begin{array}{ccc|c}
   z_i & z_+ & z_- \\
   \hline
   Q^a_i & 0 & 0 & t^a \\
   q_i & -1 & 1 & c
  \end{array}
 \right]
\ee
where the constraints $\sum_{i=1}^{r+3}Q^a_i=\sum_{i=1}^{r+3}q_i=0$ ensure that the quotient admits a Calabi--Yau structure.
From here onwards we refer to $\XIVd$ as the Calabi--Yau fourfold (CY4) defined by the symplectic quotient
\be\label{eq:CY4-def}
 \XIVd = \BC^{r+3}\times \BC_+\times\BC_- \, \sslash \, U(1)^{r+1} \,.
\ee
Equivalently, we can first define the symplectic quotient $\XIIId=\BC^{r+3}\sslash U(1)^r$ as before. Then we can define a $U(1)$-action on $\XIIId$ with moment map $\mu_q:\XIIId\to\BR$ corresponding to the vector of charges $q$, and extend this action on $\BC_+\times\BC_-$ with charges $(-1,+1)$. The resulting symplectic quotient $\XIIId\times\BC_+\times\BC_-\sslash U(1)$ is naturally diffeomorphic to $\XIVd$.

To see how $\XIVd$ is related to the symplectic cut of $\XIIId$,
we consider the Braverman map $\pi:\XIIId\times\BC_+\times\BC_-\to\BC$, defined by
\be\label{eq:Braverman-map-coordinates}
 \pi: \ (x,z_+, z_-) \mapsto z_+ z_-\,,\qquad x\in\XIIId,\ z_\pm\in \BC_{\pm}\,.
\ee
Since this map is invariant under the $U(1)$-action in \eqref{eq:Braverman-moment-map},
it follows that it descends to a map on the \emph{quotient} $\XIVd$, where it defines a fibration
\be\label{eq:Braverman-map}
 \pi:\XIVd\to\BC\,,
\ee
which we also denote as $\pi$ (by a mild abuse of notation).

\begin{figure}[!ht]
\begin{center}
\includegraphics[width=0.99\textwidth]{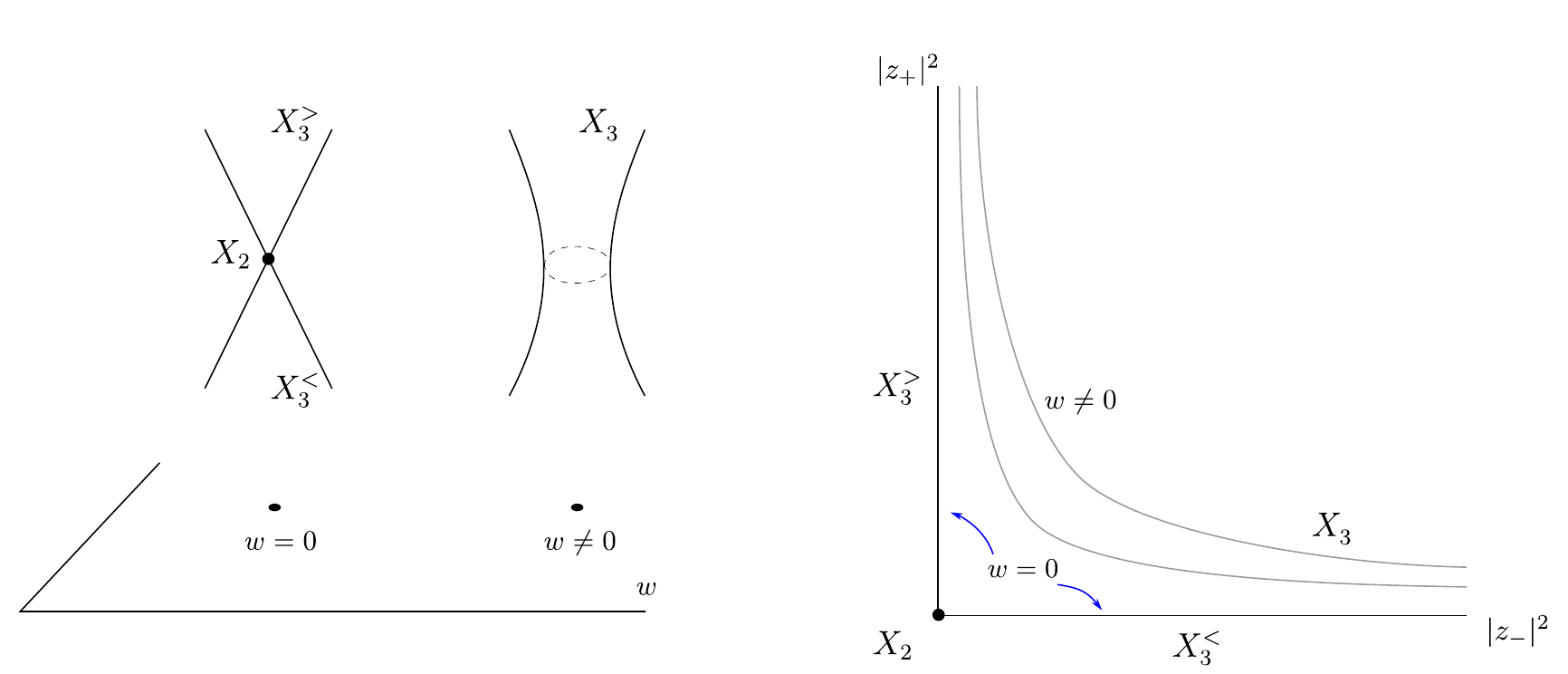}
\caption{The total space of Braverman's construction, which realizes $\XIVd$ as a fibration over the $w$-plane, with generic fiber $\XIIId$ and a degenerate fiber $\XIIId[<]\cup_{\XIId}\XIIId[>]$ over the origin.}
\label{fig:Braverman}
\end{center}
\end{figure}

The fiber of $\pi$ at a generic point $w\neq0$ is
\be
 \pi^{-1}(w)
 = \left\{(x,z_+,z_-)\in\XIIId\times\BC_+\times\BC_- \;|\;
 \mu_q(x) = c+ |z_+|^2 - |z_-|^2,\,z_+z_- = w\right\} / U(1)
\ee
Inserting the restriction $z_+z_-=w$ into the moment map condition gives $\mu_q(x)-c=|z_+|^2-|w|^2/|z_+|^2$.
Solving for $|z_+|^2$ and choosing the (globally) positive root, defines a circle in the $\BC_+$ plane,
for each $x\in\XIIId$.
The circle has nonzero radius thanks to the condition $w\neq 0$ which implies $z_\pm\neq 0$, therefore the moment map equation combined with the restriction to the fiber gives a manifold with topology $\XIIId\times S^1$.
Taking the $U(1)$ quotient reduces the circle to a point, showing that to each $w$ there corresponds a whole copy of $\XIIId$.
Therefore we conclude that the generic fiber of $\pi$ is complex isomorphic to $\XIIId$
\be
 \pi^{-1}(w)\simeq\XIIId\,,\qquad(w\neq 0)\,.
\ee

At $w=0$ the fiber is different. In this case there are two components to consider, corresponding to $z_\pm =0$ and $z_\mp\in \BC_{\mp}$ which intersect at the origin of $\BC_+\times\BC_-$. Each leads to a reduction of \eqref{eq:CY4-def} to a 3-manifold that, by definition, corresponds to one of the `half spaces' of the symplectic cut:
\begin{itemize}
\item for $z_-=0$,
\be\label{eq:half-space-toric-divisors1}
 \{(x,z_+)\in\XIIId\times\BC_+\;|\;
 \mu_q(x) = c + |z_+|^2 \} / U(1)
 \equiv\XIIId[>]\,,
\ee
\item for $z_+=0$,
\be\label{eq:half-space-toric-divisors2}
 \{(x,z_-)\in\XIIId\times\BC_-\;|\;
 \mu_q(x) = c - |z_-|^2 \} / U(1)
 \equiv\XIIId[<]\,.
\ee
\end{itemize}
Recall indeed that the half spaces $\XIIId[\lessgtr]$ are defined by the requirement that $\mu_q(x) \lessgtr c$ in $\XIIId$, see \cite{Cassia:2023uvd}.
This presentation also makes it clear that the half-spaces $\XIIId[\lessgtr]$ are complex isomorphic to toric divisors of $\XIVd$, corresponding to the reduction of the vanishing loci of the homogeneous coordinates $z_\pm$, respectively. We denote these divisors as $D_\pm$.

The reduction of the locus $z_+=z_-=0$, defines instead a complex codimension-two subspace that can be naturally identified with the reduced space $\XIId$ as
\be\label{eq:X2d-in-X4d}
 \{x\in\XIIId\;|\;\mu_q(x)=c\}/U(1)\equiv\XIId\,.
\ee
Therefore, from the viewpoint of $\XIVd$, the Calabi--Yau twofold $\XIId$ defined as in \eqref{eq:CY2-symp-quot}, is just the intersection of two toric divisors
\be
 \XIId\simeq\XIIId[<]\cap\XIIId[>]\,.
\ee
Finally, we obtain that the fiber at $w=0$ is
\be
 \pi^{-1}(0)\simeq\XIIId[<]\cup_{\XIId}\XIIId[>]\,.
\ee

To summarize, Braverman's construction defines a Calabi--Yau fourfold $\XIVd$ which admits the structure of a fibration over a complex plane $\BC$, whose generic fiber is a smooth K\"ahler manifold complex isomorphic to $\XIIId$ and with a singular fiber at the origin which is complex isomorphic to $\XIIId[<]\cup_{\XIId}\XIIId[>]$.\footnote{The isomorphism as complex manifolds is equivariant with respect to the $S^1$-action. The generic fiber is not isomorphic to $\XIIId$ as a K\"ahler manifold, see \cite[Remark~3.4]{Braverman+1999+85+98}.}
This fibration therefore realizes the symplectic cut as a degeneration of the Calabi--Yau threefold $\XIIId$.
The half-spaces of the symplectic cut are complex isomorphic to toric divisors of $\XIVd$, as made explicit in \eqref{eq:half-space-toric-divisors1} and \eqref{eq:half-space-toric-divisors2}.
Correspondingly, the common submanifold $\XIId$ arises as the intersection of these divisors as shown by \eqref{eq:X2d-in-X4d}.
The global picture is summarized by Figure~\ref{fig:Braverman}.

\subsection{Quantization of Braverman's fourfold}

We next consider the quantization of the fourfold perspective on symplectic cutting, via an \eGLSM{} with target $\XIVd$.
Later we will discuss how this relates to the quantum cut of $\XIIId$ defined in our previous work.

We define the quantum volume of $\XIVd$ as the \eGLSM{} partition function
\be
 \cF^D_{\XIVd}(t,c,\e,\e_\pm)
 = \int\frac{\dif\varphi}{2\pi\ii} \eu^{\varphi c}
 \Ga(\e_+-\varphi)\Ga(\e_-+\varphi)
 \int\prod_{a=1}^r\frac{\dif\phi_a}{2\pi\ii} \eu^{\phi_a t^a}
 \prod_{i=1}^{r+3}\Ga(\e_i+\phi_aQ^a_i + \varphi q_i)\,,
\ee
where $Q^a_i$ and $q_i$ are the rows in the extended charge matrix \eqref{eq:CY4-charges}.

Recall that $\XIVd$ is a fibration of $\XIIId$ over $\BC$, except at the origin in $\BC$ where the fiber $X_3$ degenerates to a singular space as explained in \eqref{eq:Braverman-map},
and that $\XIIId$ itself is sliced by $\XIId$ (as in \eqref{eq:classical-slicing}).
Since the locus where the generic fiber becomes singular is measure zero within $\XIVd$,
we expect to be able to decompose the quantum volume $\cF^D_{\XIVd}$ in terms of the function $\cH^D$.
In order to see this explicitly, we introduce the Fourier transform\footnote{This is defined for $\varphi\in\ii\BR$ and extended by analytic continuation.} of the quantum volume of $\XIId$
\be
 \tilde\cH^D(t,\varphi,\e)
 = \int_{\BR} \dif c \,\eu^{-\varphi c}\, \cH^D(t,c,\e)
 = \int \prod_{a=1}^r\frac{\dif\phi_a}{2\pi\ii} \eu^{\phi_a t^a}
 \prod_{i=1}^{r+3}\Ga(\e_i+\phi_a Q^a_i + \varphi q_i)\,.
\ee
Remarkably, this can be interpreted as a deformation of the quantum volume of $\XIIId$,
since the latter is recovered at $\varphi=0$,
\be
 \tilde\cH^D(t,0,\e) = \cF^D(t,\e)\,.
\ee
Through these definitions, we can express the quantum volume of $\XIVd$ directly in terms of that of $\XIId$,
\be\label{eq:F-4d-from-rho}
\begin{aligned}
 \cF^D_{\XIVd}(t,c,\e,\e_\pm)
 & = \int \frac{\dif\varphi}{2\pi\ii} \eu^{\varphi c}
 \Ga(\e_+-\varphi)\Ga(\e_-+\varphi)\,
 \tilde\cH^D(t,\varphi,\e) \\
 & = \int_{\BR} \dif c' \, \cH^D(t,c',\e) \cdot \rho(c-c',\e_\pm)\,,
\end{aligned}
\ee
where all dependence on the equivariant parameters $\e_\pm$ lies entirely within the distribution
\be\label{eq:rho}
\begin{aligned}
 \rho(c,\e_\pm) :=&
 \int_{\ii\BR} \frac{\dif\varphi}{2\pi\ii} \,\eu^{\varphi c}
 \,\Ga(\e_+ -\varphi)\Ga(\e_- +\varphi) \\
 = & \Ga(\e_++\e_-) \int\frac{\dif\varphi}{2\pi\ii}
 \,\eu^{\varphi c} \,\Beta(\e_+ -\varphi,\e_- +\varphi) \\
 = & \Ga(\e_++\e_-) \frac{\eu^{-c\e_-}}{(1+\eu^{-c})^{\e_++\e_-}}\,.
\end{aligned}
\ee
Note that this is the quantum volume of a onefold obtained by taking the symplectic quotient $\BC\cong\BC_+\times\BC_-\sslash U(1)$ with charges $(-1,+1)$, and it is a multivalued function with ramification at $\eu^{-c}\in\{0,-1,\infty\}$ therefore the computation of the monodromy must be handled with some care. This also gives a geometric interpretation to formula \eqref{eq:F-4d-from-rho}, as the computation of a 4d quantum volume in terms of a fibration over the complex line by a product of a twofold with quantum volume $\cH^D$ and a onefold with quantum volume $\rho$.

Next, recall that from the viewpoint of $\XIVd$ the half-spaces $\XIIId[\lessgtr]$ arise as divisors, as described in \eqref{eq:half-space-toric-divisors1} and \eqref{eq:half-space-toric-divisors2}.
In the framework of \eGLSM, equivariant periods of divisors, by which we mean solutions to the PF equations of $\XIVd$ with classical behavior equal to the (equivariant) volume of the corresponding divisor, can be obtained from $\cF^D_{\XIVd}$ by acting with difference operators as in \eqref{eq:equiv-period}, namely
\be
\begin{aligned}
    \Pi(D_+) & = \frac{(1-\eu^{2\pi\ii\cD_+})}{2\pi\ii} \cF^D_{\XIVd}
    = \int_{\BR} \dif c' \, \cH^D(t,c',\e)\cdot
    \frac{(1-\eu^{2\pi\ii (\e_+ - \partial_c)})}{2\pi\ii}
    \rho(c-c',\e_\pm)\,, \\
    \Pi(D_-) & = \frac{(1-\eu^{2\pi\ii\cD_-})}{2\pi\ii} \cF^D_{\XIVd}
    = \int_{\BR} \dif c' \, \cH^D(t,c',\e)\cdot
    \frac{(1-\eu^{2\pi\ii (\e_- + \partial_c)})}{2\pi\ii}
    \rho(c-c',\e_\pm)\,.
\end{aligned}
\ee
where $\cD_\pm=\e_\pm\mp\partial_c$ are differential operators analogous to those in \eqref{eq:divisor-operator}.
The periods $\Pi(D_\pm)$ are the CY4 counterparts of the functions defined in \eqref{eq:quantum-half-volumes}.
Furthermore, the intersection of these divisors corresponds to $\XIId$, as shown in \eqref{eq:X2d-in-X4d}, which implies that the equivariant period of the intersection can be obtained by acting with both difference operators simultaneously, and can be expressed as follows
\be
\label{eq:period-intersection-divisors-X4}
 \Pi(D_+\cap D_-)
 :=  \frac{(1-\eu^{2\pi\ii\cD_+})}{2\pi\ii}
 \frac{(1-\eu^{2\pi\ii\cD_-})}{2\pi\ii} \cF^D_{\XIVd}\,.
\ee
If we define the distribution $\sigma$ by
\be\label{eq:sigma-def}
\begin{aligned}
 \sigma(c,\e_\pm)
 :=&\,
 \frac{(1-\eu^{2\pi\ii (\e_+ - \partial_c)})}{2\pi\ii}
 \frac{(1-\eu^{2\pi\ii (\e_- + \partial_c)})}{2\pi\ii}
 \rho(c,\e_\pm) \\
 =& \int \frac{\dif\varphi}{2\pi\ii} \,\eu^{\varphi c} \,
 \eu^{\pi\ii(\e_+ +\e_-)} \frac{1}{\Ga(1-\e_+ +\varphi)\Ga(1-\e_- -\varphi) }\,,
\end{aligned}
\ee
we can then rewrite $\Pi(D_+\cap D_-)$ as
\be
 \Pi(D_+\cap D_-)
 = \int_{\BR} \dif c' \, \cH^D(t,c',\e)\cdot \sigma(c-c',\e_\pm)\,.
\ee

\begin{remark}
\label{rmk:monodromy-rho}
Observe that the action of the operators $\frac{(1-\eu^{2\pi\ii\cD_\pm})}{2\pi\ii}$
on a multivalued function such as $\rho(c,\e_\pm)$ is subtle as it depends on the
value of the variable $x=\eu^{-c}$ relative to the ramification points at $0,-1,\infty$.
Suppose in fact that we take $c>0$ and apply the difference operator $\frac{(1-\eu^{2\pi\ii\cD_-})}{2\pi\ii}$ to $\rho(c,\e_\pm)$.
Because we are computing the difference between $\rho(c,\e_\pm)$ and $\eu^{2\pi\ii\e_-}\rho(c+2\pi\ii,\e_\pm)$, we can think of this operator as a sort of monodromy around the ramification point at $x=0$. Near $x=0$ we can write $\rho(c,\e_\pm)$ as
\be
 \rho(c,\e_\pm)
 = \Ga(\e_++\e_-) \frac{x^{\e_-}}{(1+x)^{\e_++\e_-}}
\ee
and the monodromy is determined by the multivalued function $x^{\e_-}$, \ie{}
\be
 (1-\eu^{2\pi\ii(\e_-+\partial_c)})\, x^{\e_-}
 = (1-\eu^{2\pi\ii(\e_--\e_-)})\, x^{\e_-} = 0\,.
\ee
On the other hand, if we take $c<0$ and apply the same operator, we see that this computes the monodromy around the ramification point $x=\infty$ where the function $\rho(c,\e_\pm)$ can be written as
\be
 \rho(c,\e_\pm)
 = \Ga(\e_++\e_-) \frac{x^{-\e_+}}{(1+x^{-1})^{\e_++\e_-}}\,.
\ee
The monodromy in this case is determined by the multivalued function $x^{-\e_+}$, and we get
\be
 (1-\eu^{2\pi\ii(\e_-+\partial_c)})\, x^{-\e_+}
 = (1-\eu^{2\pi\ii(\e_-+\e_+)})\, x^{-\e_+}\,.
\ee
Clearly, we obtain two different results and the reason is that we have crossed the ramification point at $x=-1$, as depicted in Figure~\ref{fig:monodromy-rho}.

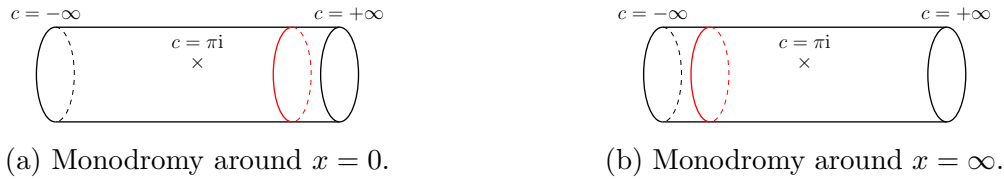
\begin{figure}[!ht]
\centering
\begin{subfigure}[b]{0.32\textwidth}
\centering
\resizebox{\textwidth}{!}{
\begin{tikzpicture}
    \draw[thick] (-3,0) -- (3,0);
    \draw[thick] (-3,-2) -- (3,-2);
    \draw[thick] (-3,0) arc[start angle=90, end angle=270, x radius=0.4cm, y radius=1cm];
    \draw[dashed] (-3,0) arc[start angle=90, end angle=-90, x radius=0.4cm, y radius=1cm];
    \draw[thick] (3,0) arc[start angle=90, end angle=270, x radius=0.4cm, y radius=1cm];
    \draw[thick] (3,0) arc[start angle=90, end angle=-90, x radius=0.4cm, y radius=1cm];
    \draw[thick,red] (2,0) arc[start angle=90, end angle=270, x radius=0.4cm, y radius=1cm];
    \draw[dashed,red] (2,0) arc[start angle=90, end angle=-90, x radius=0.4cm, y radius=1cm];
    \draw (-3.2,0) node[anchor=south]{$c=-\infty$};
    \draw (3.2,0) node[anchor=south]{$c=+\infty$};
    \draw (0,-0.3) node{$c=\pi\ii$};
    \draw (0,-0.75) node{$\times$};
\end{tikzpicture}
}
\caption{Monodromy around $x=0$.}
\end{subfigure}
\hspace*{.15\textwidth}
\begin{subfigure}[b]{0.32\textwidth}
\centering
\resizebox{\textwidth}{!}{
\begin{tikzpicture}
    \draw[thick] (-3,0) -- (3,0);
    \draw[thick] (-3,-2) -- (3,-2);
    \draw[thick] (-3,0) arc[start angle=90, end angle=270, x radius=0.4cm, y radius=1cm];
    \draw[dashed] (-3,0) arc[start angle=90, end angle=-90, x radius=0.4cm, y radius=1cm];
    \draw[thick] (3,0) arc[start angle=90, end angle=270, x radius=0.4cm, y radius=1cm];
    \draw[thick] (3,0) arc[start angle=90, end angle=-90, x radius=0.4cm, y radius=1cm];
    \draw[thick,red] (-2,0) arc[start angle=90, end angle=270, x radius=0.4cm, y radius=1cm];
    \draw[dashed,red] (-2,0) arc[start angle=90, end angle=-90, x radius=0.4cm, y radius=1cm];
    \draw (-3.2,0) node[anchor=south]{$c=-\infty$};
    \draw (3.2,0) node[anchor=south]{$c=+\infty$};
    \draw (0,-0.3) node{$c=\pi\ii$};
    \draw (0,-0.75) node{$\times$};
\end{tikzpicture}
}
\caption{Monodromy around $x=\infty$.}
\end{subfigure}
\caption{Picture of the monodromy of $\rho(c,\e_\pm)$ for $c>0$ and $c<0$, respectively.}
\label{fig:monodromy-rho}
\end{figure}
\end{remark}

\subsection{Quantum cuts of CY3 from equivariant periods of CY4}

Next we discuss how the \eGLSM{} for $\XIVd$ recovers the quantum cut of $\XIIId$.
The main issue is that equivariant (quantum) volumes and periods in the fourfold depend on additional equivariant parameters $\e_\pm$. To recover results for $\XIIId$ we therefore need to study the limit $\e_\pm\to 0$.

Observe that for small $\e_\pm$, Euler's beta function has leading order behavior\footnote{\label{foot:beta-delta}
This is a well-known property, that can be derived using the integral representation
$\Beta(a,b) = \int_0^\infty \frac{t^{a-1}}{(1+t)^{a+b}}\,\dif t$ as follows
\be
 \Beta(\ii x,-\ii x)
 = \int_0^\infty t^{\ii x-1}\, \dif t
 = \int_{-\infty}^\infty \eu^{\ii x y }\, \dif y = 2\pi\,\delta(x)\,.
\ee
}
\be
 \Beta(\e_+ -\phi,\e_- +\phi)
 = \Beta(-\phi,\phi) \left(1 + O(\e_\pm) \right)
 = 2\pi\ii\,\delta(\phi) \left(1 + O(\e_\pm)\right)
\ee
It follows then that the leading behavior of $\rho$ defined in \eqref{eq:rho} is
\be\label{eq:rho-limit}
 \rho(c,\e_\pm) = \Ga(\e_++\e_-)
 \left(1 + O(\e_\pm)\right)
\ee
and therefore the quantum volume of $\XIVd$ reduces to that of $\XIIId$ times a constant factor that depends on $\e_\pm$
\be
    \cF^D_{\XIVd}(t,c,\e,\e_\pm)
    = \int_{\BR} \dif c' \, \cH^D(t,c',\e)\cdot\rho(c-c',\e_\pm)
    = \cF^D(t,\e)\,\Ga(\e_++\e_-)\left(1+O(\e_\pm)\right)
\ee
where, in the second equality we used relation \eqref{eq:q-lebesgue-integral} between $\cH^D$ and $\cF^D$.

Similarly, we can recover equivariant half-space functions $\cF^D_{\lessgtr}$ by taking a limit on the equivariant periods of divisors in $\XIVd$.
For this purpose, observe that
\be
 \frac{(1-\eu^{2\pi\ii\cD_\pm})}{2\pi\ii}\,\rho(c,\e_\pm)
 = -\eu^{\pi\ii\e_\pm}\int_{\ii\BR}\frac{\dif\varphi}{2\pi\ii}\,
 \eu^{\varphi(c\mp\pi\ii)} \,
 \frac{\Ga(\e_\mp\pm\varphi)}{\Ga(1-\e_\pm\pm\varphi)}
 = - \frac{\Theta_H(\pm(c\mp\pi\ii))\,\eu^{\mp\e_\mp c +\pi\ii(\e_++\e_-)}}
 {(1+\eu^{\mp c})^{\e_++\e_-} \Ga(1-\e_+-\e_-)}
\ee
where $\Theta_H$ is a Heaviside theta function and the integral has been evaluated using the Jeffrey--Kirwan residue prescription.%
\footnote{The JK prescription in this case is equivalent to closing the imaginary contour with a semicircle at infinity either to the left or to the right according to the value of the coefficient of $\varphi$ in the exponential.}
Then setting $\e_\pm$ to zero gives 
\be
\label{eq:Theta-H-def}
 \lim_{\e_\pm\to0} \frac{(1-\eu^{2\pi\ii\cD_\pm})}{2\pi\ii}\,\rho(c,\e_\pm)
 = - \Theta_H(\pm c-\pi\ii)
\ee
This shows that
\be
 \lim_{\e_\pm\to 0} \Pi(D_+)
 = - \int_{\BR} \dif c' \, \cH^D(t,c',\e)\cdot\Theta_H(c-c'-\pi\ii)
 = - \cF^D_<(t,c-\pi\ii,\e)
\ee
and
\be
 \lim_{\e_\pm\to 0} \Pi(D_-)
 = - \int_{\BR} \dif c' \, \cH^D(t,c',\e)\cdot\Theta_H(c'-c-\pi\ii)
 = - \cF^D_>(t,c+\pi\ii,\e)\,.
\ee
Observe that, while the argument here is rather formal, in specific examples one can take the limit rigorously by computing the monodromy of the analytic function $\cF^D_{\XIVd}$ as explained in Remark~\ref{rmk:monodromy-rho}.

A similar statement holds for the intersection of divisors, which classically corresponds to $\XIId$. Indeed, observe that the behavior of $\sigma$ as $\e_\pm$ approach zero is
\be
\label{eq:sigma-limit}
\begin{aligned}
 \lim_{\e_\pm \to 0}
 \sigma(c,\e_\pm)
 & = \int_{\ii\BR}\frac{\dif\varphi}{2\pi\ii}\,\eu^{\varphi c} \,
 \frac{1}{\Ga(1+\varphi)\Ga(1-\varphi)}\\
 & = \int_{\ii\BR}\frac{\dif\varphi}{2\pi\ii} \,\eu^{\varphi c} \,
 \frac{\sin(\pi\varphi)}{\pi\varphi} \\
 & = \frac1{2\pi\ii}\int_{c-\pi\ii}^{c+\pi\ii}\dif c'\,
 \int_{\ii\BR}\frac{\dif\varphi}{2\pi\ii} \,\eu^{\varphi c'} \\
 & = \frac1{2\pi\ii}\int_{c-\pi\ii}^{c+\pi\ii}\dif c'\, \delta(c') \\
 & = \frac1{2\pi\ii}\Big(\Theta_H(c+\pi\ii)-\Theta_H(c-\pi\ii)\Big)\,,
\end{aligned}
\ee
therefore we obtain
\be
\begin{aligned}
 \lim_{\e_\pm \to 0} \Pi(D_+\cap D_-)
 & = \frac1{2\pi\ii}\int_{\BR} \dif c' \, \cH^D(t,c',\e)\cdot
 \Big(\Theta_H(c-c'+\pi\ii)-\Theta_H(c-c'-\pi\ii)\Big) \\
 & = \frac1{2\pi\ii} \Big(\cF_<^D(t,c+\pi\ii,\e)-\cF_<^D(t,c-\pi\ii,\e)\Big)
 \equiv W(t,c,\e)\,.
\end{aligned}
\ee
This leads to the conclusion that, in the limit $\e_\pm \to 0$, the period associated to $\XIId$ computed as intersection of divisors $D_\pm$ in $\XIVd$,
coincides with the equivariant disk potential computed from the quantum cut of $\XIIId$.
Moreover, the derivation of the limit in \eqref{eq:sigma-limit} also gives an alternative integral expression for the equivariant disk potential as
\be
 W(t,c,\e) = \oint\prod_a\frac{\dif\phi_a}{2\pi\ii}\oint\frac{\dif\varphi}{2\pi\ii}
 \,\frac{\sin(\pi\varphi)}{\pi\varphi}\,
 \eu^{\sum_a\phi_at^a+\varphi c} \prod_i\Ga(\e_i+\sum_a\phi_aQ_i^a+\varphi q_i)
\ee
which is exactly of the form \eqref{eq:cHD-def}, the only difference being the insertion of the function $\frac{\sin(\pi\varphi)}{\pi\varphi}$.\footnote{Notice that this insertion does not introduce any new poles in the integrand. Formally, we can write $W(t,c,\e)=\frac{\sin(\pi\partial_c)}{\pi\partial_c}\cH^D(t,c,\e)$ where the operator $\frac{\sin(\pi\partial_c)}{\pi\partial_c}=\sum_{n=0}^\infty\frac{(\pi\ii\partial_c)^{2n}}{(2n+1)!}$ is regarded as a power series in derivatives.}

It is important here to distinguish between the period associated to $\XIId$ regarded as a subvariety of $\XIVd$ and the quantum volume of $\XIId$ as a CY twofold.
The equivariant period associated to $\XIId=D_+\cap D_-$, contains information about the embedding into the ambient space $\XIVd$ even in the limit $\e_\pm\to 0$, since it is a solution to the PF equations of $\XIVd$, while the function $\cH^D$ is the partition function of an \eGLSM{} with target $\XIId$ without reference to any embedding into a larger space, and as such, it satisfies instead the PF equations of $\XIId$.
For this reason, the functions $\Pi(D_+\cap D_-)$ and $\cH^D$ are not the same, and instead they are related according to the formulae above.

\subsection{Extended Picard--Fuchs equations for quantum cuts}\label{sec:4d-PF-3d}

While $\XIId$, $\XIIId$ and $\XIVd$ are all Calabi--Yau manifolds by construction,
the half-spaces $\XIIId[\lessgtr]$ defined by the symplectic cut are generally not.
Additionally, the functions $\cF^D_{\lessgtr}$ are not obtained as disk partition functions
of \eGLSM{} with targets $\XIIId[\lessgtr]$, but rather they are defined as integrals of $\cH^D$. For this reasons we would not necessarily expect $\cF^D_{\lessgtr}$ to satisfy Picard--Fuchs equations for  the spaces $\XIIId[\lessgtr]$, and in fact this is not the case.
However, as we will argue shortly, the functions $\cF^D_{\lessgtr}$ do obey a certain generalization of the Picard--Fuchs equations of the Calabi--Yau threefold $\XIIId$.

This is a direct consequence of the observation that both $\cF^D_{\lessgtr}$, as well as the equivariant disk potential $W$, arise from the limit $\e_\pm\to0$ of equivariant periods of divisors (and their intersection) in $\XIVd$.

According to the general theory of \eGLSM{} developed in \cite{Cassia:2022lfj}, the quantum volume of $\XIVd$ obeys a system of equations of the form
\be
 \PF_\gamma \cdot \cF^D_{\XIVd} = 0
\ee
where $\gamma\in\BZ^{r+1}$ is a vector of integers, and
\be\label{eq:PF-operator-generic}
 \PF_\gamma:=  \prod_{\{i|\sum_a \gamma_a \tilde Q^a_i>0\}}
 \left(\cD_i\right)_{\sum_a \gamma_a \tilde Q^a_i}
 - \eu^{-\sum_a \gamma_a t^a}
 \prod_{\{i|\sum_a \gamma_a \tilde Q^a_i\leq 0\}} \left(\cD_i\right)_{-\sum_a \gamma_a \tilde Q^a_i}
\ee
where $\tilde Q$ is the charge matrix $Q$ augmented by $q$ and by the extension to $\BC_+\times \BC_-$ as in \eqref{eq:CY4-charges} and we use the convention that the index $i$ ranges from $1$ to $r+5$ where the last two values are identified with the labels $\pm$.
In particular, we have $\e_{r+4}=\e_+$, $\e_{r+5}=\e_-$ and similarly $\cD_{r+4}=\cD_{+}$,
$\cD_{r+5}=\cD_{-}$ for the corresponding divisor operators.
We adopt an analogous convention for the K\"ahler moduli and we identify $t^{r+1}\equiv c$.

Restricting to vectors of the form $\gamma = (\gamma_1,\gamma_2,\dots, \gamma_r,0)$
gives operators $\PF_\gamma$ that formally have the same structure as the Picard--Fuchs operators of $\XIIId$, with the difference that $\cD_i$ now also include derivative terms $q_i \partial_c$ for the modulus of the symplectic cut.

The same operators annihilate also solutions associated to toric divisors, and intersections thereof, because of the commutation relations
\be
 \left[\PF_\gamma,\tfrac{(1-\eu^{2\pi\ii\cD_i})}{2\pi\ii}\right]=0
\ee
with the finite difference/monodromy operators.
This implies that
\be
 \PF_\gamma\cdot\,\Pi(D_\pm) = 0\,,
 \hspace{30pt}
 \PF_\gamma\cdot\,\Pi(D_+\cap D_-) = 0\,.
\ee
Taking the limit $\e_\pm\to 0$ of these equations leads to Picard--Fuchs equations involving open string moduli for the equivariant disk potential and for the half-volumes defined by the quantum cut of $\XIIId$, namely
\be\label{eq:q-cut-PF}
 \left.\PF_\gamma\right|_{\e_\pm=0} \cdot\, \cF^D_{\lessgtr}(t,c,\e) = 0\,,
 \hspace{30pt}
 \left.\PF_\gamma\right|_{\e_\pm=0} \cdot\, W(t,c,\e) = 0\,.
\ee
We thus find that half-volumes and the equivariant disk potential obey an \emph{extension} of the Picard--Fuchs equations, where $\cD_i$ are modified by the addition of $q_i \partial_c$.
Of course, these equations also admit solutions that are independent of $c$, which correspond to $\cF^D$ and equivariant periods of intersections of divisors of $\XIIId$.
However, when considering more general solutions that do depend on $c$,
we find among these $\cF^D_\lessgtr$ and $W$.

\begin{remark}[Non-equivariant limit]
\label{rmk:inhomogeneity}
As will be shown in examples below, the equations \eqref{eq:q-cut-PF} become inhomogeneous in the non-equivariant limit.
The same property was encountered in the study of extended Picard--Fuchs equations for open Gromov--Witten invariants on the real quintic \cite{Walcher:2006rs, Walcher:2008gho, Morrison:2007bm}.
This shows that, at least in the context of toric threefolds (the examples that we consider), the inhomogeneous PF equations for disk potentials become homogeneous after turning on equivariance.
\end{remark}

\subsubsection{Relation to work of Mayr and Lerche}
Mayr and Lerche \cite{Mayr:2001xk, Lerche:2001cw} observed that the disk potentials of open topological strings in toric Calabi--Yau threefolds with a toric brane can be derived from the periods of certain CY fourfolds. Liu and Yu \cite{Liu:2021eeb,Liu:2022swc} later formalized this observation for more general CY manifolds and orbifolds.
In all explicit cases we examined, the CY fourfold constructed via Braverman's fibration matches that of Mayr and Lerche.

Our construction of the equivariant disk potential, along with our observation that it corresponds to the $\e_\pm \to 0$ limit of the equivariant period associated with an intersection of divisors in $\XIVd$, provides an alternative formalization of the Mayr--Lerche observation for general CY3s and embedded toric branes.
In particular, our construction of the fourfold $\XIVd$ is well-suited for studying equivariant Picard--Fuchs equations and quantum cohomology rings using the language of \eGLSM{}s.

Although certain aspects of Liu and Yu's fourfold construction resemble elements of Braverman's approach%
\footnote{For instance, the threefold $Y$ in \cite{Liu:2021eeb}, constructed as a partial compactification of $\XIIId$, appears closely related to one of the two half-spaces $\XIIId[\lessgtr]$. Similarly, the divisor $D$ in $Y$ seems related to the divisor $\XIId$ in $\XIIId[\lessgtr]$.},
it remains unclear whether both methods necessarily lead to the same CY fourfold $\XIVd$.

\subsection{Examples}
\subsubsection{\texorpdfstring{$\BC^3$}{C3}}\label{sec:C3-4d}

\begin{figure}[!ht]
\centering
\begin{subfigure}[b]{0.32\textwidth}
\centering
\includegraphics[width=\textwidth]{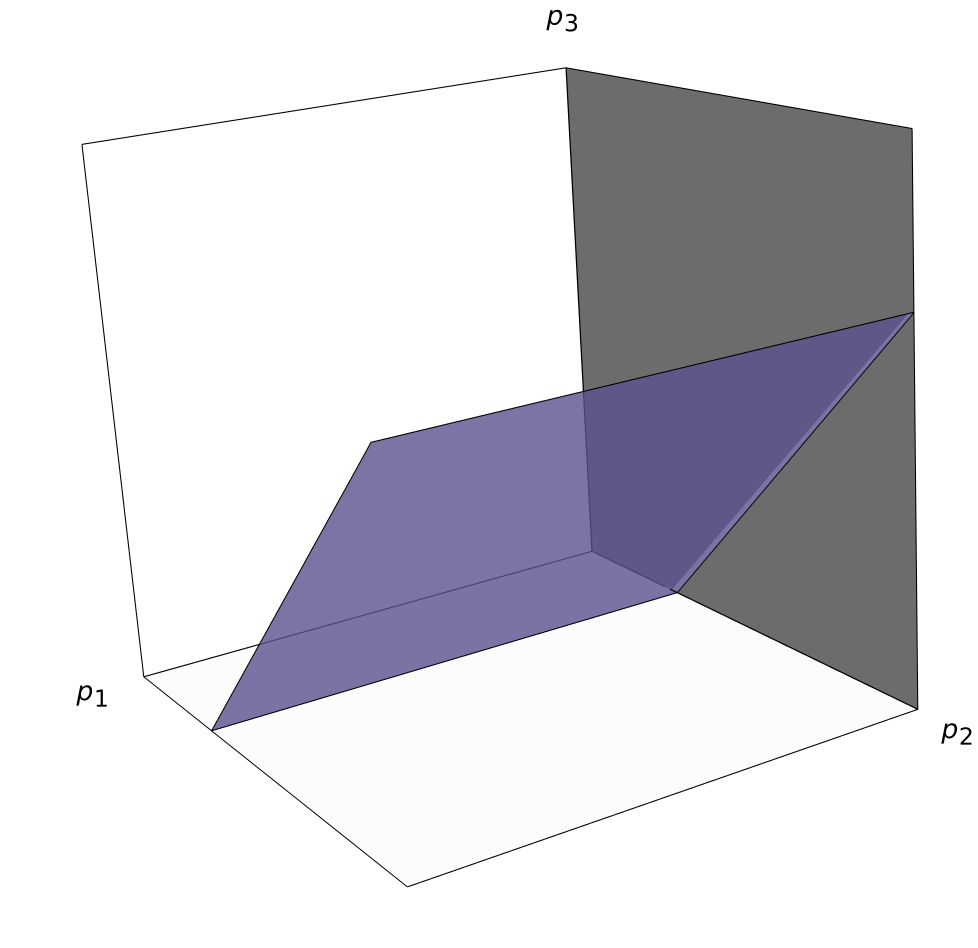}
\caption{Phase 1: $c>0$}
\end{subfigure}
\hspace*{.15\textwidth}
\begin{subfigure}[b]{0.32\textwidth}
\centering
\includegraphics[width=\textwidth]{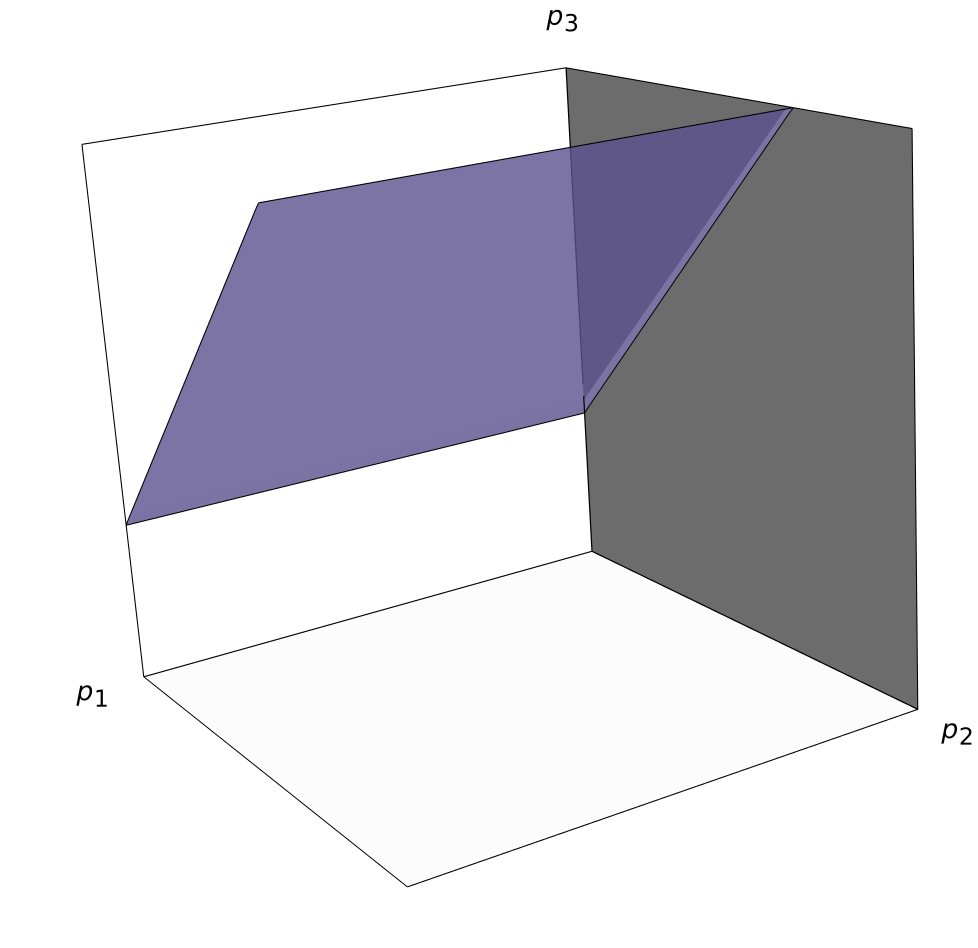}
\caption{Phase 2: $c<0$}
\end{subfigure}
\caption{Hyperplane for a symplectic cut of $\BC^3$ in two different phases. The axes here are labeled by the variables $p_i:=|z_i|^2$.}
\label{fig:C3cutphases}
\end{figure}

We reconsider the quantum cut of $\BC^3$ studied in Section~\ref{sec:C3-lauricella}. The associated Calabi--Yau fourfold is given by the symplectic quotient
\be
    \XIVd  = (\BC^3\times \BC_+\times\BC_-)\sslash U(1)
\ee
defined by the following  charge matrix
\be\label{eq:C3-single-cut-Q-matrix}
 \tilde{Q} = \left[
    \begin{array}{ccccc|c}
    z_1 & z_2 & z_3 & z_+ & z_- \\
    \hline
    0 & 1 & -1 & -1 & 1 & c
    \end{array}
 \right]
\ee
This geometry corresponds to the direct product of a resolved conifold and a complex line.
There are two phases $c\gtrless 0$ related by a flop transition, which correspond to different positions of the hyperplane of the symplectic cut in $\XIIId=\BC^3$, see Figure \ref{fig:C3cutphases}.

The quantum volume can be evaluated explicitly as follows
\be
\label{eq:F4d-C3}
\begin{aligned}
 \cF^D_{\XIVd} & (c,\e,\e_\pm)
 = \Ga(\e_1)
 \int \frac{\dif\phi}{2\pi\ii}
 \eu^{\phi c}
 \Ga(\e_+-\phi)\Ga(\e_-+\phi)
 \Ga(\e_2+\phi)\Ga(\e_3-\phi) \\
 =\,&
 \eu^{\e_3 c} \Ga(\e_1)\Ga(\e_+-\e_3)\Ga(\e_2+\e_3)\Ga(\e_3+\e_-)
 \,{}_2F_1(\e_2+\e_3,\e_3+\e_-;\e_3-\e_++1;\eu^c) \\
 &+ \eu^{\e_+ c} \Ga(\e_1)\Ga(\e_3-\e_+)\Ga(\e_2+\e_+)\Ga(\e_++\e_-)
 \, {}_2F_1(\e_2+\e_+,\e_++\e_-;\e_+-\e_3+1;\eu^c)
\end{aligned}
\ee
This expression is analytic in $c$, reflecting the observation from \cite{Cassia:2023uvd} that \eGLSM{} quantum volumes such as $\cF^D$, $\cH^D$ (and $\cF^D_{\XIVd}$)
do not feature the jumps that might be expected due to a change in the JK contour for the integral \cite{Cassia:2022lfj}.

In the limit when $\e_\pm\to 0$, the leading order behavior of the fourfold quantum volume is dominated by the last line in \eqref{eq:F4d-C3}
\be
 \lim_{\e_\pm\to0}\frac{1}{\Ga(\e_-+\e_+)}\cF^D_{\XIVd}(c,\e,\e_\pm)
 = \Ga(\e_1)\Ga(\e_2)\Ga(\e_3)
\ee
recovering the quantum volume of $\BC^3$ as expected.

The intersection of divisors has the following equivariant period
\be
\begin{aligned}
 \Pi(D_+\cap D_-)
 =& \frac{(1-\eu^{2\pi\ii(\e_+ - \partial_c)})}{2\pi\ii}
 \frac{(1-\eu^{2\pi\ii (\e_- + \partial_c)})}{2\pi\ii} \cF^D_{\XIVd}(c,\e,\e_\pm) \\
 =& \frac{(1-\eu^{2\pi\ii(\e_+ -\e_3)})}{2\pi\ii}
 \frac{(1-\eu^{2\pi\ii(\e_- +\e_3)})}{2\pi\ii}
 \Ga(\e_1) \Ga(\e_+-\e_3) \Ga(\e_2+\e_3) \Ga(\e_3+\e_-) \\
 &\times \eu^{\e_3 c}\, {}_2F_1(\e_2+\e_3,\e_3+\e_-;\e_3-\e_++1;\eu^{c})
\end{aligned}
\ee
where we assumed $c<0$.
Taking the limit $\e_\pm\to 0$ we find
\be\label{eq:W-C3-CY4}
 \lim_{\e_\pm \to 0} \Pi(D_+\cap D_-)
 = \frac{\sin(\pi\e_3)}{\pi\e_3} \Ga(\e_1) \Ga(\e_2+\e_3)\, \eu^{\e_3 c}
 \, {}_2F_1(\e_2+\e_3,\e_3,\e_3+1;\eu^{c})
 = W(c,\e)
\ee
where $W$ agrees exactly with the equivariant disk potential in \cite[eq. (3.28)]{Cassia:2023uvd} after matching conventions (see Remark~\ref{rmk:c-shift}).

In a similar way, we can compute the equivariant period of a divisor of $\XIVd$ as follows
\begin{multline}
 \Pi(D_+)
 = \frac{(1-\eu^{2\pi\ii(\e_+ -\e_3)})}{2\pi\ii}
 \Ga(\e_1) \Ga(\e_+-\e_3) \Ga(\e_2+\e_3) \Ga(\e_3+\e_-) \\
 \times \eu^{\e_3 c}\, {}_2F_1(\e_2+\e_3,\e_3+\e_-;\e_3-\e_++1;\eu^{c})
\end{multline}
which in the limit $\e_\pm\to 0$ gives
\be\label{eq:Fmin-CY4}
 \lim_{\e_\pm\to 0}\Pi(D_+)
 = -\frac{\Ga(\e_1)\Ga(\e_2+\e_3)}{\e_3}\eu^{\e_3(c-\pi\ii)}
 \, {}_2F_1(\e_2+\e_3,\e_3;\e_3+1;\eu^{c})
 = -\cF^D_<(c-\pi\ii,\e)
\ee
in agreement with \cite[eq. (3.19)]{Cassia:2023uvd}.

In this case, there are no moduli for the CY3 geometry, so there are no Picard--Fuchs equations for the CY3 $\XIIId$.
However, the CY4 carries the open string modulus $c$, and there is an associated operator \eqref{eq:PF-operator-generic}.
Here the index $a$ takes only one value corresponding to the (extended) charge vector $q^1 = (0,1,-1,-1,1)$.
Without loss of generality we set $\gamma_a=1$, and obtain
\be
 \PF_1 = \cD_2 \cD_- - \eu^{-c} \cD_3 \cD_+
\ee
where
\be
 \cD_2 = \e_2+\partial_c\,,\qquad
 \cD_3 = \e_3-\partial_c\,,\qquad
 \cD_\pm = \e_\pm \mp \partial_c\,.
\ee
The limit $\e_\pm\to 0$ gives differential equations \eqref{eq:q-cut-PF} for the equivariant disk potential \eqref{eq:W-C3-CY4} (as well as for the half-volume $\cF^D_\lessgtr$ from \eqref{eq:Fmin-CY4})
\be\label{eq:PF-W-C3}
 \left(\cD_2 + \eu^{-c}	\cD_3 \right) \partial_c \, W(c,\e) = 0\,.
\ee
It is interesting to consider the non-equivariant counterpart of this equation.
In fact, somewhat remarkably, in the limit $\e_i\to 0$ this equation becomes inhomogeneous.
Expanding \eqref{eq:PF-W-C3} in powers of $\e$ and keeping only the $O(\e^0)$ terms,
we get the equation
\be
 (1-\eu^{-c})\partial^2_c [W(c,\e)]_0
 = - (\e_2+\eu^{-c}\e_3)\partial_c [W(c,\e)]_{-1}\,.
\ee
If we define $W_\disk$ to be the instantonic part of $[W(c,\e)]_0$, namely
\be
 W_\disk = \frac{\e_3}{\e_1}\Li_2(\eu^c)\,,
\ee
we then obtain the inhomogeneous equation
\be\label{eq:C3-open-PF}
 (1-\eu^{-c})\,\partial^2_c\, W_\disk = -1
\ee
where the inhomogeneous term is generated by the terms of order $O(\e)$ in the operator $\PF_1$ acting on $[W(c,\e)]_{-1}=\frac{\e_3c-\gamma(\e_1+\e_2+\e_3)}{\e_1(\e_2+\e_3)}$. In this sense, the equivariant open-string PF equation \eqref{eq:C3-open-PF} has no counterpart in the non-equivariant setting, where it is replaced by an inhomogeneous equation instead. This is expected from earlier results on open string mirror symmetry, see Remark~\ref{rmk:inhomogeneity}.

\subsubsection{Local \texorpdfstring{$\BP^2$}{P2}}\label{sec:localP2-4d}

\begin{figure}[!ht]
\centering
\begin{subfigure}[b]{0.32\textwidth}
\centering
\includegraphics[width=\textwidth]{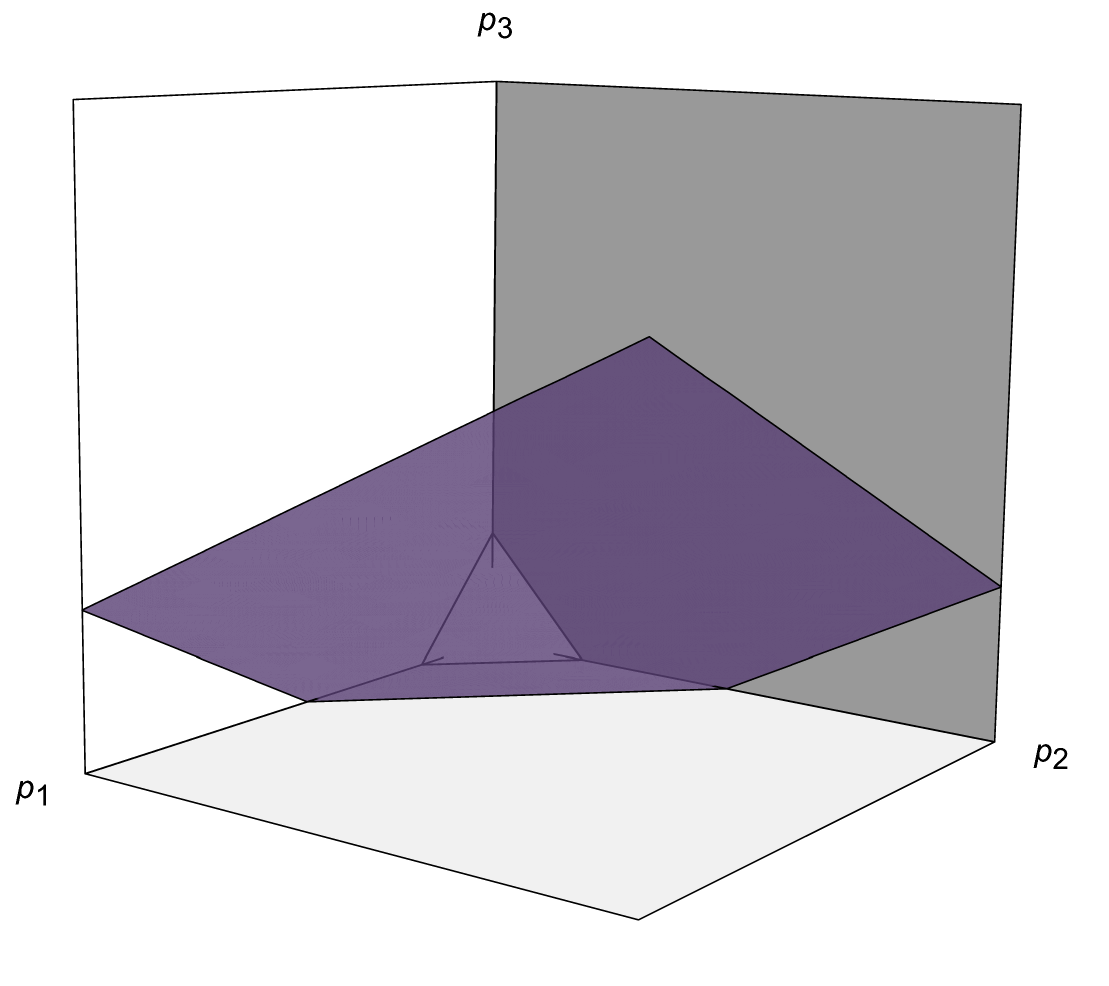}
\caption{Phase 1: $c,t>0$}
\end{subfigure}
\hfill
\begin{subfigure}[b]{0.32\textwidth}
\centering
\includegraphics[width=\textwidth]{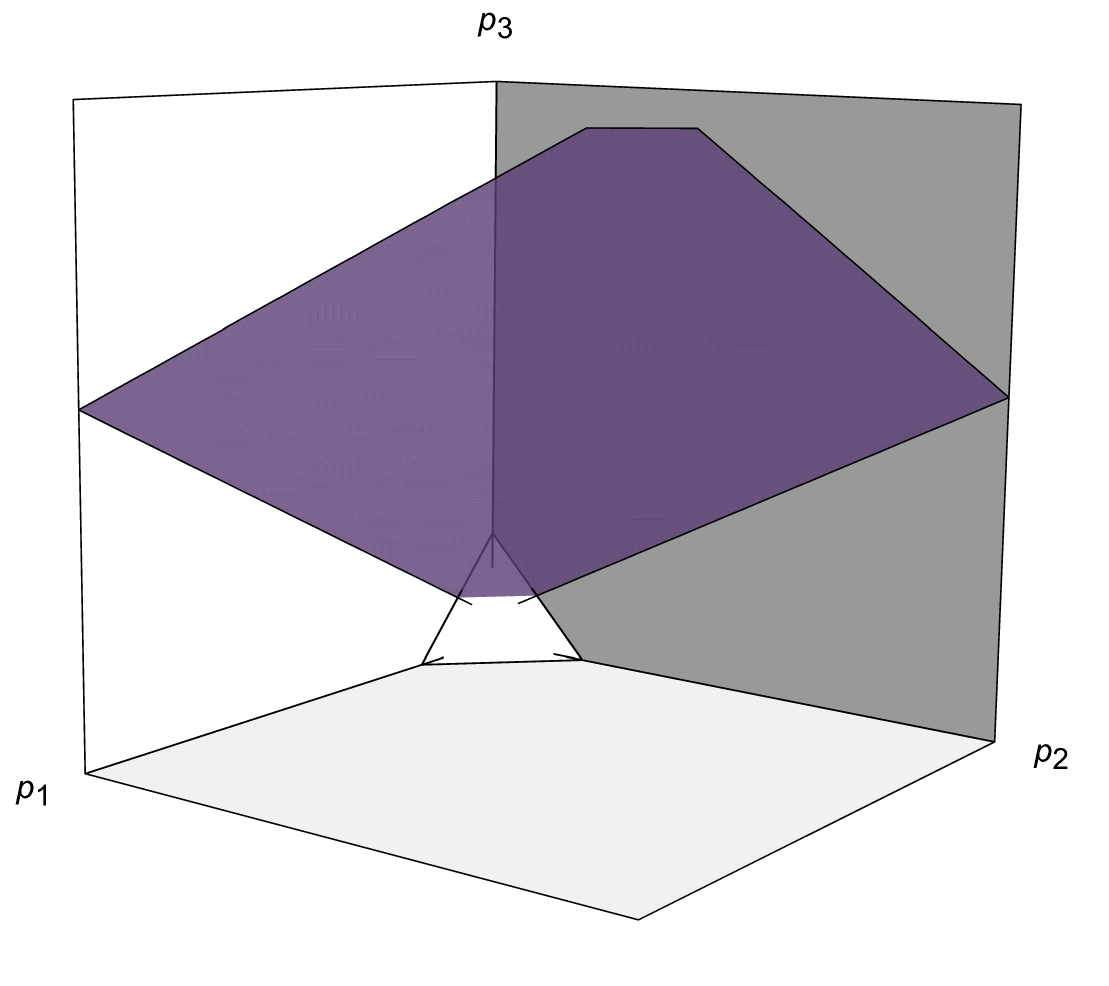}
\caption{Phase 2: $0>c>-t$}
\end{subfigure}
\hfill
\begin{subfigure}[b]{0.32\textwidth}
\centering
\includegraphics[width=\textwidth]{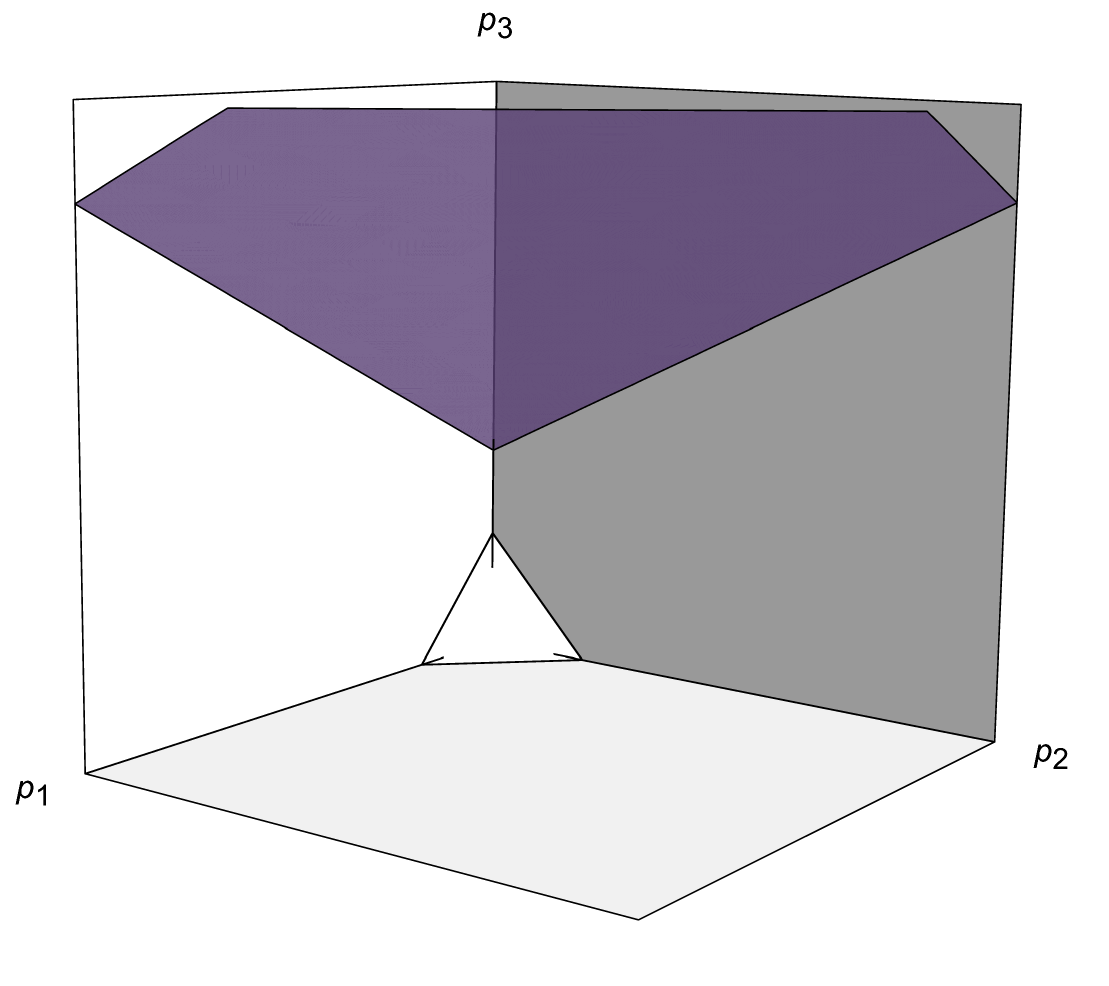}
\caption{Phase 3: $c<-t<0$}
\end{subfigure}
\caption{Hyperplane for a symplectic cut of local $\BP^2$ in three different phases with $t>0$. For $t<0$, there are instead two phases corresponding to $c>-\frac13 t$ and $c<-\frac13 t$, respectively.}
\label{fig:P2cutphases}
\end{figure}

Next, we reconsider the quantum cut of local $\BP^2$ studied in Section~\ref{sec:localP2-lauricella}. The associated Calabi--Yau fourfold is given by the symplectic quotient
\be
 \XIVd = (\BC^4\times\BC_+\times\BC_-)\sslash U(1)^2
\ee
defined by the following charge matrix
\be\label{eq:localP2-CY4-charges}
 \tilde{Q} = \left[
 \begin{array}{cccccc|c}
  z_1 & z_2 & z_3 & z_4 & z_+ & z_- \\
  \hline
  1 & 1 & 1 & -3 & 0 & 0 & t \\
  0 & 0 & -1 & 1 & -1 & 1 & c
 \end{array}
 \right]
\ee
Restricting to $t>0$, the geometry has three different phases in the moduli space of the parameter $c$, corresponding to different positions of the hyperplane of the symplectic cut
as depicted in Figure~\ref{fig:P2cutphases}.

The quantum volume of $\XIVd$ is given by the integral
\begin{multline}
\label{eq:F4d-localP2}
 \cF^D_{\XIVd}(t,c,\e,\e_\pm)
 = \int \frac{\dif\phi}{2\pi\ii} \int \frac{\dif\varphi}{2\pi\ii}
 \eu^{t\phi+c\varphi}
 \Ga(\e_1+\phi)\Ga(\e_2+\phi)\Ga(\e_3+\phi-\varphi) \\
 \times \Ga(\e_4-3\phi+\varphi)\Ga(\e_+-\varphi)\Ga(\e_-+\varphi)\,.
\end{multline}
This may be evaluated by choosing the appropriate Jeffrey--Kirwan contour for the phase of interest.
As in the case of $\BC^3$ (see Remark~\ref{rmk:analyticity}), it is expected that the quantum volumes of $\XIVd$ computed in different phases are related to each other by analytic continuation.

In the limit when $\e_\pm\to 0$ the leading order behavior of the fourfold quantum volume is obtained by
using (see footnote \ref{foot:beta-delta})
\be
 \Ga(\e_+-\varphi)\Ga(\e_-+\varphi)
 = \Ga(\e_++\e_-)\Beta(\e_+-\varphi,\e_-+\varphi)
 = \Ga(\e_++\e_-) \, 2\pi\ii\,\delta(\varphi) \, (1+O(\e_\pm))\,,
\ee
which gives
\be
 \cF^D_{\XIVd}
 = \int \frac{\dif\phi}{2\pi\ii}\eu^{t\phi}
 \Ga(\e_1+\phi)\Ga(\e_2+\phi)\Ga(\e_3+\phi)\Ga(\e_4-3\phi)
 \Ga(\e_++\e_-) \left(1+O(\e_\pm)\right)
\ee
so that, at leading order in $\e_\pm$, we recover the CY3 quantum volume $\cF^D(t,\e)$ as in Section~\ref{sec:localP2-lauricella}. Note that all dependence on $c$ automatically disappears in the limit.

Taking the limit $\e_\pm\to 0$ of the equivariant period $\Pi(D_+\cap D_-)$ associated to the intersection of divisors $D_\pm$ according to the general formula \eqref{eq:period-intersection-divisors-X4}, we obtain the equivariant disk potential $W(t,c,\e)$ given in \cite[Section~5.3]{Cassia:2023uvd} after matching conventions (see Remark~\ref{rmk:c-shift}). We leave the details of the computation to interested readers.

Next, we turn to the extended equivariant Picard--Fuchs equations.
The CY4 has both the closed string modulus $t$ and the open string modulus $c$. We can write down operators of the form \eqref{eq:PF-operator-generic}.
Here the index $a$ takes values $1,2$ corresponding, respectively, to $(t,c)$, and the extended charge matrix is given in \eqref{eq:localP2-CY4-charges}.
Taking $\gamma=(1,0), (0,1)$ and $(1,1)$ gives PF equations for $\XIVd$ involving the open modulus
\be
\begin{split}
	\PF_{(1,0)} & := \cD_1\cD_2\cD_3-\eu^{-t}\cD_4(\cD_4+1)(\cD_4+2) \\
	\PF_{(0,1)} & := \cD_4\cD_--\eu^{-c}\cD_3\cD_+ \\
	\PF_{(1,1)} & := \cD_1\cD_2 \cD_--\eu^{-c-t}\cD_4(\cD_4+1)\cD_+
\end{split}
\ee
where
\be
	\cD_{1,2} = \e_{1,2}+\partial_t\,,\qquad
	\cD_3 = \e_3+\partial_t-\partial_c\,,\qquad
	\cD_4 = \e_4-3\partial_t+\partial_c\,,\qquad
	\cD_\pm = \e_\pm \mp \partial_c\,.
\ee
The limit $\e_\pm\to 0$ gives differential equations \eqref{eq:q-cut-PF} for the equivariant disk potential
\be\label{eq:localP2-pre-inhomogeneous-PF}
\begin{split}
 \left(\cD_1\cD_2\cD_3-\eu^{-t}\cD_4(\cD_4+1)(\cD_4+2)\right) W(t,c,\e) &=0 \\
 \left(\cD_4+\eu^{-c}\cD_3\right) \partial_c\, W(t,c,\e) &=0 \\
 \left(\cD_1\cD_2+\eu^{-c-t}\cD_4(\cD_4+1)\right) \partial_c\, W(t,c,\e) &= 0
\end{split}
\ee
The first equation multiplied by $\partial_c$ gives the additional equation
\be
 \left(\cD_1\cD_2\cD_3-\eu^{-t}\cD_4(\cD_4+1)(\cD_4+2)\right) \partial_c\, W=0 \,.
\ee
Note that $\partial_c W$ is equal to $\Delta_c\cH^D$ according to
\eqref{eq:equiv-W-deriv}, with $\cH^D$ given in \eqref{eq;P2Hd3-resum-main}.
Using that
\be
 \Delta_c\,\eu^{-c} = -\eu^{-c}\Delta_c\,,
 \hspace{30pt}
 \Delta_c\,\eu^{-t} = \eu^{-t}\Delta_c\,,
\ee
the fourfold PF equations imply
\be
\begin{split}
 \Delta_c \left(\cD_1\cD_2\cD_3-\eu^{-t}\cD_4(\cD_4+1)(\cD_4+2)\right) \cH^D(t,c,\e) &=0 \\
 \Delta_c \left(\cD_4-\eu^{-c}\cD_3\right) \cH^D(t,c,\e) &=0 \\
\end{split}
\ee
which are exactly the PF equations for $\XIId$, up to the action of $\Delta_c$ from the left.

We next consider the non-equivariant limit of these PF equations. Recall expression \eqref{eq:P2-case1} for the regular part of the monodromy $\Delta_c[\cH^D]_0\propto\partial_c W_\disk\propto\log(-y_1(-x))$, with $y_1$ given in \eqref{eq:localP2-mirror-sheets}. The regular part obeys the following equations
\be\label{eq:localP2-open-PF}
\begin{aligned}
 \left((-3\partial_t+\partial_c)+\eu^{-c}(\partial_t-\partial_c)\right)
 \partial_c\,W_\disk &=	\eu^{-c} \\
 \left(\partial_t^2+\eu^{-c-t}(-3\partial_t+\partial_c)(-3\partial_t+\partial_c+1)\right)
 \partial_c\,W_\disk &=	0
\end{aligned}
\ee
The operators on the left correspond to the na\"ive non-equivariant limit of the operators in~\eqref{eq:localP2-pre-inhomogeneous-PF},
while the inhomogeneous term on the r.h.s.\ appears in the same way it
did for $\BC^3$ in the last section.
This is generated by the terms of order $O(\e)$ in the operators $\PF_\gamma$ acting on the \emph{singular} terms of order $O(\e^{-1})$ in the equivariant disk potential $W(t,c,\e)$. As in the previous example, we find that the homogeneous extended PF equations are replaced by inhomogeneous equation in the non-equivariant limit. Again this is in line with earlier results from open Gromov--Witten theory, see Remark~\ref{rmk:inhomogeneity}.

\section{Two branes, double symplectic cuts and CY5}\label{sec:double-cut}

In this section, we consider pairs of toric branes in Calabi--Yau threefolds, and model them by double symplectic cuts. We give a description of double symplectic cuts in terms of a Calabi--Yau fivefold, extending Braverman's original construction.

\subsection{Double cuts from Calabi--Yau fivefolds}

Given a toric Calabi--Yau threefold $\XIIId$, one may consider two symplectic cuts defined by charge vectors $q^1$ and $q^2$. As in the case of a single cut, each of $q^\alpha$'s defines a family of framed toric branes. The parameter space of the CY3 with branes is the K\"ahler moduli space with local coordinates $(t^a,c_\alpha)$, and consists of several chambers corresponding to the phases of the two branes and of the threefold.\footnote{To obtain the toric Lagrangian one needs to complement each $q^\alpha$ with another charge vector. This is uniquely fixed by a choice of chamber, by taking the unique hyperplane orthogonal to $(1,1,1)$ that also contains the corresponding edge of the toric diagram.}

Braverman's construction relating symplectic cuts of toric threefolds to toric fourfolds has a natural generalization to the case of double cuts.
We consider the Calabi--Yau fivefold defined by the symplectic quotient
\be\label{eq:CY5-def}
    \XVd :=
    \BC^{r+3}\times (\BC_+\times \BC_-)\times (\BC^+\times \BC^-) \ \sslash \ U(1)^{r+2}
\ee
where $U(1)^{r+2}$ acts according to the following charge matrix
\be\label{eq:CY5-charges}
 \left[
    \begin{array}{ccccc|c}
    z_i & z_+ & z_- & z^+ & z^- & \\
    \hline
    Q^a_i & 0 & 0 & 0 & 0 & t^a \\
    q^1_i & -1 & 1 & 0 & 0 & c_1 \\
    q^2_i & 0 & 0 & -1 & 1 & c_2
    \end{array}
 \right]
\ee

Similar to the fourfold case, $\XVd$ admits a fibration over $\BC^2$
\be\label{eq:Braverman-map-5d}
    \pi: \ \XVd \to \BC\times \BC
\ee
defined by
\be\label{eq:Braverman-map-coordinates-5d}
    \pi: \ (x,z_+, z_-,z^+,z^-) \mapsto (z_+ z_-, z^+ z^-)\,\qquad x\in\XIIId,\ z_\pm\in \BC_{\pm},\ z^\pm\in \BC^{\pm} \,.
\ee
The $U(1)^2$ action defined by $q^\alpha_i$ on the ambient space descends to a $U(1)^2$ action on $\XIIId\times(\BC_+\times\BC_-)\times(\BC^+\times\BC^-)$.
It follows that
\be
 \XVd\simeq\XIIId\times\BC_+\times\BC_-\times\BC^+\times\BC^-\sslash U(1)^2\,.
\ee
Next we consider fibers of \eqref{eq:Braverman-map-5d}. If $z_+ z_- = w_1$ and $z^+ z^- = w_2$ are both nonzero, the fiber is complex isomorphic to $\XIIId$
\be
 \pi^{-1}(w_1,w_2)
 = \left\{
 \begin{array}{c}
 \mu_{q^1}(x)=c_1+|z_+|^2-|z_-|^2, \\
 \mu_{q^2}(x)=c_2+|z^+|^2-|z^-|^2, \\
 z_+z_-=w_1,\;z^+z^-=w_2
 \end{array}
 \right\}\Bigg/U(1)^2\simeq\XIIId\,.
\ee
Here $\mu_{q^\alpha}$ denote the moment maps of the $U(1)^2$ action on $\XIIId$, as it descends from the action defined by $q^\alpha$ on the ambient manifold $\BC^{r+3}\times (\BC_{+}\times \BC_-) \times (\BC^{+}\times \BC^-)$, cf. \eqref{eq:Braverman-moment-map}.

To see this, we proceed in a way that is analogous to the fourfold case.
First, we observe that the map $\pi$ is invariant under the $U(1)^2$-action, so that it gives rise to a well-defined map on the quotient $\XVd$ as in \eqref{eq:Braverman-map-5d}.
Moreover, the conditions $z_+ z_-=w_1$, $z^+ z^-=w_2$ and
$\mu_{q^1}=c_1+|z_+|^2-|w_1|^2/|z_+|^2$, $\mu_{q^2}=c_2+|z^+|^2-|w_2|^2/|z^+|^2$ define,
for each $x\in\XIIId$, a torus $T^2$ with fixed values of, say $|z_+|^2$ and $|z^+|^2$.
The fact that we get a torus, \ie{} that $|z_+|,|z^+|>0$ is crucial, and it is ensured by the assumption $w_1, w_2\neq 0$.
The $U(1)^2$ quotient then reduces this torus to a point, showing that to each point in the base $(w_1,w_2)$ corresponds a copy of $\XIIId$.

The fiber degenerates when either of $w_\alpha$ vanishes. Over the locus $w_1=0, w_2\neq 0$ the fiber is complex isomorphic to the symplectic cut defined by $q^1$, while over the locus $w_1\neq 0, w_2 = 0$ is complex isomorphic to the symplectic cut defined by $q^2$.
At the origin $w_1=w_2=0$ the fiber is complex isomorphic to the double cut of $\XIIId$.

We focus on the most degenerate case, \ie{} the fiber over the origin.
The fiber has now four components corresponding to independent choices of signs in
\be
    z_\pm =0 \  \text{with}\  z_\mp\in \BC_{\mp}\,,
    \text{ and }
    z^\pm =0  \  \text{with}\  z_\mp\in \BC^{\mp}\,.
\ee
Each leads to a reduction of \eqref{eq:CY5-def} to a threefold that corresponds to a `quarter space' defined by the double symplectic cut. For example, the reduction of the locus where
$z_-=z^-=0$ corresponds to the intersection of two toric divisors, which we denote as $D_-$ and $D^-$. The intersection can be described as
\be\label{eq:quarter-space-toric-divisors}
 D_- \cap D^-
 = \left\{(x,z_+,z^+)\in\XIIId\times\BC_+\times\BC^+\;\middle|\;
 \begin{array}{l}
  \mu_{q^1}(x) = c_1 + |z_+|^2\,, \\
  \mu_{q^2}(x) = c_2 + |z^+|^2
 \end{array}
 \right\}\Big/U(1)^2 \,,
\ee
which we will denote as $\XIIId[>,>]$ to indicate that it is obtained as a reduction of the locus where $\mu_{q^1}(x) > c_1$ and $\mu_{q^2}(x) > c_2$.
The remaining three spaces $\XIIId[>,<],\XIIId[<,>],\XIIId[<,<]$ are defined in a similar way by the requirement that $\mu_{q^1}(x) \lessgtr c_1$ and $\mu_{q^2}(x) \lessgtr c_2$ in $\XIIId$ (with signs chosen according to the labels).
This presentation makes it clear that all of the $\XIIId[\lessgtr,\lessgtr]$ are obtained as intersections of toric divisors $D_\pm$ and $D^\pm$ inside of $\XVd$.

Of special interest will be the intersection locus $z_\pm=z^\pm=0$ of all four divisors, which defines a complex variety of dimension one
\be\label{eq:X1d-in-X5d}
 D_+\cap D_-\cap D^+\cap D^-
 = \left\{x\in\XIIId\;\middle|\;\mu_{q^1}(x) = c_1\,,\; \mu_{q^2}(x)
 = c_2 \right\} / U(1)^2\,.
\ee
Note that, by a standard result for symplectic quotients known as ``\emph{reduction in stages}'' \cite{MR2091310}, this variety can be viewed either as a quotient of the threefold $\XIIId\sslash U(1)^2$ or equivalently
as the quotient $\BC^{r+3}\sslash U(1)^{r+2}$.
We will then define the onefold associated to the double cut as follows
\be\label{eq:CY1-def}
    \XId := \left\{ \sum_{i=1}^{r+3} Q^a_i |z_i|^2 = t^a,\
    \sum_{i=1}^{r+3} q^1_i |z_i|^2 = c_1,\
    \sum_{i=1}^{r+3} q^2_i |z_i|^2 = c_2\right\} \Big/ U(1)^{r+2}\,.
\ee
Here both $q^\alpha$ have entries that add up to zero because of our assumptions on the type of cuts, which implies that the canonical bundle of $\XId$ is trivial, so that it is a Calabi--Yau onefold.

\subsection{Quantization of the double cut of a CY3}

We now turn to the quantization of the double cut of a toric Calabi--Yau threefold.
First, we define the quantum volume of the Calabi--Yau onefold defined in \eqref{eq:CY1-def} as the equivariant disk partition function of an \eGLSM{}
\be\label{eq:KD-def}
 \cK^D(t,c_1,c_2,\e) =
 \int \frac{\dif\varphi_1}{2\pi\ii}
 \int \frac{\dif\varphi_2}{2\pi\ii}
 \int \prod_{a=1}^r\frac{\dif\phi_a}{2\pi\ii}
 \eu^{\varphi_1c_1+\varphi_2c_2+\sum_a\phi_a t^a}
 \prod_{i=1}^{r+3}\Ga(\e_i+\sum_a\phi_aQ^a_i+\varphi_1q^1_i+\varphi_2q^2_i)\,.
\ee
This integral can be evaluated explicitly, and the result is
\be\label{eq:KD-B-int}
\begin{aligned}
 \cK^D(t,c_1,c_2,\e)
 &= \Ga\Big({\sum_i}\e_i\Big)\, \eu^{-t\cdot M\cdot\e}
 \frac{x_1^{A_1\cdot\e}x_2^{A_2\cdot\e}}
 {H(x_1,x_2,z)^{\sum_i\e_i}}\,,
\end{aligned}
\ee
where $x_\alpha=\eu^{-c_\alpha}$ and $z_a=\eu^{-t^a}$ while $A_\alpha\cdot\e$ and $t\cdot M\cdot\e$ are certain linear functions of $\e_i$ (and $t^a$).
The function in the denominator is a complex power of $H(x_1,x_2,z)$, which is a Laurent polynomial in $x_\alpha$ and $z_a$. It is related to the Hori--Vafa mirror curve of the Calabi--Yau threefold $\XIIId$ by a simple change of coordinates.
More details and a derivation of this formula are given in Appendix~\ref{sec:B-model-stuff}.

Via the function $\cK^D$, we can further define the four `quarter space' functions associated to $\XIIId[\lessgtr,\lessgtr]$ as defined by the double cut procedure.
They are as follows
\be\label{eq:double-cut-quarter-volumes}
\begin{aligned}
 \cF^D_{<,<}
 & = \int_{-\infty}^{c_1} \dif c'_1\, \int_{-\infty}^{c_2} \dif c'_2\,
 \cK^D(t,c'_1,c'_2,\e)\,, \\
 \cF^D_{>,<}
 & = \int_{c_1}^{\infty} \dif c'_1\, \int_{-\infty}^{c_2} \dif c'_2\,
 \cK^D(t,c'_1,c'_2,\e)\,,
\end{aligned}
\qquad
\begin{aligned}
 \cF^D_{<,>}
 & = \int_{-\infty}^{c_1} \dif c'_1\, \int_{c_2}^{\infty} \dif c'_2\,
 \cK^D(t,c'_1,c'_2\e)\,, \\
 \cF^D_{>,>}
 & = \int_{c_1}^{\infty} \dif c'_1\, \int_{c_2}^{\infty} \dif c'_2\,
 \cK^D(t,c'_1,c'_2,\e)\,.
\end{aligned}
\ee
For later convenience, we also introduce the double Fourier transform of $\cK^D$
\be
\begin{aligned}
 \tilde\cK^D(t,\varphi_1,\varphi_2,\e)
 & = \int_{\BR} \dif c_1 \,\eu^{-\varphi_1 c_1}
 \int_{\BR} \dif c_2 \,\eu^{-\varphi_2 c_2}\,
 \cK^D(t,c_1,c_2,\e)\qquad (\varphi_\alpha\in\ii\BR) \\
 & = \int \prod_{a=1}^{r}\frac{\dif\phi_a}{2\pi\ii} \eu^{\sum_a\phi_at^a}
 \prod_{i=1}^{r+3}\Ga(\e_i+\sum_a\phi_aQ^a_i+\varphi_1q^1_i+\varphi_2q^2_i)
\end{aligned}
\ee
This can be viewed as a deformation of quantum volumes encountered previously, in the following sense
\be
 \tilde\cK^D(t,\varphi_1,0,\e) = \tilde\cH_1^D(t,\varphi_1,\e)\,,
 \hspace{30pt}
 \tilde\cK^D(t,0,\varphi_2,\e) = \tilde\cH_2^D(t,\varphi_2,\e)\,,
\ee
\be
 \tilde\cK^D(t,0,0,\e) = \cF^D(t,\e)\,.
\ee
The latter expression can be viewed as a statement that $\cK^D$ is also a quantum Lebesgue measure for $\XIIId$
\be\label{eq:FD-B-int}
 \cF^D(t,\e) = \int_{\BR} \dif c_1 \int_{\BR} \dif c_2 \, \cK^D(t,c_1,c_2,\e)\,.
\ee
In fact, also $\cH^D$ can be expressed in terms of $\cK^D$
\be\label{eq:HD-B-int}
\begin{split}
 \cH_1^D(t,c_1,\e) & = \int_{\BR} \dif c_2 \, \cK^D(t,c_1,c_2,\e)\,, \\
 \cH_2^D(t,c_2,\e) & = \int_{\BR} \dif c_1 \, \cK^D(t,c_1,c_2,\e)\,.
\end{split}
\ee
In this sense, $\cK^D$ is the most fundamental in the hierarchy of quantum volumes.

Finally, we remark also that the regular term in the non-equivariant expansion of $\cK^D$ can be computed straightforwardly as
\be
 \left[\cK^D(t,c_1,c_2,\e)\right]_0 = P_{\cK}(t,c) - \log H(x_1,x_2,z)
\ee
where $P_{\cK}(t,c):=-(t\cdot M\cdot\e+c\cdot A\cdot\e)/\sum_i\e_i-\gamma$ is a polynomial in $t^a$ and $c_\alpha$.

\subsection{Quantization of the CY5}

Next, we consider the quantization of the fivefold construction.
The quantum volume of  $\XVd$ is given by the equivariant disk partition function with a space-filling brane
\begin{multline}
    \cF^D_{\XVd}(t,c_1,c_2,\e,\e_\pm,\e^\pm)
    = \int \frac{\dif\varphi_1}{2\pi\ii}
    \Ga(\e_+ -\varphi_1)\Ga(\e_- +\varphi_1)
    \int \frac{\dif\varphi_2}{2\pi\ii}
    \Ga(\e^+ -\varphi_2)\Ga(\e^- +\varphi_2)
    \\
    \times \int \prod_{a=1}^{r}\frac{\dif\phi_a}{2\pi\ii}
    \eu^{c_1\varphi_1+c_2\varphi_2+ \sum_a\phi_at^a}
    \prod_{i=1}^{r+3}\Ga(\e_i+\sum_{a=1}^r\phi_aQ^a_i+\varphi_1q^1_i+\varphi_2q^2_i)
\end{multline}
This can be rewritten in terms of $\cK^D$ in following suggestive way
\be\label{eq:KD-5d-FD}
\begin{aligned}
    & \cF^D_{\XVd}(t,c_1,c_2,\e,\e_\pm,\e^\pm) \\
    & = \int \frac{\dif\varphi_1}{2\pi\ii}
    \eu^{c_1\varphi_1}
    \Ga(\e_+ -\varphi_1)\Ga(\e_- +\varphi_1)
    \int \frac{\dif\varphi_2}{2\pi\ii}
    \eu^{c_2\varphi_2}
    \Ga(\e^+ -\varphi_2)\Ga(\e^- +\varphi_2)
    \tilde\cK^D(t,\varphi_1,\varphi_2,\e) \\
    & = \int_{\BR} \dif c'_1 \int_{\BR} \dif c'_2\
    \cK^D(t,c_1',c_2',\e)\cdot
    \rho(c_1-c_1',\e_\pm)\, \rho(c_2-c_2',\e^\pm)\,.
\end{aligned}
\ee
where $\rho$ was defined in \eqref{eq:rho}. Note that $\cK^D$ does not carry any dependence on $\e_\pm,\e^\pm$.
Acting with divisor operators we may define the equivariant periods
\begin{multline}
\label{eq:double-cut-quarter-volumes-5d}
    \Pi(D_+\cap D^+) =
    \frac{(1-\eu^{2\pi\ii(\e_+ - \partial_{c_1})})}{2\pi\ii}
    \frac{(1-\eu^{2\pi\ii(\e^+ - \partial_{c_2})})}{2\pi\ii}
    \cF^D_{\XVd} \\
    = \int_{\BR} \dif c_1' \, \int_{\BR} \dif c_2' \, \cK^D(t,c'_1,c'_2,\e)
    \cdot
    \frac{(1-\eu^{2\pi\ii (\e_+ - \partial_{c_1})})}{2\pi\ii}
    \frac{(1-\eu^{2\pi\ii (\e^+ - \partial_{c_2})})}{2\pi\ii}
    \rho(c_1-c_1',\e_\pm)\,
    \rho(c_2-c_2',\e^\pm)
\end{multline}
and similarly for $\Pi(D_-\cap D^+)$, $\Pi(D_+\cap D^-)$, $\Pi(D_-\cap D^-)$.
By construction, these are solutions of the PF equations of $\XVd$.

It follows from \eqref{eq:rho-limit} that in the limit $\e_\pm, \e^\pm \to 0$ the quantum volume of the Calabi--Yau fivefold reduces to the quantum volume of the threefold $\XIIId$, times divergent factors related to the additional diagonal subspaces of $(\BC_+\times\BC_-)\times (\BC^+\times \BC^-)$,
\be
\begin{aligned}
    \cF^D_{\XVd}(t,c_1,c_2,\e,\e_\pm,\e^\pm)
    & =
    \cF^D(t,\e)\,
    \Ga(\e_++\e_-)\,
    \Ga(\e^++\e^-)
    \left( 1 + O(\e_\pm)+ O(\e^\pm) \right)\,.
\end{aligned}
\ee
Recalling \eqref{eq:KD-5d-FD}, this recovers the relation between $\cF^D$ and $\cK^D$ for the CY3 stated in \eqref{eq:FD-B-int}.

Indeed one may also consider the partial limits where equivariance is turned off only along part of the extra dimensions. This recovers the Calabi--Yau fourfolds associated to each of the two cuts individually. Let $\XIVd^{(1)}$ denote the fourfold associated with the cut $q^1_i$, and $\XIVd^{(2)}$ that associated with the cut $q^2_i$, then we have the relations
\be
\begin{aligned}
 \cF^D_{\XVd}(t,c_1,c_2,\e,\e_\pm,\e^\pm)
 & = \cF_{\XIVd^{(1)}}^D(t,\e,\e_\pm)\,\Ga(\e^++\e^-)\left(1+O(\e^\pm)\right)
\end{aligned}
\ee
\be
\begin{aligned}
 \cF^D_{\XVd}(t,c_1,c_2,\e,\e_\pm,\e^\pm)
 & = \cF_{\XIVd^{(2)}}^D(t,\e,\e^\pm)\,\Ga(\e_++\e_-)\left(1+O(\e_\pm)\right)
\end{aligned}
\ee
Finally, taking the limit $\e_\pm,\e^\pm \to 0$ of \eqref{eq:double-cut-quarter-volumes-5d} gives the shifted quarter volumes of $\XIIId$ computed earlier in \eqref{eq:double-cut-quarter-volumes}. To see this, we recall the identity \eqref{eq:Theta-H-def} which gives
\be\label{eq:Fminmin-from-KD}
\begin{aligned}
 \lim_{\e_\pm,\e^\pm\to 0}
 \Pi(D_+\cap D^+)
 & = \int_{\BR} \dif c_1' \, \int_{\BR} \dif c_2' \, \cK^D(t,c'_1,c'_2,\e)
 \cdot \Theta_H(c_1-c_1'-\pi\ii)\,\Theta_H(c_2-c_2'-\pi\ii) \\
 & = \cF^D_{<,<}(t,c_1-\pi\ii,c_2-\pi\ii,\e)\,.
\end{aligned}
\ee

\subsection{Extended Picard--Fuchs equations for double cuts}\label{sec:ext-PF--CY5}

The equivariant periods associated to divisors of the CY5, and their intersections \eqref{eq:double-cut-quarter-volumes-5d}, satisfy the Picard--Fuchs equations for the \eGLSM{} describing $\XVd$.
Taking the 3d limit gives a two-parameter extension of the Picard--Fuchs equations of the CY3, whose solutions include the `quarter volumes' $\cF^D_{\lessgtr,\lessgtr}$ defined in \eqref{eq:double-cut-quarter-volumes}.
These doubly-extended Picard--Fuchs equations represent a generalization of the extended Picard--Fuchs equations obtained in the previous section for a single symplectic cut, to the case of two cuts.

The logic is similar to the one developed in Section~\ref{sec:4d-PF-3d}.
The quantum volume of $\XVd$ obeys a system of equations of the form
\be
    \PF_\gamma \cdot \cF^D_{\XVd} = 0
\ee
where $\PF_\gamma$ is an operator of the form \eqref{eq:PF-operator-generic}
with $\tilde Q$ denoting now the charge matrix $Q$ augmented by $q^1,q^2$ and by the extension to $\BC_+\times\BC_-\times\BC^+\times\BC^-$ as in \eqref{eq:CY5-charges}.

Restricting to vectors of the form $\gamma = (\gamma_1,\gamma_2,\dots,\gamma_r,0,0)$
gives operators $\PF_\gamma$ that formally have the same structure as the Picard--Fuchs operators of $\XIIId$, with the difference that $\cD_i$ now also include contributions $q^1_i\partial/\partial c_1+q^2_i\partial/\partial c_2$ from the moduli of the double symplectic cut.

The same operators annihilate also equivariant periods of divisors, and intersections thereof. In particular, the commutation relations
\be
 \left[\PF_\gamma\,,\, (1-\eu^{2\pi\ii\cD_\pm})\right] = 0
 = \left[\PF_\gamma \,,\, (1-\eu^{2\pi\ii\cD^\pm})\right]
\ee
imply that $\PF_\gamma$ annihilates $\Pi(D_\pm\cap D^\pm)$.

Taking the limit $\e_\pm,\e^\pm\to 0$ of these equations leads to Picard--Fuchs equations involving open string moduli for the volumes of the four `quarter spaces'
\be\label{eq:q-cut-PF-2}
 \left.\PF_\gamma\right|_{\e_\pm=0} \cdot \cF^D_{\lessgtr,\lessgtr}(t,c_1,c_2,\e) = 0\,.
\ee
These equations represent a generalization of the Picard--Fuchs equations for $\XIIId$, which involve dependence on $c_1, c_2$.
The solutions that are independent of $c_\alpha$ correspond to $\cF^D$ and to the standard equivariant periods of $\XIIId$.
However, among the solutions with nontrivial dependence on $c_\alpha$ we find the quarter periods $\cF^D_{\lessgtr,\lessgtr}$.

\subsection{Examples}

\subsubsection{\texorpdfstring{$\BC^3$}{C3}}\label{sec:C3-5d}

\begin{figure}[!ht]
\centering
\begin{subfigure}[b]{0.32\textwidth}
\centering
\includegraphics[width=\textwidth]{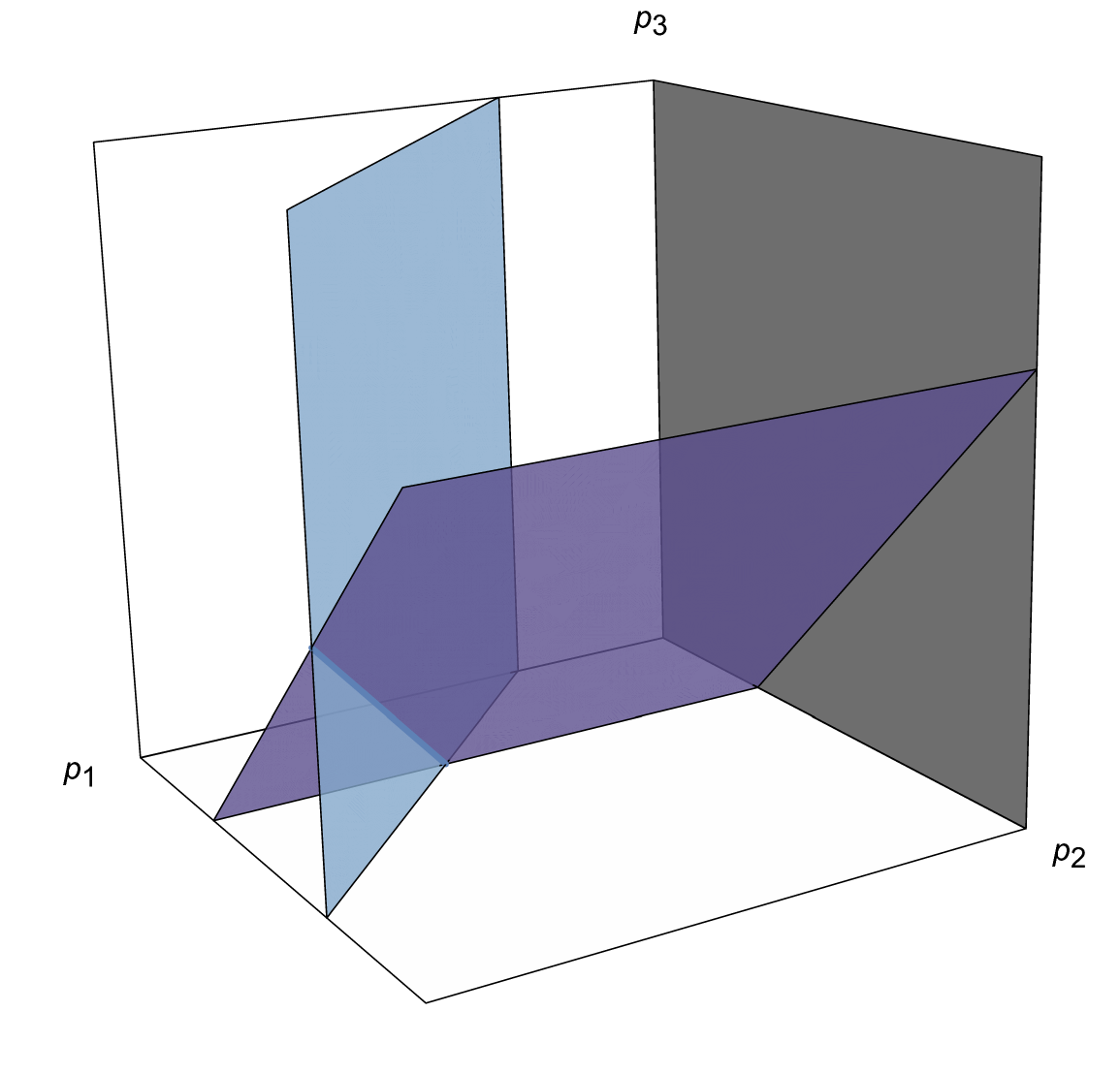}
\caption{Phase 1: $c_1,c_2>0$}
\end{subfigure}
\hfill
\begin{subfigure}[b]{0.32\textwidth}
\centering
\includegraphics[width=\textwidth]{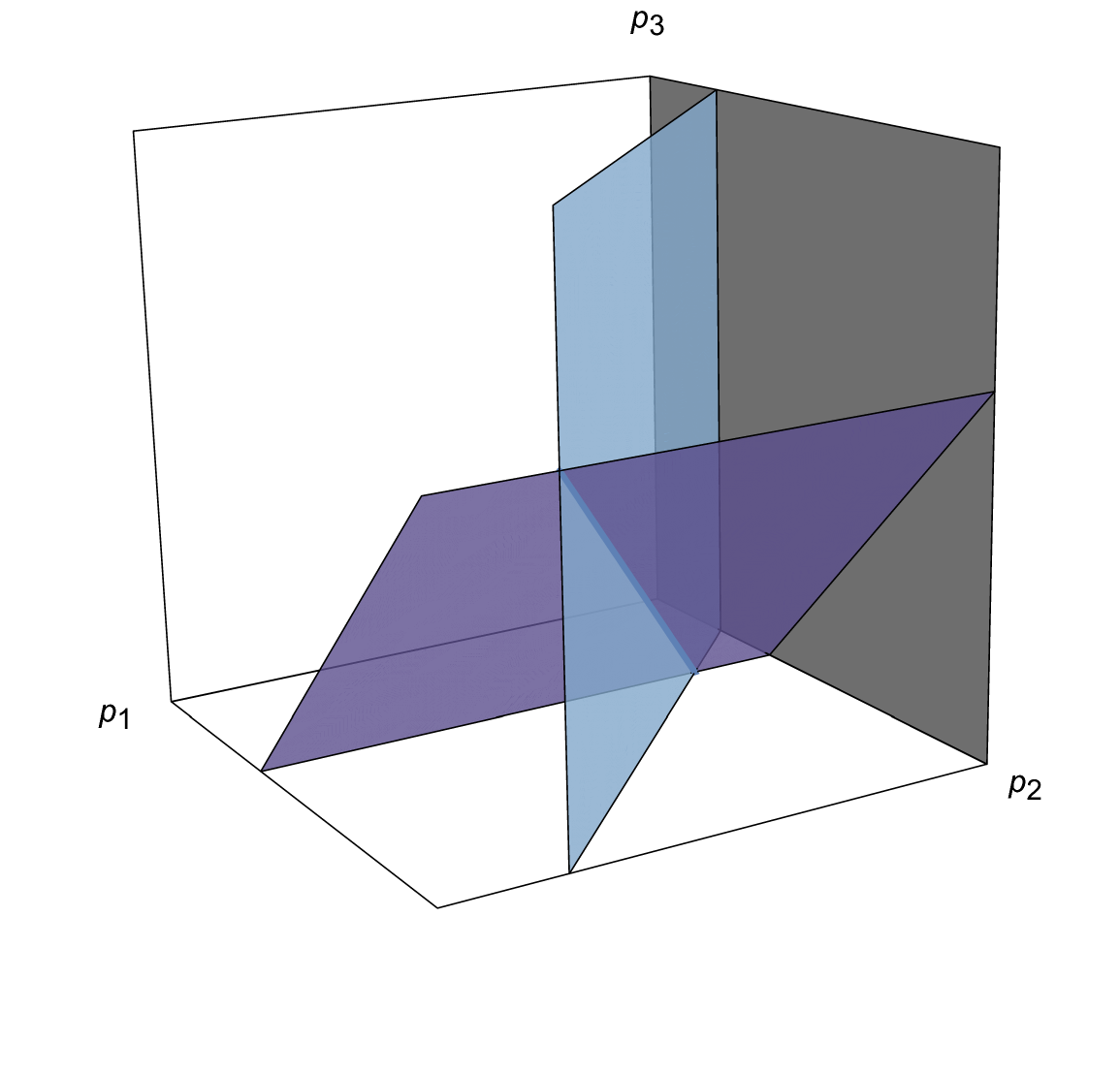}
\caption{Phase 2: $c_1>-c_2>0$}
\end{subfigure}
\hfill
\begin{subfigure}[b]{0.32\textwidth}
\centering
\includegraphics[width=\textwidth]{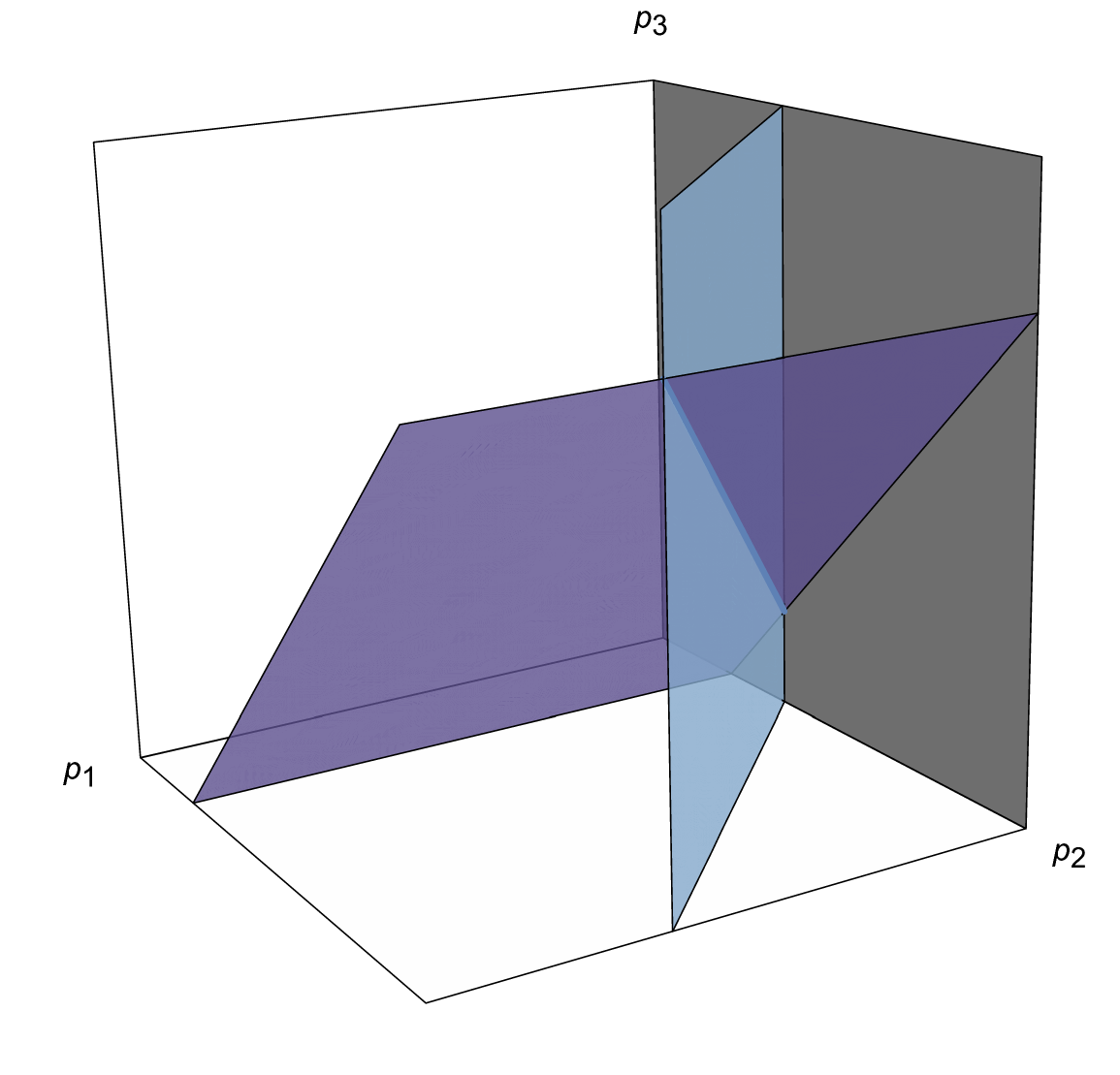}
\caption{Phase 3: $-c_2>c_1>0$}
\end{subfigure}
\\
\begin{subfigure}[b]{0.32\textwidth}
\centering
\includegraphics[width=\textwidth]{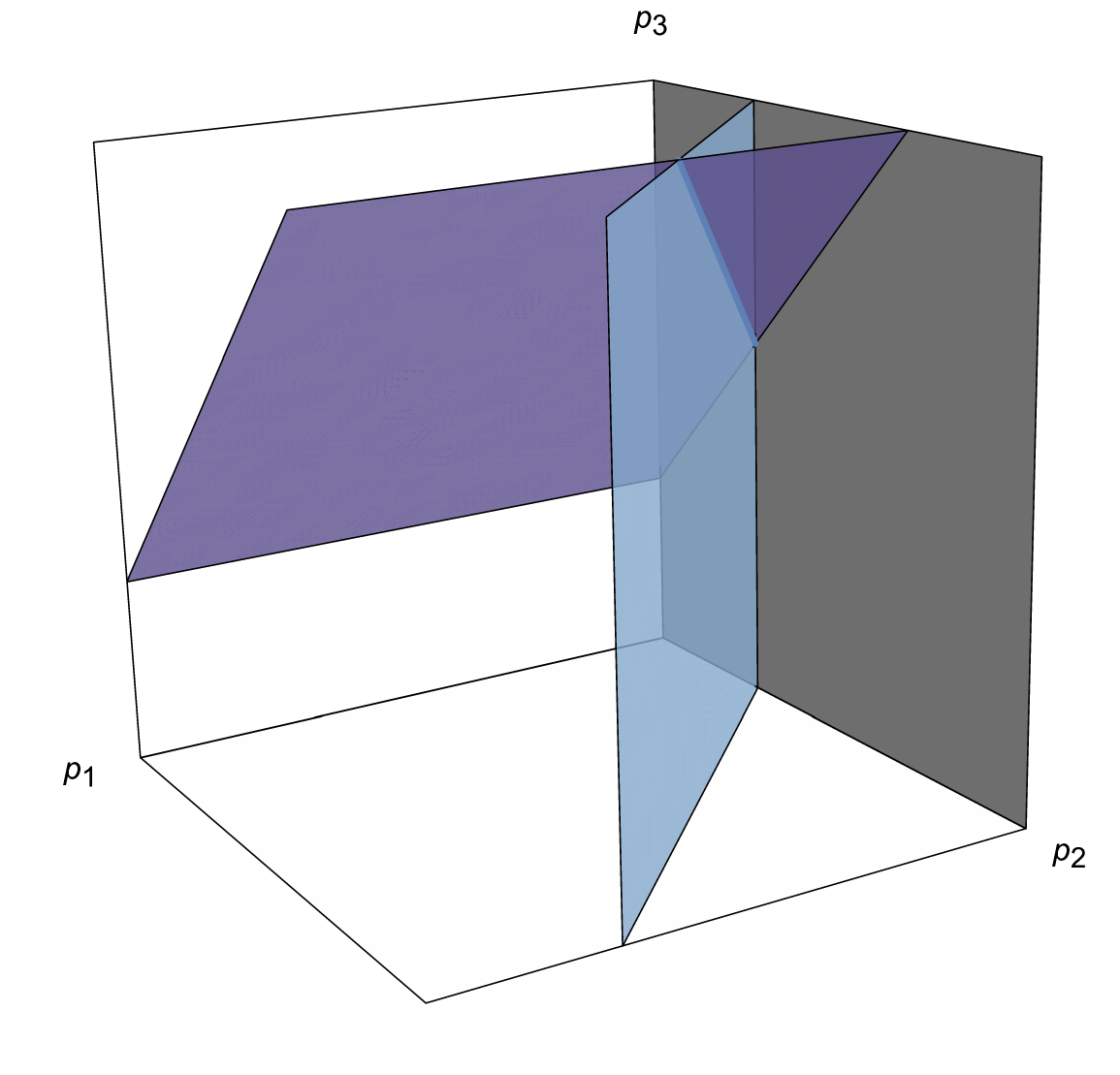}
\caption{Phase 4: $c_1,c_2<0$}
\end{subfigure}
\hspace*{.1\textwidth}
\begin{subfigure}[b]{0.32\textwidth}
\centering
\includegraphics[width=\textwidth]{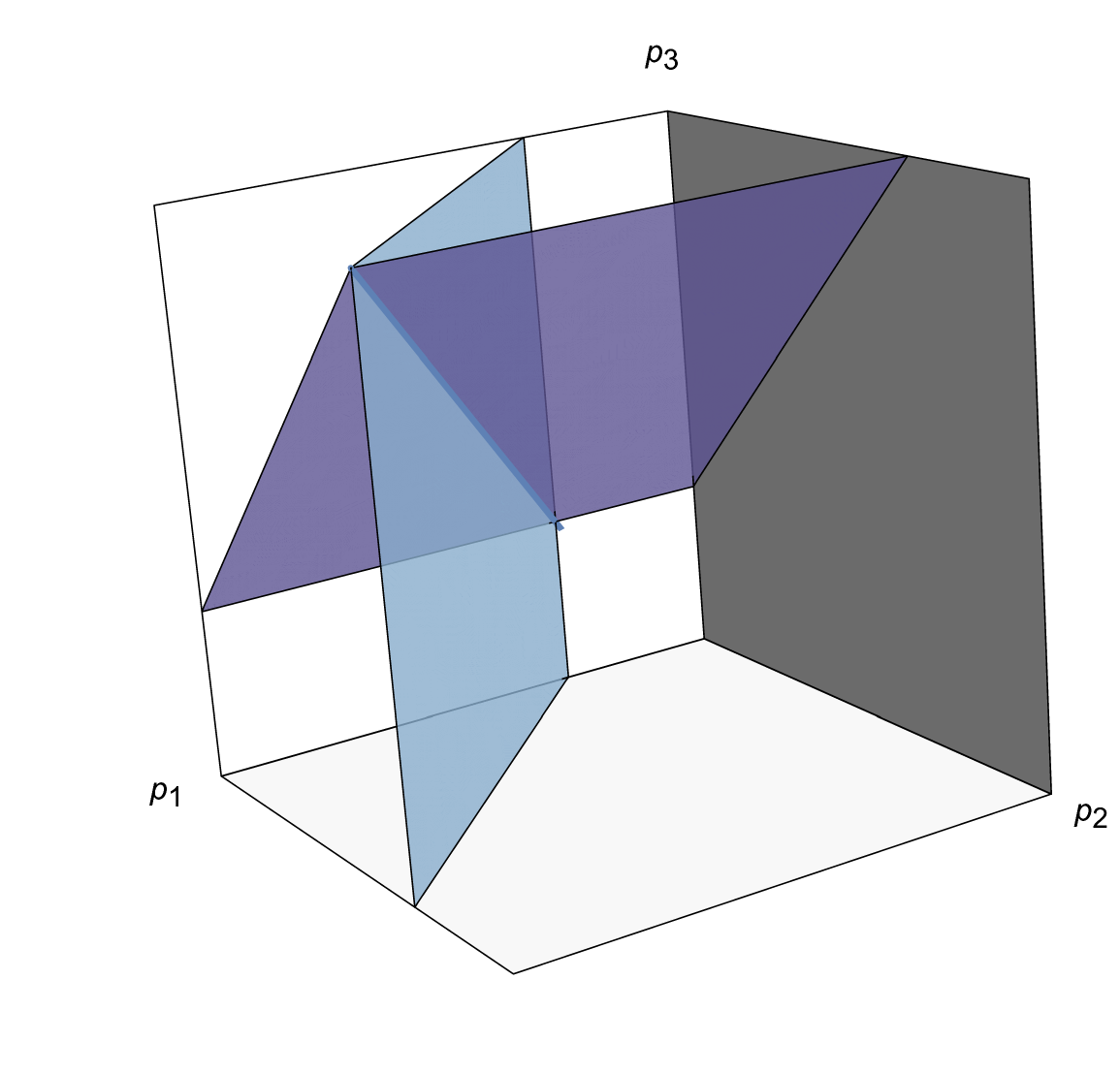}
\caption{Phase 5: $c_2>0, c_1<0$}
\end{subfigure}
\caption{Hyperplanes for a double symplectic cut of $\BC^3$ in five different phases.}
\label{fig:C3doublecutphases}
\end{figure}

\begin{figure}[!ht]
\begin{center}
\includegraphics[width=0.85\textwidth]{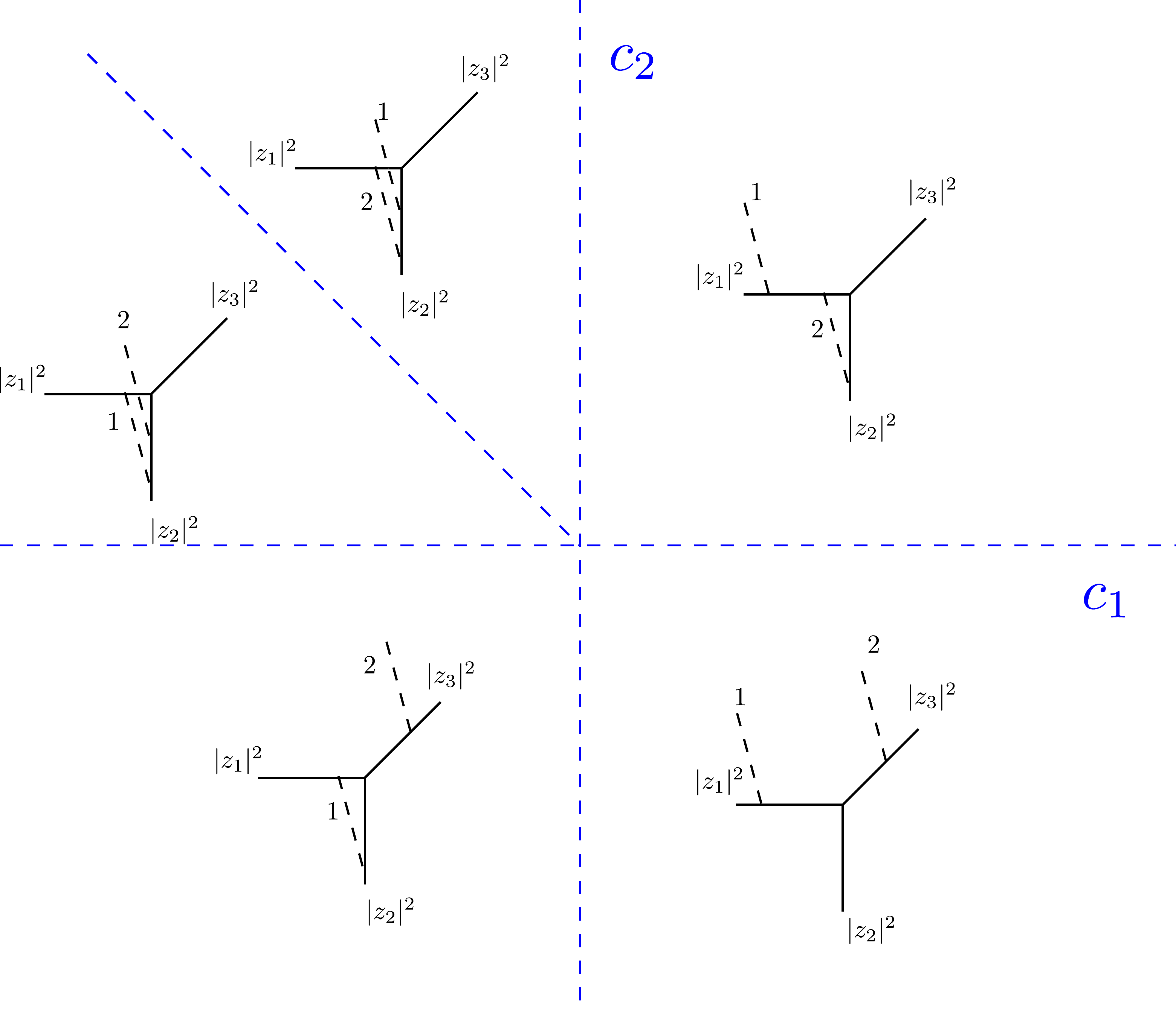}
\caption{Phases of the geometry \eqref{eq:C3-double-cut-Q-matrix}. }
\label{fig:C3-CY5-phases}
\end{center}
\end{figure}

We consider a double cut of $\BC^3$
\be
 \XVd = (\BC^3 \times \BC_+\times \BC_- \times \BC^+\times \BC^-)\sslash U(1)^2
\ee
defined by the following charge matrix
\be\label{eq:C3-double-cut-Q-matrix}
 \left[
    \begin{array}{ccccccc|c}
    z_1 & z_2 & z_3 & z_+ & z_- & z^+ & z^- \\
    \hline
    0 &  1 & -1 & -1 & 1 &  0 & 0 & c_1 \\
    1 & -1 &  0 &  0 & 0 & -1 & 1 & c_2 \\
    \end{array}
 \right]
\ee
Different phases of the geometry correspond to different positions of the two hyperplanes defining the symplectic cuts, see Figure \ref{fig:C3doublecutphases}. An alternative way to think about phases is to recall that they also describe relative positions of the two branes in the associated CY3, see Figure~\ref{fig:C3-CY5-phases}.

The quantum volume of $\XVd$ is given by the integral expression
\begin{multline}
\label{eq:C3-F5d}
 \cF^D_{\XVd}(c_1,c_2,\e) =
 \int \frac{\dif\varphi_1}{2\pi\ii} \int \frac{\dif\varphi_2}{2\pi\ii}
 \eu^{c_1\varphi_1+c_2\varphi_2}
 \Ga(\e_+ -\varphi_1)\Ga(\e_- +\varphi_1)
 \Ga(\e^+ -\varphi_2)\Ga(\e^- +\varphi_2) \\
 \times
 \Ga(\e_1+\varphi_2)\Ga(\e_2+\varphi_1-\varphi_2) \Ga(\e_3-\varphi_1) \,.
\end{multline}
The integral can be evaluated in several ways.
One of them is to sum over poles selected by an appropriate Jeffrey--Kirwan contour.
The choice of contour depends on the phase of the geometry in the moduli space of $c_1$, $c_2$ (in this case there are no closed string moduli $t^a$).
Although the contour changes discontinuously with the phase, the quantum volumes obtained by different contours are expected to be related to each other by analytic continuation. See Remark~\ref{rmk:analyticity}.

Another approach is to switch to integral representations of $\Ga$ functions, which leads to $B$-model-like integrals, such as those discussed in Appendix~\ref{app:technical-identities}.

The double cut of the associated CY3 defines a CY1 which we denote as $\XId$,
whose quantum volume is given by formula \eqref{eq:KD-B-int}, which in this case reads
\be\label{eq:KD-C3}
\begin{aligned}
 \cK^D(c_1,c_2,\e)
 & = \Ga(\e_1+\e_2+\e_3)
 \frac{x_1^{\e_1+\e_2}x_2^{\e_1}}{(1+x_1(1+x_2))^{\e_1+\e_2+\e_3}}\,.
\end{aligned}
\ee
The details computation are given in Appendix~\ref{sec:B-model-stuff-example}.

\subsubsection{Local \texorpdfstring{$\BP^2$}{P2}}\label{sec:localP2-5d}

\begin{figure}[!ht]
\centering
\begin{subfigure}[b]{0.32\textwidth}
\centering
\includegraphics[width=\textwidth]{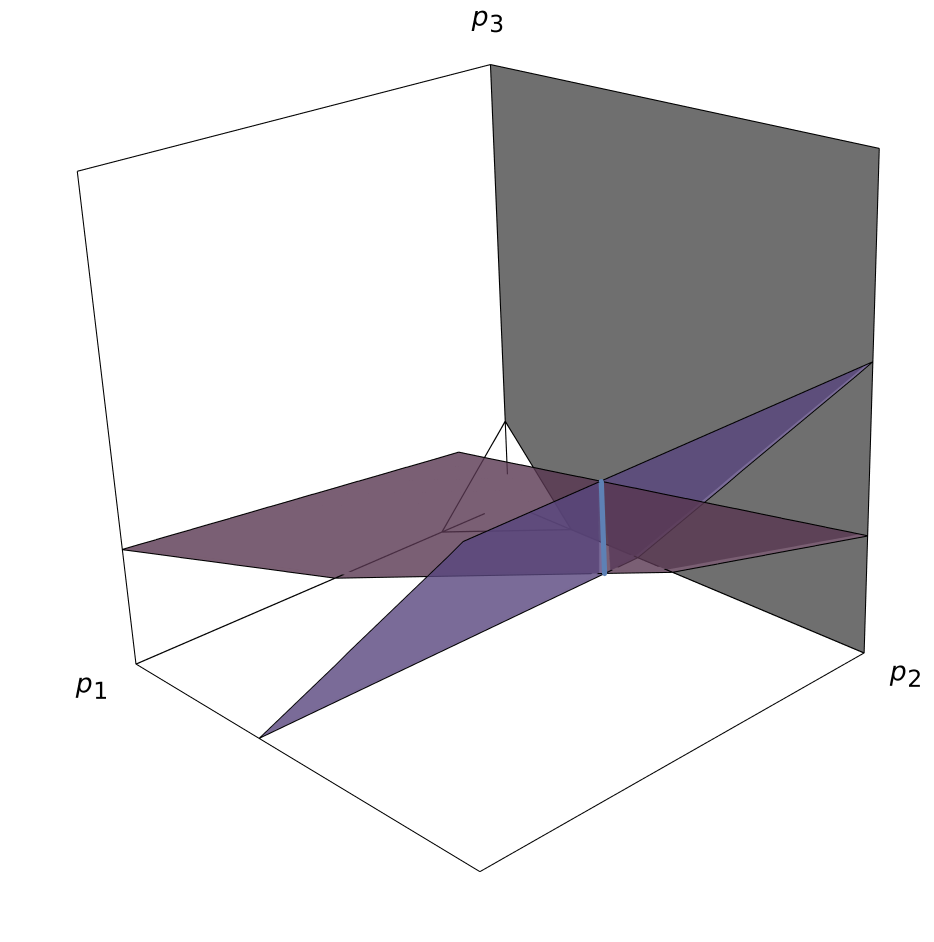}
\caption{Phase 1: $c_1=0.5, c_2=2$}
\end{subfigure}
\hfill
\begin{subfigure}[b]{0.32\textwidth}
\centering
\includegraphics[width=\textwidth]{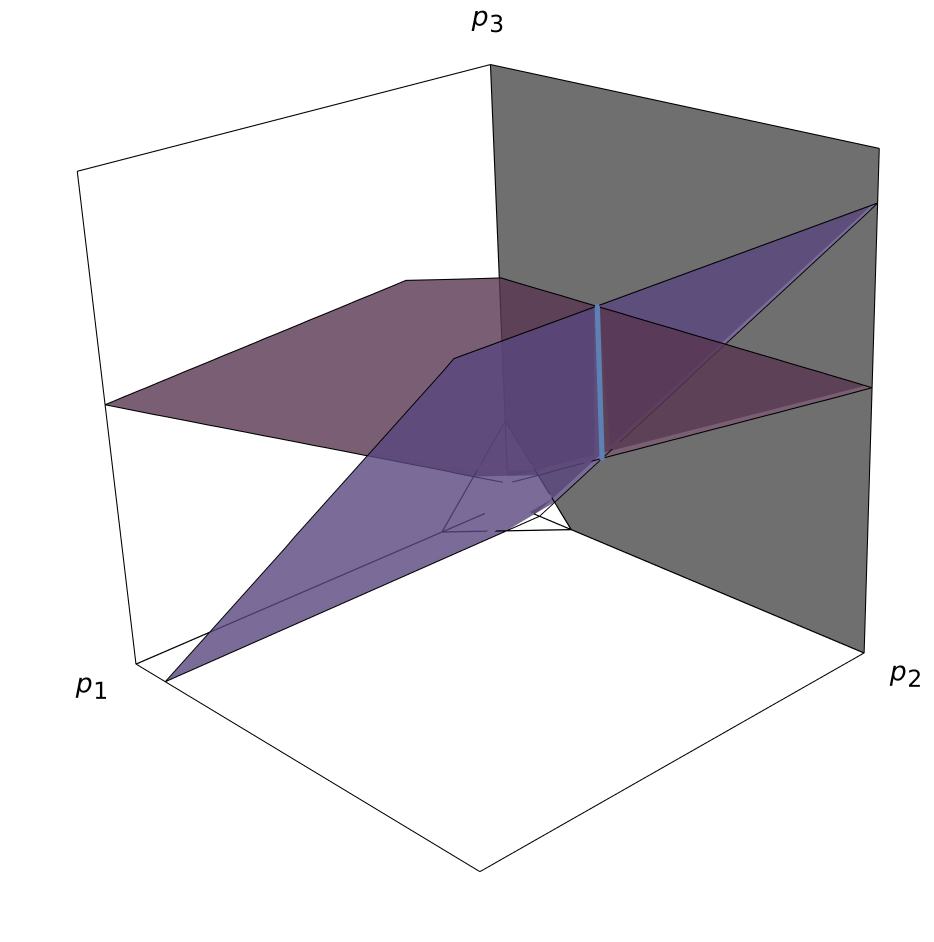}
\caption{Phase 2: $c_1=-0.5, c_2=0.5$}
\end{subfigure}
\hfill
\begin{subfigure}[b]{0.32\textwidth}
\centering
\includegraphics[width=\textwidth]{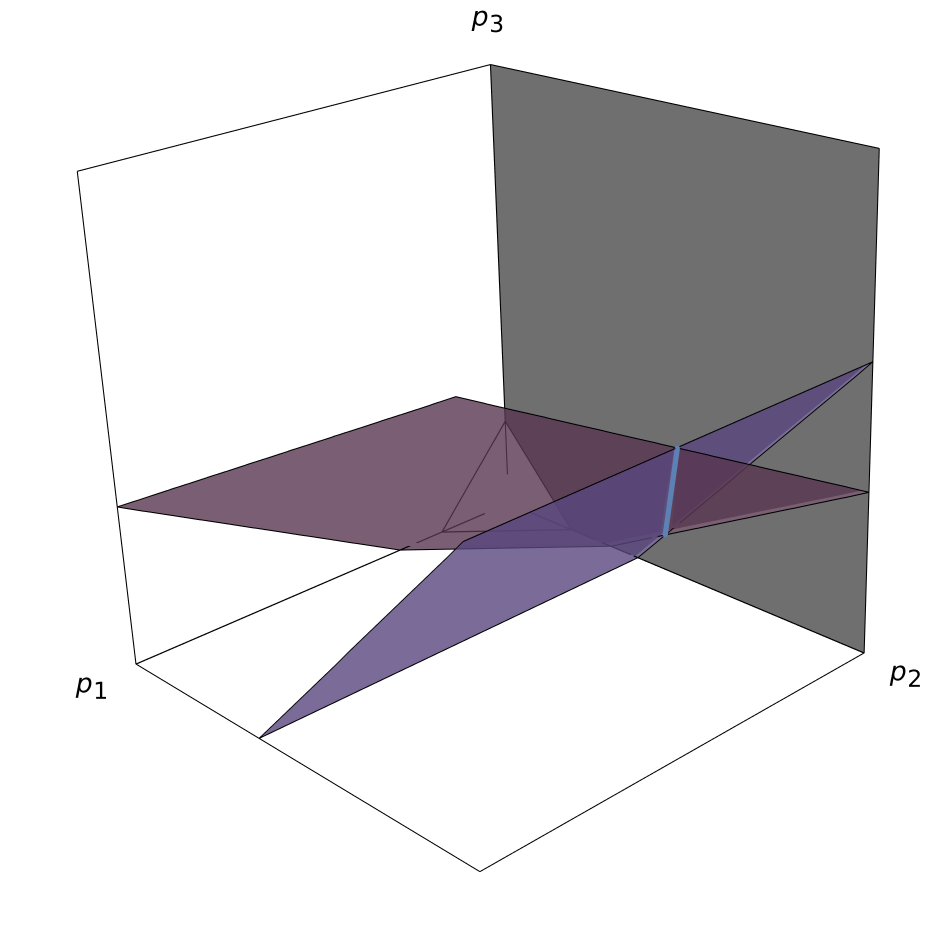}
\caption{Phase 3: $c_1=0.2, c_2=2$}
\end{subfigure}
\\
\begin{subfigure}[b]{0.32\textwidth}
\centering
\includegraphics[width=\textwidth]{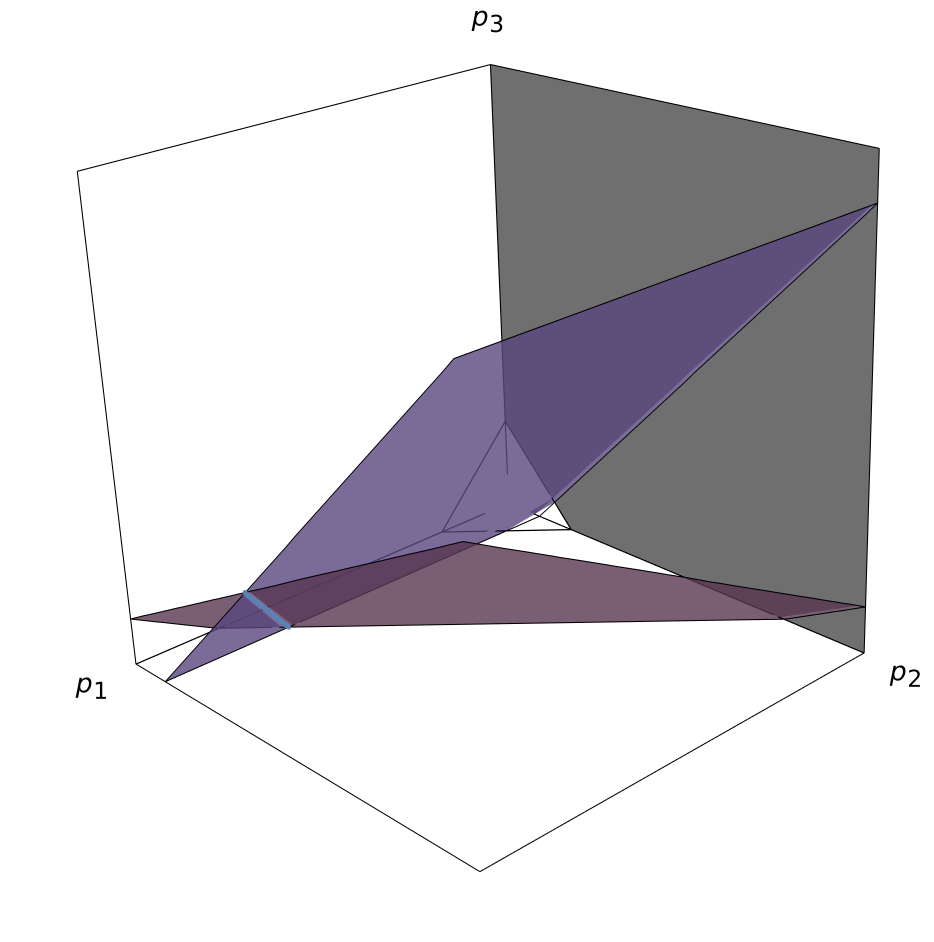}
\caption{Phase 4: $c_1=1, c_2=0.5$}
\end{subfigure}
\hspace*{.1\textwidth}
\begin{subfigure}[b]{0.32\textwidth}
\centering
\includegraphics[width=\textwidth]{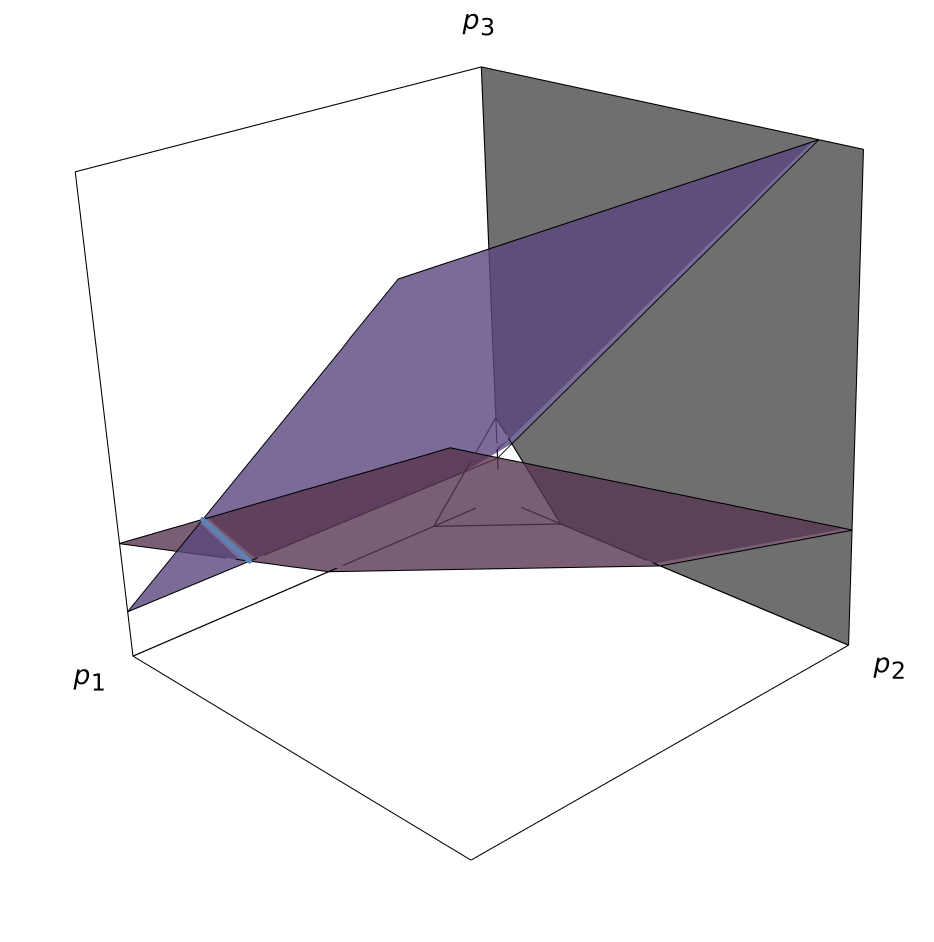}
\caption{Phase 5: $c_1=0.5, c_2=-0.5$}
\end{subfigure}
\caption{Hyperplanes for a double symplectic cut of local $\BP^3$ in five different phases, all at $t=1$.}
\label{fig:P2doublecutphases}
\end{figure}

An example of Calabi--Yau fivefold describing a double symplectic cut of local $\BP^2$ is given by the quotient
\be
 \XVd = (\BC^4 \times \BC_+\times \BC_- \times \BC^+\times \BC^-)\sslash U(1)^3
\ee
defined by the following charge matrix
\be
 \left[
    \begin{array}{cccccccc|c}
    z_1 & z_2 & z_3 & z_4 & z_+ & z_- & z^+ & z^- \\
    \hline
    1 & 1 &  1 & -3 &  0 & 0 &  0 & 0 & t \\
    0 & 0 & -1 &  1 & -1 & 1 &  0 & 0 & c_1 \\
    0 & 1 & -1 &  0 &  0 & 0 & -1 & 1 & c_2 \\
    \end{array}
 \right]\,.
\ee
The CY5 quantum volume is given by the integral
\begin{multline}
    \cF^D_{\XVd}(t,c,\e)
    =
    \int \frac{\dif\phi}{2\pi\ii} \int \frac{\dif\varphi_1}{2\pi\ii}
    \int \frac{\dif\varphi_2}{2\pi\ii}
    \eu^{t\phi + c_1\varphi_1+c_2\varphi_2}
    \Ga(\e_1+\phi)
    \Ga(\e_2+\phi+\varphi_2)
    \Ga(\e_3+\phi-\varphi_1-\varphi_2) \\
    \times
    \Ga(\e_4-3\phi+\varphi_1)
    \Ga(\e_+ -\varphi_1)\Ga(\e_- +\varphi_1)
    \Ga(\e^+ -\varphi_2)\Ga(\e^- +\varphi_2) \,.
\end{multline}
The phase structure in this example is more complicated, each phase corresponding to a configuration of hyperplanes for the double symplectic cut, leading to different choices of contours for the integral. See Figure~\ref{fig:P2doublecutphases} for some examples of possible phases.

This double cut of local $\BP^2$ defines a CY onefold $\XId$ with quantum volume given by \be
\begin{aligned}
 \cK^D(t,c_1,c_2,\e)
 & = \Ga(\e_1+\e_2+\e_3+\e_4)
 \frac{ z^{\e_1} x_1^{2\e_1-\e_2-\e_3} x_2^{-\e_1+\e_2} }
 {(1+ x_1^{-1} + z x_1^2 x_2^{-1} + x_1^{-1} x_2)^{\e_1+\e_2+\e_3+\e_4}}\,.
\end{aligned}
\ee
See Section~\ref{sec:B-model-stuff-example-P2} for details on the derivation.

\section{Summary and conclusions}\label{sec:conclusions}

In this work, we developed the relation between symplectic cuts and disk potentials first observed in \cite{Cassia:2023uvd} in several directions.

Our first main result is a general expression for the quantum Lebesgue measure $\cH^D$ in terms of Lauricella's hypergeometric function of type $D$, given in \eqref{eq:cHD-FD}.
We then showed that, under certain technical assumptions, the monodromy of its regular part is related to the derivative of the toric brane's disk potential by formula \eqref{eq:delta-cHD-0-final-main}. This makes contact with expectations from open string mirror symmetry  \cite{Aganagic:2000gs, Aganagic:2001nx} and shows that the correspondence observed in \cite{Cassia:2023uvd} holds in larger generality, including in the context of toric geometries not considered there.
We expect the correspondence to hold for all (CY) symplectic cuts of both smooth and orbifold toric Calabi--Yau threefolds, and that this may be proven either by studying monodromies of Lauricella's function or by a careful analysis of the $B$-model-like integral representation for $\cH^D$ given in \eqref{eq:HD-B-int-main}.

Another central point of this work hinges on the description of symplectic cuts in terms of higher-dimensional geometries.
Based on Braverman's observation that the symplectic cut of a Calabi--Yau threefold
can be described via an associated Calabi--Yau fourfold, we have showed that equivariant periods of the CY4's divisors and their intersections descend, in a partially non-equivariant limit, to the functions $\cF^D_{\lessgtr}$ associated to the half-spaces of the symplectic cut, and to the equivariant disk potential $W(t,c,\e)$. An interesting consequence of this correspondence is that the equivariant disk potential must obey extended Picard--Fuchs equations. We have verified this in concrete examples, and found that in the non-equivariant limit the extended PF equations become inhomogeneous.

Finally, we have further generalized our construction by considering double symplectic cuts, which correspond to the maximal allowed number of simultaneous CY cuts of a Calabi--Yau threefold. We have shown that their geometry is encoded by a Calabi--Yau fivefold, whose equivariant periods correspond to deformations of functions $\cF^D_{\lessgtr,\lessgtr}$ associated with the four subspaces defined by the double cut.

\subsubsection*{Open problems}

To complete our description of the quantum symplectic cut, it would be desirable to develop an appropriate notion of its quantum cohomology ring.
The Calabi--Yau fourfold description of the cut provides a natural approach to this problem, since the CY4 comes with its own Picard--Fuchs equations, and since the various strata of the cut arise as toric divisors as in the case of $\XIIId[\lessgtr]$, or as intersections of divisors as in the case of $\XIId$.
In particular, we have showed that in the limit $\e_\pm\to 0$ the CY4 Picard--Fuchs operators descend to extended Picard--Fuchs operators for the CY3 which annihilate $\cF^D_{\gtrless}$ and the equivariant disk potential $W(t,c,\e)$.
One may speculate that these may provide a description of the equivariant quantum cohomology ring for $\XIIId[<]\cup_{\XIId}\XIIId[>]$.
It would be interesting to make this expectation more precise. In particular one may ask whether the extended PF equations completely characterize the involved equivariant periods, and what would constitute appropriate initial conditions.
Similarly, one may ask whether the extended Picard--Fuchs equations arising from the CY5 can be regarded as a description of the quantum cohomology ring for the double cut of $\XIIId$.

A related question is how to interpret the symplectic cut procedure from the perspective of local mirror symmetry. In particular, it would be interesting to clarify the role of the CY fourfold $\XIVd$ on the mirror-dual side.
We expect that manipulations of equivariant $B$-model integrals, such as those discussed in Appendix~\ref{app:technical-identities}, may provide insight into this direction.

Another question raised in this work is whether double cuts encode more than just the disk potentials of two toric branes—specifically, whether they also capture the annulus potential. A natural candidate for this additional information is the CY1 quantum volume $\cK^D(t,c_1,c_2,\e)$, which depends on the open string moduli of both Lagrangians. It would be interesting to investigate whether its analytic structure encodes annuli stretching between branes.
A hint that this might be the case comes from its relation to the mirror curve, as given in \eqref{eq:KD-B-int-intro}. 
A direct approach might be to study the GLSM on a circle with suitable boundary conditions at the two spatial boundaries, suggesting that candidates for the annulus potential might be the Witten index, or certain generalizations thereof \cite{Cecotti:1992qh}.\footnote{We thank the anonymous referee for suggesting a possible role of the Witten index in relation to annulus potentials.} 
It would be very interesting to explore this question further, but we leave it to future work.

\appendix
\addtocontents{toc}{\protect\setcounter{tocdepth}{1}}

\section{The Lauricella hypergeometric of type \texorpdfstring{$D$}{D}}\label{app:lauricella}

\subsection{Series and integral definitions}
The Lauricella hypergeometric function $F_D^{(n)}$ can be defined as follows
(see \cite{Bezrodnykh_2018} for further details)
 \be\label{eq:FD-def}
 F^{(n)}_D (a,b_1,\dots,b_n,c;x_1,\dots,x_n)
 = \sum_{i_1,\dots,i_n=0}^\infty
 \frac{(a)_{i_1+\dots+i_n} (b_1)_{i_1} \cdots (b_n)_{i_n}}
 {(c)_{i_1+\dots+i_n} i_1! \cdots i_n!} x_1^{i_1} \cdots x_n^{i_n}
\ee
where
\be
\label{eq:pochhammer}
 (z)_n = \prod_{k=0}^{n-1}(z+k) = \frac{\Ga(z+n)}{\Ga(z)}
\ee
is the Pochhammer symbol.
The definition in \eqref{eq:FD-def} holds in the region $|x_i|<1$,
and the function $F^{(n)}_D$ is defined by analytic continuation elsewhere.

For $\mathrm{Re}(c)>\mathrm{Re}(a)>0$, we also have the following integral representation
\begin{multline}
\label{eq:Lauricella-integral}
 F^{(n)}_D (a,b_1,\dots,b_n,c;x_1,\dots,x_n) \\
 = \frac{\Ga(c)}{\Ga(a)\Ga(c-a)}
 \int_0^1 t^{a-1} (1-t)^{c-a-1} (1-x_1 t)^{-b_1} \cdots (1-x_n t)^{-b_n}~\dif t\,.
\end{multline}
Observe that when $n=1$ the Lauricella function reduces to the Gauss hypergeometric function
\be
\label{eq:Lauricella2Gauss}
 F^{(1)}_D (a,b,c;x) = {}_2F_1(a,b,c;x)\,.
\ee

\subsection{Useful properties}
Changing variables under the integral in \eqref{eq:Lauricella-integral} by $t= 1- \tilde{t}$ we obtain
\be
 \int_0^1 (1-\tilde{t})^{a-1} \tilde{t}^{c-a-1} (1-x_1 + x_1 \tilde{t})^{-b_1}
 \cdots (1-x_n + x_n \tilde{t})^{-b_n}~\dif\tilde{t}
\ee
and thus we have
\be
 F^{(n)}_D (a,b_1,\dots,b_n,c;x_1,\dots,x_n) = \prod_{i=1}^n (1-x_i)^{-b_i}
 F^{(n)}_D \Big(c-a,b_1,\dots,b_n,c;\frac{x_1}{x_1-1},\dots,\frac{x_n}{x_n-1}\Big)\,.
\ee
If we set $x_i = 1+ y_i^{-1}$ we have the identity
\begin{multline}
\label{eq:FDLauricella-identity1}
 F^{(n)}_D (a,b_1,\dots,b_n,c;1+y^{-1}_1,\dots,1+y^{-1}_n) \\
 = \prod_{i=1}^n (-y_i^{-1})^{-b_i} F^{(n)}_D (c-a,b_1,\dots,b_n,c;1+y_1,\dots,1+y_n)
\end{multline}

Now, let us explain how the above properties are related to the study of the integral
\be
 I(\alpha,\beta;y_1,\dots,y_n)
 :=\int_0^\infty \frac{y^{\alpha-1}}{\prod_{i=1}^n (y-y_i)^\beta}\,\dif y
\ee
that appears in the general form of the quantum Lebegue measure \eqref{eq:HD-B-int-main}.

Notice that if we change the variables as $y=\frac{w}{1-w}$,
we can show that this integral is a special case of the that in \eqref{eq:Lauricella-integral},
and we obtain
\be
\label{eq:Lauricella-y-integral}
 I(\alpha,\beta;y_1,\dots,y_n)
 = \Big(\prod_{i=1}^n (-y_i)^{-\beta}\Big)
 \frac{\Ga(\alpha)\Ga(\beta n-\alpha)}{\Ga(\beta n)}
 F^{(n)}_D (\alpha,\beta,\dots,\beta,\beta n;1+y_1^{-1},\dots,1+y_n^{-1})\,.
\ee
From the identity \eqref{eq:FDLauricella-identity1}, it then follows that
\be
 I(\alpha,\beta;y_1,\dots,y_n)
 = \Big(\prod_{i=1}^n (-y_i^{-1})^{-\beta}\Big)
 I(\beta n-\alpha,\beta;y_1^{-1},\dots,y_n^{-1})\,.
\ee

\section{Quantum volumes as equivariant \texorpdfstring{$B$}{B}-model integrals}
\label{app:technical-identities}

In this Appendix, we collect certain expressions for the three different types of quantum equivariant volumes that play a central role in our work, namely $\cF^D,\cH^D$ and $\cK^D$. 
We show that all of these admit expressions that are strongly reminiscent of $B$-model integrals, by manipulations that parallel those of the derivation of Hori--Vafa mirror curves of toric Calabi--Yau threefolds. This observation points to a mirror interpretation of symplectic cuts, where each of the quantum volumes involved maps to a certain integral in the mirror geometry.
More details will be provided below, we postpone a physical interpretation of these identities to a separate publication.

\subsection{Symplectic quotient operators}

Euler's $\Ga$ function is the building block for integral expressions of $\cF^D, \cH^D$ and $\cK^D$ arising from the viewpoint of GLSMs, see \eqref{eq:disk-general-contour}, \eqref{eq:cHD-def} and \eqref{eq:KD-def}, respectively.

Let us first recall the Euler integral representation
\be
\label{eq:integral-representation-gamma}
	\Ga(z) = \int_0^\infty \eu^{-u} u^{z-1} \dif u\,.
\ee
A distinctive property of this presentation is that the shift operator $\hat{u}:=\exp\frac{\partial}{\partial z}$ acts in the following way
\be
 f(\hat{u})\cdot\Ga(z) = \int_0^\infty \eu^{-u} u^{z-1} f(u) \dif u\,,
\ee
for any algebraic function $f(u)$.

Let $\Psi(\e_1,\dots,\e_n)$ be a function of $n$ variables $\e_i$, and let $q=(q_1,\dots,q_n)\in\BZ^n$ be a vector of integer charges. Then we define the \emph{symplectic quotient operator} $\cG_q(t)$ associated to the charge vector $q$ and with moment map parameter $t\in\BR$,
as the following formal operator
\be\label{eq:Q-operator}
\begin{aligned}
 \cG_q(t)\cdot\Psi(\e_1,\dots,\e_n)
 :=& \,\delta\Big(t+\sum_{i=1}^n q_i\frac{\partial}{\partial\e_i}\Big)\cdot
 \Psi(\e_1,\dots,\e_n) \\
 =& \int_{-\ii\infty}^{+\ii\infty} \frac{\dif\phi}{2\pi\ii}
 \, \eu^{\phi(t+\sum_i q_i \frac{\partial}{\partial\e_i})} \Psi(\e_1,\dots,\e_n) \\
 =& \int_{-\ii\infty}^{+\ii\infty} \frac{\dif\phi}{2\pi\ii}
 \, \eu^{\phi t} \Psi(\e_1+q_1\phi,\dots,\e_n+q_n\phi)\,.
\end{aligned}
\ee
Suppose $Q$ is the charge matrix associated to a toric quotient $\BC^n\sslash U(1)^r$ with K\"ahler parameters $t^a$, then we can define $r$ commuting quotient operators $\cG_{Q^a}(t^a)$, one for each row of the matrix $Q$.
The quantum equivariant volumes can then be obtained in a simple way in terms of the action of the operators $\cG_{Q^a}(t^a)$ on a product of $\Ga$ functions in the $\e_i$ parameters.
In fact, when acting on a product of $\Ga$ functions, we can rewrite $\cG_q(t)$ in terms of the operators $\hat{u}$ as
\be
 \cG_q(t)\cdot\prod_i\Ga(\e_i)
 = \delta\Big(\log\Big(\eu^t\prod_i\hat{u}_i{}^{q_i}\Big)\Big)\cdot\prod_i\Ga(\e_i)
\ee

The formal inverse operator to $\cG_q(t)$ corresponds to integrating over the moment map parameter $t$,
\be\label{eq:Q-inv-operator}
	\cG_q(t)^{-1}\cdot f(t,\e) := \int_{\BR} \dif t\,f(t,\e)\,.
\ee

\subsection{Relations among quantum equivariant volumes}

Starting with the equivariant quantum volume of $\BC^{r+3}$, which is just a product of $\Ga$ functions $\prod_{i=1}^{r+3} \Ga(\e_i)$, we can express the symplectic quotient \eqref{eq:CY3-symp-quot} as
\be
 \cF^D(t,\e)
 = \Big( \prod_{a=1}^r \cG_{Q^a}(t^a) \Big) \cdot \prod_{i=1}^{r+3} \Ga(\e_i)
\ee

Similarly, for a symplectic cut, if we denote by $q^\alpha_i$ the charges associated to the moment maps of the affine hyperplanes \eqref{eq:toric-brane-hyperplanes} and by $c_\alpha$ the associated open moduli,
we can then express the quantum Lebesgue measure \eqref{eq:cHD-def} as
\be
\label{eq:operatorHD}
	\cH^D(t,c_\alpha,\e)
	= \cG_{q^\alpha}(c_\alpha) \cdot \cF^D(t,\e)\,.
\ee
With two cuts we obtain the function $\cK^D$ by applying the quotient operator twice
\be
 \cK^D(t,c_1,c_2,\e) :=
 \Big( \prod_{\alpha=1,2} \cG_{q^\alpha}(c_\alpha) \Big) \cdot \cF^D(t,\e)
\ee
These relations can be inverted by integrating over the moment map moduli using the operator \eqref{eq:Q-inv-operator}. We find
\be\label{eq:q-vol-hierarchy}
\begin{aligned}
 \int_{\BR}\dif c_2\, \cK^D(t,c_1,c_2,\e) &= \cH^D(t,c_1,\e) \\
 \int_{\BR}\dif c_1\, \cK^D(t,c_1,c_2,\e) &= \cH^D(t,c_2,\e) \\
 \int_{\BR^2}\dif c_1\,\dif c_2\,\cK^D(t,c_1,c_2,\e) &= \cF^D(t,\e) \\
 \int_{\BR^r}\dif t^1 \cdots \dif t^r\, \cF^D(t,\e) &= \prod_{i=1}^{r+3} \Ga(\e_i)\,.
\end{aligned}
\ee

\subsection{Equivariant \texorpdfstring{$B$}{B}-model integral for \texorpdfstring{$\cK^D$}{KD}}
\label{sec:B-model-stuff}

We now give explicit integral formulae for the quantum equivariant volumes of the CY3 and of its cuts.
These fit into the hierarchy given in \eqref{eq:q-vol-hierarchy}, with the most fundamental object being $\cK^D$, which will be out starting point.

We begin by rewriting \eqref{eq:KD-def} by means of the integral representation \eqref{eq:integral-representation-gamma} of Euler's $\Ga$ function and in terms of symplectic quotient operators \eqref{eq:Q-operator}
\begin{multline}
 \cK^D(t,c_1,c_2,\e)
 = \prod_{\alpha=1,2}\int_{\ii\BR}\frac{\dif\varphi_\alpha}{2\pi\ii}
 \eu^{c_\alpha\varphi_\alpha}
 \prod_{a=1}^r\int_{\ii\BR}\frac{\dif\phi_a}{2\pi\ii}
 \eu^{t^a\phi_a}
 \prod_{i=1}^{r+3} \int_0^\infty \frac{\dif u_i}{u_i} \,
 u_i^{\e_i + \sum_a Q^a_i\,\phi_a+\sum_\alpha q^\alpha_i\,\varphi_\alpha}\,\eu^{-u_i} \\
 = \prod_{i=1}^{r+3}\int_0^\infty\frac{\dif u_i}{u_i}\,u_i^{\e_i}\,\eu^{-u_i}
 \,\prod_{a=1}^r\delta\Big(t^a+\sum_i Q^a_i\log u_i\Big)
 \prod_{\alpha=1,2} \delta\Big(c_\alpha+\sum_i q^\alpha_i \log u_i\Big)\,.
\end{multline}
We can write this more uniformly by packaging moment map conditions together as follows.
Let $\tilde{Q}$ be the matrix with $(r+2)\times(r+3)$ matrix obtained by stacking $Q^a_i$
and $q^\alpha_i$ on top of each other. Similarly, let $\tilde t = (t^1,\dots t^r,c_1,c_2)$. Then we can rewrite $\cK^D$ as
\be
\label{eq:KD-appendix-B-model}
 \cK^D(\tilde t,\e)
 = \prod_{i=1}^{r+3} \int_0^\infty \frac{\dif u_i}{u_i}
 \, u_i^{\e_i} \, \eu^{-u_i}\,
 \prod_{a=1}^{r+2} \delta\Big(\tilde t^a + \sum_i \tilde Q^a_i \log u_i\Big) \,.
\ee
To carry out the integral over $r+3$ variables with $r+2$ constraints, we single out a choice of $r+2$ variables $u_i$. This is equivalent to a choice of index $j$ for the only integration variable that is left after restriction to the support of the Dirac $\delta$ functions. Let $\tilde{Q}(j)$ denote the square matrix obtained by removing the $j$-th column from $\tilde{Q}$. We shall require that $j$ is chosen in such a way that the matrix $\tilde{Q}(j)$ is invertible.\footnote{This is always possible by an appropriate choice of $j$, because of the assumption that the two hyperplanes intersect non-trivially inside of the toric polytope of~$\XIIId$.}
The moment map constraints enforced by the $r+2$ $\delta$-functions allow us to express all other $u_i$ for $i\neq j$ in terms of the chosen coordinate $u_j$ and of the moduli $\tilde{t}$,
\be
 \log u_i
 = -\sum_{a=1}^{r+2}
 \left[
 \tilde{Q}(j)^{-1}\right]^i_a(\tilde{t}^a+\tilde{Q}^a_j\log u_j)
 \hspace{30pt}\forall i\neq j\,,
\ee
or in terms of exponentiated moment map moduli $\tilde z_a:=\eu^{-\tilde t^a}$
\be
 u_i = \prod_{a=1}^{r+2}
 \left(\tilde z_a^{-1} u_j^{\tilde Q^a_j}\right)^{-\left[\tilde{Q}(j)^{-1}\right]^i_a} \,.
\ee
This allows us to reduce the integral to a single variable
\begin{multline}
 \cK^D(\tilde{t},\e)
 = \int_0^\infty \,\frac{\dif u_j}{u_j^{1-\e_j}}
 \left(\prod_{a=1}^{r+2}
 (\tilde z_a^{-1}u_j^{\tilde Q^a_j})^{-\sum_{i\neq j}\left[\tilde{Q}(j)^{-1}\right]^i_a\e_i}\right)
 \eu^{-u_j-\sum_{i\neq j}\prod_{a=1}^{r+2}
 (\tilde z_a^{-1}u_j^{\tilde{Q}^a_j})^{-\left[\tilde{Q}(j)^{-1}\right]^i_a}} \\
 = \eu^{-\sum_{a=1}^{r+2}
 \sum_{i\neq j}\tilde t^a\left[\tilde{Q}(j)^{-1}\right]^i_a\e_i}
 \int_0^\infty\frac{\dif u_j}{u_j}\,
 u_j^{\e_j-\sum_{a=1}^{r+2}\sum_{i\neq j}\tilde Q^a_j\left[\tilde{Q}(j)^{-1}\right]^i_a\e_i}
 \eu^{-u_j-\sum_{i\neq j}\prod_{a=1}^{r+2}(\tilde z_a^{-1}u_j^{\tilde{Q}^a_j})^{-\left[\tilde{Q}(j)^{-1}\right]^i_a}}\,.
\end{multline}
To proceed we note two properties of the matrix $R^i_k:=\sum_{a=1}^{r+2}\left[\tilde{Q}(j)^{-1}\right]^i_a\tilde{Q}_k^a$ :
\begin{itemize}
 \item $[R(j)]^i_k=\delta^i_k$, where $R(j)$ is the square matrix obtained from $R$ by removing the $j$-th column. This trivially follows by noting that $R(j)=\tilde{Q}(j)^{-1}\tilde{Q}(j)$.
 \item $R^i_j=-1$ for every $i=1,\dots,r+2$. This follows from the CY condition on the matrix of charges $\tilde{Q}$, namely we have $\tilde{Q}_j^a=-\sum_{k\neq j}\tilde{Q}_k^a$ and by the previous point, we deduce
\be
 R^i_j = \sum_{a=1}^{r+2}\left[\tilde{Q}(j)^{-1}\right]^i_a\tilde{Q}_j^a
 =-\sum_{k\neq j} \sum_{a=1}^{r+2}\left[\tilde{Q}(j)^{-1}\right]^i_a\tilde{Q}_k^a
 =-\sum_{k\neq j} \delta^i_k = -1
\ee
\end{itemize}
Using these facts, the above expression simplifies to
\be
\begin{aligned}
 \cK^D(\tilde{t},\e)
 &= \eu^{-\sum_{a=1}^{r+2}\sum_{i\neq j}\tilde t^a\left[\tilde{Q}(j)^{-1}\right]^i_a\e_i}
 \int_0^\infty\frac{\dif u_j}{u_j}\,u_j^{\sum_{i=1}^{r+3}\e_i}\,
 \exp\left(-u_j\left(1+\sum_{i\neq j}\prod_{a=1}^{r+2}\tilde z_a^{\left[\tilde{Q}(j)^{-1}\right]^i_a}\right)\right)
\end{aligned}
\ee
Let us define
\be\label{eq:technical-defs1}
 H(x_1,x_2,z) :=
 1+\sum_{i\neq j}\prod_{a=1}^{r+2}\tilde z_a^{\left[\tilde{Q}(j)^{-1}\right]^i_a}
\ee
and
\be\label{eq:technical-defs2}
 A_\alpha^i :=\left\{
 \begin{array}{ll}
 \left[\tilde{Q}(j)^{-1}\right]^i_{r+\alpha} & \text{if $i\neq j$} \\
 0 & \text{if $i=j$}
 \end{array}\right.\,,
 \hspace{30pt}
 M_a^i :=\left\{
 \begin{array}{ll}
 \left[\tilde{Q}(j)^{-1}\right]^i_a & \text{if $i\neq j$, $a=1,\dots,r$} \\
 0 & \text{if $i=j$}
 \end{array}\right.
\ee
where $x_\alpha=\eu^{-c_\alpha}=\tilde z_{r+\alpha}$ for $\alpha=1,2$, and $z^a=\eu^{-t^a}=\tilde z^a $ for $a=1,\dots,r$.
Note that these definitions depend on the choice of $j$ made at the beginning.
This leads us to the final expression for $\cK^D$
\be\label{eq:KD-B-int-app}
\begin{aligned}
 \cK^D(t,c_1,c_2,\e)
 & = \eu^{-t\cdot M\cdot\e} x_1^{A_1\cdot\e} x_2^{A_2\cdot\e}
 \int_0^\infty\frac{\dif u_j}{u_j}\,u_j^{\sum_{i=1}^{r+3}\e_i}
 \eu^{-u_j H(x_1,x_2,z)} \\
 & = \Ga\Big(\sum_{i=1}^{r+3}\e_i\Big) \eu^{-t\cdot M\cdot\e}
 \frac{x_1^{A_1\cdot\e} x_2^{A_2\cdot\e}}{H(x_1,x_2,z)^{\sum_{i=1}^{r+3}\e_i}}
\end{aligned}
\ee
where we used the shorthands $A_\alpha\cdot\e=\sum_{i=1}^{r+3}A_\alpha^i\e_i$ and
$t\cdot M\cdot\e=\sum_{i=1}^{r+3}\sum_{a=1}^{r+2}t^aM_a^i\e_i$.

\subsubsection{Example: double cut of \texorpdfstring{$\BC^3$}{C3}}
\label{sec:B-model-stuff-example}

In the case of a double cut of $\BC^3$ defined as in Section~\ref{sec:C3-5d},
the symplectic quotient $\XId$ is associated to the charge matrix
\be
\tilde Q = \left[
\begin{array}{ccc}
 0 & 1 & -1 \\
 1 & -1 & 0 \\
\end{array}
\right]
\ee
which corresponds to the matrix obtained after removing the last four columns from \eqref{eq:C3-double-cut-Q-matrix}. This precisely corresponds to the fact that $\XId$ is the intersection of the last four toric divisors inside of $\XVd$.

The computation of $\cK^D$ as a $B$-model integral requires us to make a choice of how to solve the moment map constraints. We choose to solve w.r.t.\ the first two variables $u_1$ and $u_2$ in \eqref{eq:KD-appendix-B-model} as functions of $u_3$.
As discussed in the previous section, this is equivalent to the choice of index $j=3$ and reduced matrix
\be
\tilde Q(3) =
\left[
\begin{array}{cc}
 0 & 1 \\
 1 & -1 \\
\end{array}
\right].
\ee
Substituting into \eqref{eq:technical-defs1} and \eqref{eq:technical-defs2},
we obtain the identifications
\be
 A_1\cdot\e=\e_1+\e_2\,,
 \hspace{30pt}
 A_2\cdot\e=\e_1\,,
 \hspace{30pt}
 M=0
 \hspace{30pt}\text{and}\hspace{30pt}
 H(x_1,x_2)=1+x_1+x_1x_2\,.
\ee
Finally, the quantum volume of $\XId$ can be computed via \eqref{eq:KD-B-int-app}, which gives
\be
\begin{aligned}
 \cK^D(c_1,c_2,\e)
 & = \Ga(\e_1+\e_2+\e_3)
 \frac{x_1^{\e_1+\e_2}x_2^{\e_1}}{(1+x_1(1+x_2))^{\e_1+\e_2+\e_3}}\,.
\end{aligned}
\ee

\subsubsection{Example: double cut of local \texorpdfstring{$\BP^2$}{P2}}
\label{sec:B-model-stuff-example-P2}

The double cut of local $\BP^2$ chosen as in Section~\ref{sec:localP2-5d} leads to a CY onefold $\XId$ with charge matrix
\be
\tilde Q =
 \left[
    \begin{array}{cccc|c}
    z_1 & z_2 & z_3 & z_4 \\
    \hline
    1 & 1 &  1 & -3 & t \\
    0 & 0 & -1 &  1 & c_1 \\
    0 & 1 & -1 &  0 & c_2 \\
    \end{array}
 \right]\,.
\ee
Choosing to solve the moment map constraints w.r.t.\ the first three variables $u_i$ as functions of $u_4$, we obtain the following identifications of parameters:
\be
 A_1\cdot\e=2\e_1-\e_2-\e_3\,,
 \hspace{30pt}
 A_2\cdot\e=-\e_1+\e_2\,,
 \hspace{30pt}
 t\cdot M\cdot\e=t\e_1
\ee
and
\be
 H(x_1,x_2,z)=1+x_1^{-1}+zx_1^2x_2^{-1}+x_1^{-1}x_2\,.
\ee
Plugging in \eqref{eq:KD-B-int-app} we obtain the quantum volume of $\XId$.

\subsection{\texorpdfstring{$H(x_1,x_2)$}{H(x1,x2)} vs the Hori--Vafa mirror curve}
\label{sec:H-Hori--Vafa}

We recall the standard definition of the Hori--Vafa mirror curve of a toric CY threefold. Let $\XIIId$ be a toric quotient as in \eqref{eq:CY3-symp-quot}, then the mirror geometry can be described in terms of $r+3$ complex coordinates $Y^i\in\BC^\times$ subject to the constraints
\be
\label{eq:mirror-curve-constr}
 \sum_{i=1}^{r+3} Q_i^a Y^i = -t^a\,,
 \hspace{30pt}
 a = 1,\dots,r\,,
\ee
whose solutions form a three-dimensional family (see \cite{Katz:1996fh,Hori:2000kt}). Because of the CY condition on $Q_i^a$, the space of solution is invariant under a simultaneous translation of all the coordinates as
\be
 Y^i \mapsto Y^i + s
\ee 
therefore we can quotient by this action to reduce the space of solutions to a two-dimensional family parametrized by two coordinates $x,y\in\BC^\times$. Solving explicitly the constraints w.r.t.\ the $Y^i\cong Y^i(x,y)$ as functions of $x$ and $y$, we can define the mirror CY manifold as the complex hypersurface
\be
\label{eq:mirror-hypersurface}
 UV = \sum_{i=1}^{r+3} \eu^{Y^i(x,y)} =: F(x,y)\,,
\ee
where $U,V\in\BC$ are auxiliary complex variables.
Because $F(x,y)$ depends on an explicit choice of coordinates for the two-dimensional family of solutions, we have an ambiguity in defining this hypersurface, related to the underlying freedom in the way we parametrize the variables $Y^i$ as functions of $x$ and $y$.
This ambiguity is of the same nature as the one observed in solving the Dirac-delta constraints in \eqref{eq:KD-appendix-B-model}. In fact, we can formally identify the variables $Y^i=\log u_i$, where $u_i$ are the integration variables in a $B$-model integral representation of the quantum volume of $\XIIId$,
\be
 \cF^D(t,\e)
 = \prod_{i=1}^{r+3} \int_0^\infty \frac{\dif u_i}{u_i}
 \, u_i^{\e_i} \, \eu^{-u_i}\,
 \prod_{a=1}^r \delta\Big(t^a + \sum_i Q^a_i \log u_i\Big) \,,
\ee
so that the constraints in \eqref{eq:mirror-curve-constr} match with the moment map equations defining $\XIIId$.

Similarly, a Lagrangian toric brane $L$ in $X_3$ is mapped under mirror symmetry \cite{Hori:2000ic,Aganagic:2001nx} to a mirror brane given by the equation
\be
 U = 0 = F(x,y)\,,
\ee
which defines a holomorphic submanifold of the hypersurface \eqref{eq:mirror-hypersurface}. Its moduli space is complex one-dimensional and it corresponds to the curve
\be
 F(x,y)=0
\ee
which is known as the \emph{mirror curve}.
In general, the open string modulus $c$ of the brane can be identified with a function of the two coordinates $x,y$, so that both $x$ and $y$ are completely fixed by these two equations. In practice, the mirror curve is defined by solving the moment map constraints in such a way that $x$ can be directly identified with $\eu^{-c}$, while $y=y(x)$ becomes implicitly defined as a solution to $F(x,y)=0$. The specific data associated to the location and orientation of the toric brane $L$ are then encoded by the explicit form of the function $F(x,y)$.

The polynomial $H(x_1, x_2)$ appearing in the formula for $\cK^D$ is related to, but does not quite coincide with, the Hori--Vafa mirror curve $F(x,y)$.
More precisely, since $\cK^D$ is defined by a double cut, which corresponds a pair of toric Lagrangians, there are \emph{two} mirror curves that describe the moduli spaces of branes $L_1$ and $L_2$ associated to the cut.

First of all, the branes may come with independent framings and different choices of coordinates. Therefore their moduli spaces would be given by different expressions $F_1(x_1, y_1)=0$ and $F_2(x_2, y_2)=0$, with $x_\alpha=\eu^{-c_\alpha}$, $\alpha=1,2$. Neither of these functions is equal to $H(x_1,x_2)$ on general grounds.
However, all three polynomials are related, since they all come from resolving the moment map constraints, on the universal curve $\sum_{i=1}^{r+3} \eu^{Y^i} = 0$. We illustrate this point with an explicit computation.

\paragraph{An example.}
For illustration, consider the double cut of $\BC^3$ in Section~\ref{sec:C3-5d}.
The symplectic cuts are specified by the charge matrix \eqref{eq:C3-double-cut-Q-matrix}, whose restriction to the three coordinates of $\BC^3$ is
\be
 \left[
 \begin{array}{ccc|c}
  z_1 & z_2 & z_3 & \\
  \hline
  0 & 1 & -1 & c_1 \\
  1 & -1 & 0 & c_2 \\
 \end{array}
 \right]
\ee
Let us work in the phase $c_1,c_2>0$. This means that the first Lagrangian ends on the first leg of the toric diagram, at $(p_1, p_2, p_3)=(c_1,0,0)$ where $p_i:=|z_i|^2$, and the second Lagrangian ends on the second leg at $(p_1, p_2, p_3)=(0,c_2,0)$.
Following \cite{Aganagic:2000gs}, we introduce a complexification of the $p_i$ coordinates
\be
 u_i = \exp(-p_i + \ii\theta_i)\,, \hspace{30pt} i=1,2,3
\ee
and use the moment map equations
\be
 p_2 - p_3 = c_1\,,
 \hspace{30pt}
 p_1 - p_2 = c_2\,,
\ee
to identify the open string moduli as follows
\be
 x_1 := \eu^{-c_1} = \frac{u_2}{u_3}\,,
 \hspace{30pt}
 x_2 := \eu^{-c_2} = \frac{u_1}{u_2}\,.
\ee
The phase determines the homogeneous hyperplanes for both toric Lagrangians: we have respectively $p_1-p_3=0$ for brane 1 and $p_3-p_2=0$ for brane 2.
Therefore we identify dual variables $y_\alpha$ with
\be
 y_1 = \frac{u_1}{u_3}\,,
 \hspace{30pt}
 y_2 = \frac{u_3}{u_2}\,.
\ee
The mirror curves for each brane are obtained by Hori--Vafa as follows
\be
\begin{split}
 u_1+u_2+u_3 & = u_3 (1+y_1+x_1) =: u_3 F_1(x_1, y_1) \\
 & = u_2 (1+y_2+x_2) =: u_2 F_2(x_2, y_2) \\
\end{split}
\ee
In this case, both branes are taken in the same framing, and for this reason the polynomials $F_1(x,y)$ and $F_2(x,y)$ coincide as functions of two variables $(x,y)$, however more generally this is not the case.

Notice that the relation to ambient coordinates $u_i$ implies the relation
\be
 x_2 = y_1x_1^{-1} \,,
 \hspace{30pt}
 y_2 = x_1^{-1}
\ee
which corresponds to the action of $TS\in SL(2,\BZ)$ on $\BC^\times\times\BC^\times$. This is the cubic root of the identity which permutes toric Lagrangians on the three toric legs of $\BC^3$, and therefore naturally relates the two branes (when they are in the same framing, as in this case.)

The curves should be contrasted with the formula for $\cK^D$ given in \eqref{eq:KD-C3}.
We rederive that result here by hand: if we single out column 3 in the matrix $\tilde Q^a_i$, it means we express $u_1,u_2$ in terms of $u_3$ by using the moment map constraints
\be
\begin{split}
 & p_2-p_3 = -\log u_2+\log u_3 = c_1
 \quad\Rightarrow\quad
 u_2 = u_3\,\eu^{-c_1} = u_3 x_2 \\
 & p_1-p_2 = -\log u_1+\log u_2 = c_2
 \quad\Rightarrow\quad
 u_1 = u_2\,\eu^{-c_2} = u_3 x_1 x_2\,.
\end{split}
\ee
The Hori--Vafa universal curve can be written as
\be
 u_1+u_2+u_3 = u_3(x_1 x_2+x_2+1) =: u_3 H(x_1,x_2)
\ee
where $H(x_1,x_2)$ coincides with the polynomial in the denominator of $\cK^D$ as in \eqref{eq:KD-C3}.

Since $H(x_1,x_2)$ comes from the universal curve $u_1+u_2+u_3$, it should coincide with $F_1,F_2$.
This is indeed the case, in fact, using the relations between $y_\alpha,x_\alpha$ given above by the ambient variables $u_i$, we find that
\be\label{eq:H-mirror-curves}
 H(x_1,x_2) = F_1(x_1,y_1) = y_2^{-1} F_2(x_2,y_2)\,.
\ee
Therefore, we learn that we might have written $\cK^D$ in terms of either $F_1$ or $F_2$.
It is clear that this kind of simple relation holds more generally for all toric geometries,
because all three polynomials are just rewritings of the universal curve $\sum_{i=1}^{r+3}u_i=0$.

\subsection{Equivariant \texorpdfstring{$B$}{B}-model integrals for \texorpdfstring{$\cH^D$}{HD} and \texorpdfstring{$\cF^D$}{FD}}\label{sec:B-model-integrals}

From \eqref{eq:KD-B-int-app}, we can immediately obtain $\cH^D$ and $\cF^D$ by integration \eqref{eq:q-vol-hierarchy}
\be\label{eq:HD-B-int-app}
 \cH^D(t,c_1,\e)
 = \Ga\Big(\sum_i\e_i\Big)\,\eu^{-t\cdot M\cdot\e}\,
 x_1^{A_1\cdot\e} \int_0^\infty \frac{\dif x_2}{x_2}
 \frac{x_2^{A_2\cdot\e}}{H(x_1,x_2,z)^{\sum_i\e_i}}\,,
\ee
and
\be
 \cF^D(t,\e)
 = \Ga\Big(\sum_i\e_i\Big)\,\eu^{-t\cdot M\cdot\e}
 \int_0^\infty\frac{\dif x_1}{x_1}\int_0^\infty\frac{\dif x_2}{x_2}
 \frac{x_1^{A_1\cdot\e}x_2^{A_2\cdot\e}}{H(x_1,x_2,z)^{\sum_i\e_i}}\,.
\ee
From the discussion of Appendix~\ref{sec:H-Hori--Vafa}, it follows that we can replace $H(x,y,z)$ by the mirror curve $F(x,y)$ of either brane defining the double cut (or of the brane defining the single cut associated to $\cH^D$).
The formula remains the same, except that $A_\alpha,M$ get shifted by some factors, see \eqref{eq:H-mirror-curves} for an example.

\section{Monodromy of \texorpdfstring{$F_D^{(n)}$}{FD(n)}}\label{app:Lauricella-D}

In this appendix we compute the regular (\ie{} order $O(\e^0)$) term in the monodromy of $\cH^D$. Schematically, if
\be
 \cH^D(t,c,\xi\e)
 = [\cH^D]_{-2}\,\xi^{-2} + [\cH^D]_{-1}\,\xi^{-1} + [\cH^D]_0\,\xi^0 + O(\xi)
\ee
represent different terms in the $\e$-expansion of $\cH^D$, the monodromy can be computed term by term
\be
 \Delta_c\,\cH^D(t,x,\xi\e)
 = \sum_{d=-2}^{\infty} \Delta_c\,[\cH^D(t,c,\e)]_d\,\xi^d\,.
\ee
Here $\Delta_c$ is the monodromy operator defined as in \eqref{eq:monodromy-operator}.
In particular, we will be interested in the regular terms $[\cH^D]_0$ and $\Delta_c\,[\cH^D]_0$.

\subsection{Non-equivariant expansion of the quantum Lebesgue measure}
\label{app:eps-exp-HD}

It follows from \eqref{eq:cHD-FD} that $\cH^D$ is the product of three distinct pieces,
each with a different leading order behavior
\be
\label{eq:cHD-expansion}
\begin{aligned}
 \cH^D(t,c,\e) = &
 \frac{\eu^{-t\cdot M\cdot\e}\,x^{A_1\cdot\e}}{H(x,0,z)^{\sum_i\e_i}}
 \times \frac{\Ga(\sum_i\e_i) \Ga(A_2\cdot\e)
 \Ga(k\sum_i\e_i-A_2\cdot\e)}{\Ga(k\sum_i\e_i)} \\
 & \times
 F^{(k)}_D\Big(A_2\cdot\e,\sum_i\e_i,\dots,\sum_i\e_i,k\sum_i\e_i;
 1+y_1^{-1},\dots,1+y_k^{-1}\Big)\,.
\end{aligned}
\ee
The first factor has the following expansion
\be
 \left[\frac{\eu^{-t\cdot M\cdot\e}\,x^{A_1\cdot\e}}{H(x,0,z)^{\sum_i\e_i}}\right]_n
 = \frac{(-1)^n}{n!} \Big(t\cdot M\cdot\e+c\,A_1\cdot\e+\log H(x,0,z)\sum_i\e_i\Big)^n
\ee
where we recall that $H(x,0,z)=\prod_{j=1}^k(-y_j(x,z))$.
Similarly, the ratio of $\Ga$ functions can be expanded as
\begin{multline}
 \frac{\Ga(\sum_i\e_i)\Ga(A_2\cdot\e)\Ga(k\sum_i\e_i-A_2\cdot\e)}{\Ga(k\sum_i\e_i)}
 = \underbrace{\frac{k}{A_2\cdot\e\left(k\sum_i\e_i-A_2\cdot\e\right)}}_{O(\e^{-2})}\;
 \underbrace{-\frac{\gamma k\sum_i\e_i}{A_2\cdot\e\left(k\sum_i\e_i-A_2\cdot\e\right)}}_{O(\e^{-1})} \\
 + \underbrace{\frac{k}{12}\left(\frac{(6\gamma^2+\pi^2)\left(\sum_i\e_i\right)^2}
 {A_2\cdot\e\left(k\sum_i\e_i-A_2\cdot\e\right)}-2\pi^2\right)}_{O(\e^0)} + O(\e)\,.
\end{multline}
The non-equivariant expansion of $F_D^{(k)}$ is more involved and relies on a choice of power series representation.
In the following, we will use the power series representation from \eqref{eq:FD-def} and expand to second order in the parameters $a,b_i,c$.
It is important to observe that use of this series expansion implies that the following results hold only when all the arguments $x_i$ are strictly contained within the unit disk.

First, observe that the Pochhammer symbols appearing in the coefficients can be expanded as
\be
 (x)_n = x(x+1)\cdots(x+i-1)
 = x \Ga(n) \left(1+xH_{n-1}+O(x^2)\right)\,,
 \hspace{30pt}
 n\geq 1
\ee
where
\be
 H_n = \sum_{i=1}^n \frac{1}{i}
\ee
is the $n$-th harmonic number.
Clearly, we have
\be
 \left[F^{(k)}_D\right]_0 = 1\,.
\ee
At order one in the parameters, the only contribution comes from the terms of the sum \eqref{eq:FD-def} for which only one of the indices $(i_1,\dots,i_k)$ is not zero, namely
\be
 \left[F^{(k)}_D\right]_1
 = \sum_{j=1}^k \sum_{i_j=1}^\infty
 \left[\frac{(a)_{i_j}(b_j)_{i_j}}{(c)_{i_j}\,i_j!}\right]_1 x_j^{i_j}
 = \sum_{j=1}^k \frac{ab_j}{c} \sum_{i_j=1}^\infty \frac{x_j^{i_j}}{i_j}
 = -\frac{a}{c}\sum_{j=1}^k b_j\log(1-x_j)\,.
\ee
At order two, there are two contributions: one coming from terms where one of the indices $(i_1,\dots,i_k)$ is not zero and one coming from terms where two indices are not zero, namely
\be
\begin{aligned}
 \left[F^{(k)}_D\right]_2
 &= \sum_{j=1}^k \sum_{i_j=1}^\infty
 \left[\frac{(a)_{i_j}(b_j)_{i_j}}{(c)_{i_j}\,{i_j}!}\right]_2 x_j^{i_j}
 + \sum_{1\leq j_1 < j_2\leq k} \sum_{i_{j_1},i_{j_2}=1}^\infty
 \left[\frac{(a)_{i_{j_1}+i_{j_2}}(b_{j_1})_{i_{j_1}}(b_{j_2})_{i_{j_2}}}
 {(c)_{i_{j_1}+i_{j_2}}\,i_{j_1}!i_{j_2}!}\right]_2
 x_{j_1}^{i_{j_1}} x_{j_2}^{i_{j_2}} \\
 &= \sum_{j=1}^k \frac{ab_j(a+b_j-c)}{c}
 \sum_{i_j=1}^\infty\frac{x_j^{i_j}}{i_j} H_{{i_j}-1}
 + \sum_{1\leq j_1 < j_2\leq k} \frac{ab_{j_1}b_{j_2}}{c}
 \sum_{i_{j_1}=1}^\infty \frac{x_{j_1}^{i_{j_1}}}{i_{j_1}}
 \sum_{i_{j_2}=1}^\infty \frac{x_{j_2}^{i_{j_2}}}{i_{j_2}} \\
 &= \sum_{j=1}^k \frac{ab_j(a+b_j-c)}{c} \frac12\log^2(1-x_j)
 + \sum_{1\leq j_1 < j_2\leq k} \frac{ab_{j_1}b_{j_2}}{c}
 \log(1-x_{j_1}) \log(1-x_{j_2}) \\
 &= \frac{a}{2c}
 \left( (a-c)\sum_{j=1}^k b_j\log^2(1-x_j)
 +\Big(\sum_{j=1}^k b_j\log(1-x_j)\Big)^2 \right)\,.
\end{aligned}
\ee
Putting all of this together and specializing the parameters and arguments of the Lauricella as
\be
 a=A_2\cdot\e\,,
 \hspace{30pt}
 b_j = \sum_i\e_i\,,
 \hspace{30pt}
 c = k\sum_i\e_i\,,
 \hspace{30pt}
 x_j = 1+y_j^{-1}\,,
\ee
we finally obtain the non-equivariant expansion of $\cH^D$ up to order zero in $\e$ as desired.

\subsection{Monodromy of the quantum Lebesgue measure}\label{app:monodromy-HD}

We now compute the monodromy of the regular part of $\cH^D$. In order to do so, we make use of the following observation: if we assume that the roots $y_j(x)$ behave as
\be
 y_j(x) \sim x^{\alpha_j}
\ee
for $x\to0$, then we can deduce that the monodromy of each $y_j(x)$ as a multivalued function of $x$ is the same as that of $x^{\alpha_j}$. Therefore we obtain
\be
 y_j(\eu^{-(c\pm\pi\ii)}) = \eu^{\mp\pi\ii\alpha_j} y_j(-\eu^{-c})\,.
\ee
Applying this result systematically to every term in the expansion of $\cH^D$,
we find that the only regular instantonic contributions to the monodromy come from terms of the form
\be
 \Delta_c\left(
 \left[\frac{\eu^{-t\cdot M\cdot\e}\,x^{A_1\cdot\e}}{H(x,0,z)^{\sum_i\e_i}}\right]_{n_1}
 \times \left[\frac{\Ga(\sum_i\e_i) \Ga(A_2\cdot\e)
 \Ga(k\sum_i\e_i-A_2\cdot\e)}{\Ga(k\sum_i\e_i)}\right]_{-2}
 \times \left[F^{(k)}_D\right]_{n_2}\right)
\ee
where $n_1,n_2$ are positive integers such that $n_1+n_2=2$, and the middle term in the product is just a constant on which $\Delta_c$ acts trivially.

From an explicit computation of the monodromy of these three terms, we find
\be
 \Delta_c\left(\left[\frac{\eu^{-t\cdot M\cdot\e}\,x^{A_1\cdot\e}}
 {H(x,0,z)^{\sum_i\e_i}}\right]_2
 \times \left[F^{(k)}_D\right]_0\right)
 = \sum_i\e_i\Big(A_1\cdot\e-\sum_i\e_i\sum_{j=1}^k\alpha_j\Big)
 \sum_{j=1}^k\log(-y_j(-x)) + \dots
\ee
\be
 \Delta_c\left(\left[\frac{\eu^{-t\cdot M\cdot\e}\,x^{A_1\cdot\e}}
 {H(x,0,z)^{\sum_i\e_i}}\right]_1
 \times \left[F^{(k)}_D\right]_1\right)
 = -\frac{A_2\cdot\e}{k} \Big(A_1\cdot\e-2\sum_i\e_i\sum_{j=1}^k\alpha_j\Big)
 \sum_{j=1}^k\log(-y_j(-x)) + \dots
\ee
\begin{multline}
 \Delta_c\left(\left[\frac{\eu^{-t\cdot M\cdot\e}\,x^{A_1\cdot\e}}{H(x,0,z)^{\sum_i\e_i}}\right]_0
 \times \left[F^{(k)}_D\right]_2\right)
 = \frac{A_2\cdot\e}{k}\Big(k\sum_i\e_i-A_2\cdot\e\Big)\sum_{j=1}^k\alpha_j\log(-y_j(-x)) \\
 -\frac{A_2\cdot\e}{k}\sum_i\e_i\Big(\sum_{j=1}^k\alpha_j\Big)\sum_{j=1}^k\log(-y_j(-x))
 + \dots
\end{multline}
Summing them all together and multiplying by the constant, we finally conclude that
\be
 \Delta_c\left[\cH^D\right]_0
 = \frac1{A_2\cdot\e} \left(A_1\cdot\e-\sum_i\e_i\sum_{j=1}^k\alpha_j\right)
 \sum_{j=1}^k\log(-y_j(-x))
 + \sum_{j=1}^k \alpha_j\log(-y_j(-x))
 + \dots
\ee
where, as usual, we omit dependence on $z$ in the roots and we neglect all terms which are polynomial in $t^a$ and $c$.

We should remark here that derivation of the formula for the monodromy of $\cH^D$ makes use of the explicit series expansion \eqref{eq:FD-def} for the Lauricella function $F_D^{(n)}$ in \eqref{eq:cHD-FD}, whose domain of convergence is the product of unit disks
\be\label{eq:validity}
    |1+y_j^{-1}| < 1 \qquad  j=1,\dots, k\,.
\ee
In particular, this condition must hold for the asymptotic values of the roots $y_j(x)$ within some region of the large volume regime, as in \eqref{eq:yi-asympt}.

Whenever condition \eqref{eq:validity} is not satisfied, we find ourselves in a region of parameters space where the series expansion \eqref{eq:FD-def} is no longer valid. In this case, we need to compute the series expansion of $F_D^{(n)}$ in the new region by means of analytic continuation formulas, see \eg{} \cite{Bezrodnykh_2018} and references therein.
Once an appropriate series expansion has been found, we can then repeat the steps in this Appendix to compute the correct formula for the monodromy compatible with the values of the parameters.
In \cite{Cassia:2023uvd}, we address one example when this issue arises in the case of local $\BP^2$, where the equivariant disk potential is computed in a phase in which \eqref{eq:validity} is violated. In that instance, the Lauricella function specializes to a simpler hypergeometric series, whose analytic continuations are well-known.

\printbibliography

\end{document}